%% file: Paper1.tex
\documentclass[a4paper]{article}
\usepackage{epsfig}
\usepackage{amssymb}
\usepackage{latexsym}
\usepackage{cite}
\begin{document}

\title{\bf {\Large  Reflexive Numbers and Berger Graphs from Calabi--Yau Spaces}}

\author{{L.~N.~Lipatov${}^{1,2}$, A.~Sabio~Vera\thanks{Alexander von Humboldt Research Fellow}~${}^{3}$,
V.~N.~Velizhanin${}^{1}$,
G.~G.~Volkov${}^{1,4,5,6}$}\\[4.5ex]
{\it $^1$ Petersburg Nuclear Physics Institute, Gatchina,
188 300 St. Petersburg, Russia}\\
{\it $^2$ Laboratoire de Physique Th{\'e}orique et Hautes Energies,}\\
{\it Universite Pierre et Marie Curie,}\\
{\it  4, place Jussieu, F - 75252 PARIS Cedex 05, France}  \\
{\it $^3$ II. Institut f{\"u}r Theoretische Physik, Universit{\"a}t Hamburg,} \\
{\it Luruper Chaussee 149, D--22761 Hamburg, Germany}\\
{\it $^4$ Instituto de F{\' \i}sica Te{\' o}rica UAM/CSIC, C--XVI, \&
Departamento de F{\' \i}sica Te{\' o}rica,}\\
{\it C--XI, Universidad Aut{\' o}noma de Madrid, E--28049, Madrid, Spain}\\
{\it $^5$ Laboratoire d'Annecy--Le--Vieux de Physique Th{\' e}orique,}\\
 {\it  LAPTH (CNRS UMR 5108), B.P. 110, Annecy--Le--Vieux, F-74941,
France,}\\
{\it $^6$ Theory Division, Physics Department, CERN, 1211 Geneva 23,
Switzerland}}

\maketitle

\vspace{-13cm}
\begin{flushright}
{\small CERN--PH--TH/2005--XXX\\
DESY--05--XXX\\
IFT--UAM/CSIC--05--XXX\\
LAPTH--1082/04\\
PNPI--TH--/2005--XXX}
\end{flushright}

\vspace{10cm}
\begin{abstract}
We review the  Batyrev approach to Calabi-Yau spaces based on 
reflexive weight vectors. The Universal ${\rm CY}$ algebra
gives a possibility to construct the corresponding reflexive
numbers in a recursive way. A physical interpretation of the
Batyrev expression for the Calabi-Yau manifolds is presented.
Important classes of these manifolds are related to the
simple-laced and quasi-simple-laced numbers. We discuss the
classification and recurrence relations for them in the framework of
quantum field theory methods. A relation between the reflexive
numbers and the so-called Berger graphs is studied.
In this correspondence the r{\^ o}le played by the generalized
Coxeter labels is highlighted. {Sets of positive roots are
investigated in order to connect them to possible new algebraic
structures stemming from the Berger matrices}.

\end{abstract}

\input{Introduction.tex}

\input{FirstPart.tex}

\input{SecondPart.tex}

\input{Conclusions.tex}

\subsubsection*{Acknowledgments}
\label{sec:acknowledgements} We are grateful to U Aglietti, E {\'
A}lvarez, L {\' A}lvarez--Gaum{\' e}, I Antoniadis, P Aurenche, J
Bartels, G B{\'e}langer, J Ellis, B Gavela, G Girardi, C
G{\'o}mez, R Kerner, P Sorba and JB Zuber for support and helpful
discussions. We also would like to thank the Laboratoire
d'Annecy--Le--Vieux de Physique Th{\'e}orique (LAPTH), DESY,
University of Hamburg, CERN, Universidad Aut{\' o}noma de Madrid,
Laboratoire de Physique Th{\'e}orique et Hautes Energies (LPTHE)
and Petersburg Nuclear Physics Institute for their hospitality.
This work was supported by the Alexander von Humboldt Foundation
and by INTAS and RFBR grant 04-02-17094, RSGSS-1124.2003.2.

\input{Appendix.tex}

\input{Bibliography.tex}
\end{document}

%% file: Introduction.tex
\section{Introduction}
The Calabi-Yau manifolds belong to an interesting class of the Riemann spaces
\cite{Berger, Cal, Yau, Hub, Mart}. They are used in physics for the
compactification of extra dimensions. For example, the proof of the duality
between the string theories IIA, IIB and the heterotic $E_8 \times 
E_8$ model was based on such compactifications. The CY spaces 
are related to the Lie 
and Kac-Moody algebras. New symmetries based on ternary, quaternary, etc 
operations are investigated now in physics and mathematics  
 \cite{Nambu,  Takhtajan, Filippov, Kern, CIZ, FZ,
 Traubenberg} and  
appear also  in the CY$_n$ geometry \cite{V, TV, ETV, LSVV}. They
will be discussed also below in terms of the so-called Berger graphs.
Our goal here is to study the  reflexivity property of CY$_n$ spaces
through the theory of numbers, recurrence relations and 
quantum field theory methods.


A CY space can be realized as an algebraic variety ${\mathcal M}$
in a weighted projective space
${\rm CP}^{n-1} (\overrightarrow{k})$ \cite{Hub, Mart}
where the weight vector reads
$\overrightarrow{k} = (k_1, \dots, k_{n})$. This variety is defined by
\begin{eqnarray}
{\mathcal M} \equiv (\left\{x_1, \dots, x_{n} \right\} \in
{\rm CP}^{n-1} (\overrightarrow{k}):
{\mathcal P} (x_1, \dots, x_{n}) \equiv
\sum_{\scriptsize{\overrightarrow{m}}} c_{\scriptsize{\overrightarrow{m}}}
x^{\scriptsize{\overrightarrow{m}}} = 0 ),
\label{Batyr}
\end{eqnarray}
i.e., {as} the zero locus of a quasi--homogeneous polynomial of
degree $d_k = \sum_{i=1}^{n} k_i$, with the monomials being
$x^{\scriptsize{\overrightarrow{m}}} \equiv x_1^{m_1} \cdots x_{n}^{m_{n}}$. The points in
CP$^{n-1}$ satisfy the property of projective invariance
$\left\{x_1, \dots, x_{n}\right\} \approx
\left\{\lambda^{k_1} x_1, \dots, \lambda^{k_{n}} x_{n}\right\}$ leading to the
constraint
$\overrightarrow{m} \cdot \overrightarrow{k} = d_k$. The sum in the above
expression is performed over all solutions $\overrightarrow{m}$ of this equation and {the} coefficients
$ c_{\scriptsize{\overrightarrow{m}}}$ are arbitrary complex numbers. 

The most important   constraint 
for a CY
candidate is
the condition of reflexivity of the vector $\overrightarrow{k}$, which can 
be defined
in terms of the Batyrev reflexive polyhedra \cite{Bat}.
Let us consider this condition in more detail. We can construct
the vector
$\overrightarrow{m}' = \overrightarrow{m} - \overrightarrow{1}$ where $\overrightarrow{m}' \cdot
\overrightarrow{k} = 0$. Then it is convenient to define the lattice
\begin{eqnarray}
\Lambda &=& \left\{\overrightarrow{m}' \in {\mathbb Z}^{n}: \overrightarrow{m}' \cdot \overrightarrow{k} = 0\right\}
\end{eqnarray}
with basis $\left\{e_i\right\}$. The dual of this lattice, $\Lambda^*$, has
the basis $\left\{e_i^*\right\}$ with the orthonormality condition
$e_i \cdot e_j^* = \delta_{ij}$. We define a polyhedron $\Delta$
as the convex hull of the lattice $\Lambda$ and the polyhedron
$\Delta^*$ as the convex hull of the dual lattice $\Lambda^*$.
The reflexivity condition means that the polyhedron $\Delta$ is integer, its
origin $\overrightarrow{0}$ is the only interior point, and its dual $\Delta^*$ is
also integer and contains only one interior point.
In this case the vector $\overrightarrow{k}$ is considered to be reflexive.
Using this criteria of reflexive vectors Batyrev proved the Mirror duality
of CY$_{3,4}$ spaces~\cite{Bat}. Namely, for each CY, ${\mathcal M}$, there exists a Mirror
CY partner, ${\mathcal M}^*$. This symmetry helped establish the duality 
between the
type IIA and IIB string theories.

The correspondence between CY spaces and reflexive polyhedra led the way for
their classification. In particular, for the case CY$_2 = K_3$
the 4319
three--dimensional polyhedra were found in Ref.~\cite{KS1, KS2}.
Among them, 95 can be
described with a single reflexive vector. The algorithm,
constructed in Ref.~\cite{KS1, KS2}, {generated}  
473~800~776 reflexive four--dimensional
polyhedra in the case of CY$_3$ spaces. A subclass of this large number can
be described by one reflexive vector. Namely, 184~026 polyhedra
belong to this subclass~\cite{KS1,KS2} (see also~\cite{AENV1, AENV2, AENV3, AENV4}).

Recently, an alternative to this classification was developed
using some properties of the theory of reflexive vectors. This new approach
was named ``Universal Calabi--Yau Algebra'' (UCYA)\cite{AENV1}.
One of its main
results is that all reflexive vectors of
dimension $n$ can be obtained from the reflexive vectors with lower
dimension $1, \dots, n-1$. {Consequently}, every reflexive vector of dimension $n$ can be 
constructed
from the simplest reflexive vector $\overrightarrow{k}=(1)$.
The key observation to realize  this program was to
use the concept of the $r$--arity composition law (with $r = 2, \dots , n$)
for the subclass
which can be described by a unique reflexive vector
(in general CY spaces this subclass corresponds to the so--called level one).
Using this composition law, it was shown how the
level one CY$_n$ space can be
obtained from its slices of lower dimension $r = 1, \dots n-1$, generating
in this way the $r$--arity slice classification. For
example, the 2--arity
composition law in $K_3$ space gives us 90 out of the 95 reflexive vectors. 
These 90 vectors were unified in 22 chains having
the same CY$_1$-slice in $K_3$. 
Four of the remaining reflexive weight vectors can
be obtained with the $3$--arity and the last one with the $4$--arity. In CY$_3$ a
similar $2$--arity classification produces 4242 chains having the same
$K_3$ slice of CY$_3$ \cite{AENV1,AENV2}.

To be more specific, the 22 chains in $K_3$ are generated by taking the
reflexive vectors (1), (1,1), (1,1,1), (1,1,2) and (1,2,3) and extending them
to dimension 4 by including additional zero components, i.e.,
(0,0,0,1), (0,0,1,1), (0,1,1,1), (0,1,1,2), (0,1,2,3) with all permutations of
their components.
In the 2--arity construction one should take all possible pairs of two
extended vectors and select those ``good pairs'' which have
a reflexive polyhedron in the intersection of
the corresponding slices $\overrightarrow{m}$. 
The selected vectors can be added with integer
coefficients.
For example, from 50 possible extended vectors we can take the
pair $\overrightarrow{k}_a^{\rm ext}=(0,1,1,1)$ and 
$\overrightarrow{k}_b^{\rm ext} = (1,0,0,0)$.
Their intersection is defined as the
solution of two constraints $\overrightarrow{m} \cdot
\overrightarrow{k}_a^{\rm ext} = d_{k_a} = 3$ and $\overrightarrow{m} \cdot
\overrightarrow{k}_b^{\rm ext} = d_{k_b} = 1$. These equations can be also
written as
\begin{eqnarray}
\Lambda &=& \left\{\overrightarrow{m}' \in {\mathbb Z}^{4}: \overrightarrow{m}' \cdot \overrightarrow{k}_a^{\rm ext}
= \overrightarrow{m}' \cdot \overrightarrow{k}_b^{\rm ext} = 0 \right\},
\end{eqnarray}
where $\overrightarrow{m}' = \overrightarrow{m} - \overrightarrow{1}$. The
lattice of solutions for  $\overrightarrow{m}'$
corresponds to a two dimensional reflexive polyhedron or CY$_1$, and, therefore,
according to UCYA, the linear combination
\begin{eqnarray}
m_a \overrightarrow{k}_a^{\rm ext} + m_b \overrightarrow{k}_b^{\rm ext}
\end{eqnarray}
with $m_a \leq d_{k_b}$ and $m_b \leq d_{k_a}$ of these two vectors forms
the chain with the eldest vector
having unit coefficients $m_a = m_b = 1$. 

Another simple example is the
combination of two vectors, $\overrightarrow{k}_m$ with dimension $m$ and
$\overrightarrow{k}_n$ with dimension $n$,
in the form $(\overrightarrow{k}_m,\overrightarrow{0}_n)
+(\overrightarrow{0}_m,\overrightarrow{k}_n)$
which is
always an eldest reflexive vector of dimension $m+n$. In particular, 
by adding $\overrightarrow{k}_b=(0,1,1,1)$ and $\overrightarrow{k}_a=(1,0,0,0)$
with certain coefficients, we obtain the three vectors
(1,1,1,1), (2,1,1,1) and (3,1,1,1). The intersection of the slices for
the vectors $\overrightarrow{k}_b$ and $\overrightarrow{k}_a$
produces a two-dimensional
reflexive slice. This slice  divides the corresponding three-dimensional $K_3$
polyhedrons in two parts. On the left and right sides of the slice the
set of points at the edges forms affine Dynkin diagrams. For example,
in the reflexive polyhedra corresponding to
the vector (1,1,1,1) from the (1,0,0,0)
side we obtain the diagram for the algebra $A_{11}^{(1)}$ and from the (0,1,1,1) side
we have the graph of
$E_6^{(1)}$. The (2,1,1,1) and (3,1,1,1) members of this chain contain different 
Dynkin graphs. This property is
universal
and valid for all 22 chains. It is a  generalization of the results
of Candelas and Font~\cite{CF,Bersh,PS,KV} who found a dictionary
for the Dynkin graphs of the Cartan--Lie
algebra in the  case of the Weierstrass slice using the type IIA and
heterotic $E_8 \times E_8$ string duality. In the $K_3$ case a
correspondence between extended reflexive vectors and Dynkin graphs was
found~\cite{AENV1, AENV2, AENV4}, for
example,
\begin{eqnarray}
(0,0,0,1) &\rightarrow& {\rm A}_r^{(1)} \nonumber\\
(0,0,1,1) &\rightarrow& {\rm D}_r^{(1)} \nonumber\\
(0,1,1,1) &\rightarrow& {\rm E}_6^{(1)} \\
(0,1,1,2) &\rightarrow& {\rm E}_7^{(1)} \nonumber\\
(0,1,2,3) &\rightarrow& {\rm E}_8^{(1)}.\nonumber
\label{links}
\end{eqnarray}
\begin{table}
\centering
\scriptsize
\vspace{.05in}
\begin{tabular}{|l|l|l||l|l|l|}
\hline
$   { \aleph} $&$ { {\vec k_{5ex}}^{(i)}} $&${ G(Gal)}
$&$ { \aleph} $&$ { {\vec k_{5ex}}^{(i)}} $&${ G(Gal)}$\\
\hline\hline
$   { i}      $&$  {\bf  (0, 0, 0, 0, 1)[1]}          $&$ 5
$&$ { 46}     $&$  {    (0, 2, 3, 4, 7)[16]}          $&$ 120
$\\ \hline
$   { ii}     $&$  {\bf  (0, 0, 0, 1, 1)[2]}          $&$ 10
$&$ { 47}     $&$  {    (0 ,2 ,3 ,4 ,9)[18]}          $&$ 120
$\\ \hline
$   { iii}    $&$  {\bf  (0, 0, 1, 1, 1)[3]}          $&$ 10
$&$ { 48}     $&$  {    (0 ,2 ,3 ,5, 5)[15]}          $&$ 60
$\\ \hline
$   { iv}     $&$  {\bf  (0, 0, 1, 1 ,2)[4]}          $&$ 30
$&$ { 49}     $&$  {    (0 ,2 ,3 ,5 ,7)[17]}          $&$ 120
$\\ \hline
$   { v}      $&$  {\bf  (0 ,0 ,1 ,2 ,3)[6]}          $&$ 60
$&$ { 50}     $&$  {    (0 ,2 ,3 ,5 ,8)[18]}          $&$ 120
$\\ \hline
$   { 1}      $&$  {\bf  (0 ,1 ,1 ,1 ,1)[4]}          $&$ 5
$&$ { 51}     $&$  {   (0 ,2 ,3 ,5 ,10)[20]}          $&$ 120
$\\ \hline
$   { 2}      $&$  {     (0 ,1 ,1 ,1 ,2)[5]}          $&$ 20
$&$ { 52}     $&$  {    (0 ,2 ,3 ,7 ,9)[21]}          $&$ 120
$\\ \hline
$   { 3}      $&$  {\bf  (0, 1, 1, 1, 3)[6]}          $&$ 20
$&$ { 53}     $&$  {   (0 ,2 ,3 ,7 ,12)[24]}          $&$ 120
$\\ \hline
$   { 4}      $&$  {\bf  (0 ,1 ,1 ,2 ,2)[6]}          $&$ 30
$&$ { 54}     $&$  {   (0 ,2 ,3 ,8 ,11)[24]}          $&$ 120
$\\ \hline
$   { 5}      $&$  {     (0 ,1 ,1 ,2 ,3)[7]}          $&$ 60
$&$ { 55}     $&$  {    (0 ,2 ,3 ,4 ,7)[16]}          $&$ 120
$\\ \hline
$   { 6}      $&$  {\bf  (0 ,1 ,1 ,2 ,4)[8]}          $&$ 60
$&$ { 56}     $&$ {\bf (0 ,2 ,3 ,10 ,15)[30]}         $&$ 120
$\\ \hline
$   { 7}      $&$  {     (0 ,1 ,1 ,3 ,4)[9]}          $&$ 60
$&$ { 57}     $&$  {     (0 ,2 ,4 ,5 ,9)[20]}         $&$ 120
$\\ \hline
$   { 8}      $&$  {     (0 ,1 ,1 ,3 ,5)[10]}         $&$ 60
$&$ { 58}     $&$  {    (0 ,2 ,4 ,5 ,11)[22]}         $&$ 120
$\\ \hline
$   { 9}      $&$  {\bf  (0, 1, 1 ,4 ,6)[12]}         $&$ 60
$&$ { 59}     $&$  {     (0 ,2 ,5 ,6 ,7)[20]}         $&$ 120
$\\ \hline
$   { 10}     $&$  {      (0 ,1 ,2 ,2 ,3)[8]}         $&$ 60
$&$ { 60}     $&$  {     (0 ,2 ,5 ,6 ,13)[26]}         $&$ 120
$\\ \hline
$   { 11}     $&$  {\bf  (0 ,1 ,2 ,2 ,5)[10]}         $&$ 60
$&$ { 61}     $&$  {    (0 ,2 ,5 ,9 ,11)[27]}         $&$ 120
$\\ \hline
$   { 12}     $&$  {      (0 ,1 ,2 ,3 ,3)[9]}         $&$ 60
$&$ { 62}     $&$  {    (0 ,2 ,5 ,9 ,16)[32]}         $&$ 120
$\\ \hline
$   { 13}     $&$  {     (0 ,1 ,2, 3, 4)[10]}         $&$ 120
$&$ { 63}     $&$  {   (0 ,2 ,5 ,14 ,21)[42]}         $&$ 120
$\\ \hline
$   { 14}     $&$  {    (0 ,1 ,2 ,3 ,5)[11]}           $&$    120
$&$ { 64}     $&$  {    (0, 2 ,6 ,7 ,15)[30]}          $&$    120
$\\ \hline
$   { 15}     $&$  {\bf    (0 ,1 ,2 ,3 ,6)[12]}           $&$    120
$&$ { 65}     $&$  {    (0 ,3 ,3 ,4 ,5)[15]}           $&$    60
$\\ \hline
$   { 16}     $&$  {    (0, 1 ,2 ,4 ,5)[12]}           $&$    120
$&$ { 66}     $&$  {    (0 ,3 ,4 ,5 ,6)[18]}           $&$    120
$\\ \hline
$   { 17}     $&$  {    (0 ,1 ,2 ,4, 7)[14]}           $&$    120
$&$ { 67}     $&$      {(0, 3, 4, 5, 7)[19]}           $&$    120
$\\ \hline
$   { 18}     $&$  {    (0 ,1 ,2 ,5 ,7)[15]}           $&$    120
$&$ { 68}     $&$  {    (0 ,3 ,4 ,5 ,8)[20]}           $&$    120
$\\ \hline
$   { 19}     $&$  {    (0 ,1 ,2 ,5 ,8)[16]}           $&$    120
$&$ { 69}     $&$  {   (0 ,3 ,4 ,5 ,12)[24]}           $&$    120
$\\ \hline
$   { 20}     $&$  {\bf (0 ,1 ,2 ,6 ,9)[18]}           $&$    120
$&$ { 70}     $&$  {   (0 ,3 ,4 ,7 ,10)[24]}           $&$    120
$\\ \hline
$   { 21}     $&$  {\bf    (0 ,1 ,3 ,4, 4)[12]}           $&$    60
$&$ { 71}     $&$  {   (0 ,3 ,4 ,7, 14)[28]}           $&$    120
$\\ \hline
$   { 22}     $&$  {    (0 ,1 ,3 ,4 ,5)[13]}           $&$    120
$&$ { 72}     $&$  {  (0 ,3 ,4 ,10 ,13)[30]}           $&$    120
$\\ \hline
$   { 23}     $&$  {    (0 ,1 ,3 ,4 ,7)[15]}           $&$    120
$&$ { 73}     $&$  {  (0 ,3 ,4 ,10 ,17)[24]}           $&$    120
$\\ \hline
$   { 24}     $&$  {    (0 ,1 ,3 ,4 ,8)[16]}           $&$    120
$&$ { 74}     $&$  {   (0 ,3, 4 ,11 ,18)[36]}          $&$    120
$\\ \hline
$   { 25}     $&$  {    (0, 1, 3, 5, 6)[15]}           $&$    120
$&$ { 75}     $&$  {  (0 ,3 ,4 ,14 ,21)[42]}           $&$    120
$\\ \hline
$   { 26}     $&$  {    (0 ,1, 3, 5, 9)[18]}           $&$    120
$&$ { 76}     $&$  {    (0 ,3 ,5 ,6 ,7)[21]}           $&$    120
$\\ \hline
$   { 27}     $&$  {   (0 ,1, 3, 7 ,10)[21]}           $&$    120
$&$ { 77}     $&$  {  (0 ,3 ,5 ,11 ,14)[33]}           $&$    120

$\\ \hline
$   { 28}     $&$  {   (0, 1 ,3 ,7 ,11)[22]}           $&$    120
$&$ { 78}     $&$  {  (0 ,3 ,5 ,11 ,19)[38]}           $&$    120
$\\ \hline
$   { 29}     $&$ {\bf (0 ,1 ,3 ,8 ,12)[24]}           $&$    120
$&$ { 79}     $&$ {   (0 ,3 ,5 ,16 ,24)[48]}           $&$    120
$\\ \hline
$   { 30}     $&$ {     (0 ,1 ,4 ,5 ,6)[16]}           $&$    120
$&$ { 80}     $&$ {     (0 ,3 ,6 ,7 ,8)[24]}           $&$    120
$\\ \hline
$   { 31}     $&$ {\bf (0 ,1 ,4 ,5 ,10)[20]}           $&$    120
$&$ { 81}     $&$  {    (0 ,4 ,5 ,6 ,9)[24]}           $&$    120
$\\ \hline
$   { 32}     $&$  {    (0 ,1 ,4 ,6 ,7)[18]}           $&$    120
$&$ { 82}     $&$  {   (0 ,4 ,5 ,6 ,15)[30]}           $&$    120
$\\ \hline
$   { 33}     $&$  {   (0 ,1 ,4 ,6 ,11)[22]}           $&$    120
$&$ { 83}     $&$  {    (0 ,4 ,5 ,7 ,9)[25]}           $&$    120
$\\ \hline
$   { 34}     $&$  {   (0 ,1 ,4 ,9 ,14)[28]}           $&$    120
$&$ { 84}     $&$  {   (0 ,4 ,5 ,7 ,16)[32]}           $&$    120
$\\ \hline
$   { 35}     $&$  {  (0 ,1 ,4 ,10, 15)[30]}           $&$    120
$&$ { 85}     $&$  {  (0 ,4 ,5 ,13 ,22)[44]}           $&$    120
$\\ \hline
$   { 36}     $&$  {    (0 ,1 ,5 ,7 ,8)[21]}           $&$    120
$&$ { 86}     $&$  {  (0 ,4 ,5 ,18 ,27)[54]}           $&$    120
$\\ \hline
$   { 37}     $&$ {    (0 ,1 ,5 ,7 ,13)[26]}           $&$    120
$&$ { 87}     $&$  {   (0 ,4 ,6 ,7 ,11)[28]}           $&$    120
$\\ \hline
$   { 38}     $&$  {  (0 ,1 ,5 ,12, 18)[36]}           $&$    120
$&$ { 88}     $&$  {   (0 ,4 ,6 ,7 ,17)[34]}           $&$    120
$\\ \hline
$   { 39}     $&$  {    (0 ,1 ,6 ,8, 9)[24]}           $&$    120
$&$ { 89}     $&$  {    (0 ,5 ,6 ,7 ,9)[27]}           $&$    120
$\\ \hline
$   { 40}     $&$  {   (0 ,1 ,6, 8, 15)[30]}           $&$    120
$&$ { 90}     $&$  {   (0 ,5, 6 ,8 ,11)[30]}           $&$    120
$\\ \hline
$   { 41}     $&$ {\bf (0 ,1, 6, 14, 21)[42]}          $&$    120
$&$ { 91}     $&$ {    (0 ,5 ,6 ,8 ,19)[38]}           $&$    120
$\\ \hline
$   { 42}     $&$  {    (0 ,2 ,2 ,3 ,5)[12]}           $&$    60
$&$ { 92}     $&$  {  (0 ,5 ,6 ,22 ,33)[66]}           $&$    120
$\\ \hline
$   { 43}     $&$  {    (0 ,2 ,2, 3, 7)[14]}           $&$    60
$&$ { 93}     $&$  {   (0 ,5 ,7 ,8 ,20)[40]}           $&$    120
$\\ \hline
$   { 44}     $&$ { \bf (0 ,2 ,3 ,3 ,4)[12]}           $&$    60
$&$ { 94}     $&$  {  (0 ,7 ,8 ,10 ,25)[50]}           $&$    120
$\\ \hline
$   { 45}     $&$     {(0, 2, 3, 4, 5 )[14]}          $&$    120
$&$ { 95}    $&$  {    (0 ,7 ,8 ,9 ,12)[36]}          $&$    120
$\\ \hline
\end{tabular}
\normalsize
\caption{The 100  distinct types of five--dimensional
extended projective vectors used to construct CY$_3$ spaces. The
order of their  permutation symmetry groups is also shown.
 Including these permutations, the total number of extended vectors is 10~270.
The simply--laced vectors (1+1+3+14=19) are highlighted with bold face.}
\label{Table100}
\end{table}

Note that the maximal Coxeter label of the graphs
at the right hand side of this correspondence coincides with the degree of the
reflexive vectors at the left hand side. We shall discuss this point
later. Our scheme offers the possibility of
constructing new graphs for CY spaces in any dimension. For example, in
$K_3$ all 4242 graphs for the reflexive numbers of the level one 
can be obtained following the {above--mentioned} approach~\cite{AENV1}.
Our analysis allows to classify the structure
of these large CY spaces in terms of number theory
and to construct the Berger graphs.
These Berger graphs might correspond to unknown
symmetries lying beyond Cartan--Lie algebras~\cite{V}.

The number of algebraic CY$_n$ varieties is very large and grows very
rapidly
with the dimension $n$ of the space. A similar situation occurs
with the number of reflexive weight vectors. For example, the number of
eldest reflexive vectors of 2--arity  is  1, 2, 22 and 4242~\cite{AENV1}
for dimensions $n=2,\,3,\,4$ and $5$, respectively.
To obtain the last number 4242 for n=5 using the arity construction
we need 100  extended reflexive vectors (and all permutations of their
components) (see Table~\ref{Table100}). 
An important remark
is that all reflexive weight vectors can be considered as
new types of numbers because in the framework of UCYA
the arithmetic of their adding and subtracting is well defined.
In the ``tree'' classification of CY spaces the trunk line of
the reflexive weight
numbers corresponds to those with unit components, i.e., (1), (1,1),
$(1,1,1), \dots$.
An interesting wider subclass is the so--called ``simply--laced'' numbers.
A simply--laced number $\overrightarrow{k}= (k_1, \cdots, k_n)$
with degree $d = \sum_{i=1}^{n} k_i$ is defined such that
\begin{eqnarray}
\frac{d}{k_i} &\in& {\mathbb Z}^+ \, \, {\rm and} \, \, d > k_i.
\end{eqnarray}
For these numbers there is a simple way of constructing the corresponding
affine Dynkin and Berger graphs together with their Coxeter labels. The
Cartan and Berger matrices of these graphs are symmetric. In the well known
Cartan case they correspond to the $ADE$ series of simply--laced algebras.
In dimensions $n = 1, 2, 3$ the numbers $(1),(1,1),(1,1,1),(1,1,2),(1,2,3)$
are simply--laced. For $n=4$ among all 95 
reflexive numbers 14 are simply--laced, as it
is shown in Table~\ref{Table100}~\cite{V,TV, ETV}. The remaining 81
correspond to the so-called quasi--simply--laced case. Before constructing these
graphs in the next section we proceed to review the concept of reflexivity
and relate it to techniques used in the functional approach {to Quantum 
Field Theory}.

%% file: FirstPart.tex
\section{From Reflexive Numbers to Quantum Field Theory Methods}

In this section we reconsider for CY spaces the condition of
reflexivity proposed firstly by Batyrev~\cite{Bat}. We shall do it
in a new approach where the reflexive numbers are studied starting
from the simply--laced case with a subsequent generalization to
quasi--simply--laced cases. The properties of these reflexive
numbers turn out to be very interesting.

\subsection{Geometrical Construction for Reflexivity}
Let us start by recalling the definition of the degree $d_k$ of a weight vector
$\overrightarrow{k}$
\begin{eqnarray}
d_k &=& \sum_{i=1}^{n} k_i\,,
\label{reflvectdk}
\end{eqnarray}
where $k_i$ are positive integer numbers.
It is convenient to normalize this vector as follows
\begin{equation}
l_i \equiv k_i / d_k \,.
\end{equation}
Then Eq.~(\ref{reflvectdk}) looks simpler
\begin{equation}
\sum_{i=1}^{n}l_{i}=1\,,\,\,l_i<1\,.
\label{ls}
\end{equation}
The numbers $l_{i}$ are positive regular ratios. An additional
constraint stems from the existence of solutions of the equation
$\overrightarrow{m} \cdot \overrightarrow{k} = d_k$ which now
reads
\begin{eqnarray}
\overrightarrow{m} \cdot \overrightarrow{l} &=& 1.
\label{ml1}
\end{eqnarray}
This last equation has the solutions for the monoms
$\left\{\overrightarrow{m}_{i}\right\}$ and an independent
 set of them with $i=1,2,...,n$
can be written as a matrix $M$. At this point let us note that
$\overrightarrow{m} = \overrightarrow{1}$ ($1_i=1$ for
$i=1,2,...,n$) also satisfies Eq.~(\ref{ml1}) due to
Eq.~(\ref{ls}). Relation (\ref{ml1}) for $\overrightarrow{m}_i$
can be presented as the set of equations
\begin{equation}
(M\,l)_{i}\,=1_{i}\,,
\label{Ml1}
\end{equation}
where $M$ is a $n\times n$ matrix constructed from the non--negative integer
numbers $m_{ij}$
\begin{equation}
(M\,)_{ij}\,=m_{ij}\,\ (m_{ij}=0,1,2,...)\,.
\end{equation}

The vectors $\overrightarrow{m}_{i}$ for $i=1,2,...,n$ correspond to $n$
points in the $n$--dimensional space with non--negative integer components $%
m_{ij}$. These vectors are considered to be linearly independent,
which means that
\begin{equation}
\det M \ne 0 \,.\label{det}
\end{equation}
All possible non-negative integer vectors $\overrightarrow{m}$
produce a lattice. {For a given $\overrightarrow{l}$ all the
solutions to Eq.~(\ref{ml1}) define a slice of this lattice}. {The
subset of them $\left\{\overrightarrow{m}_{i}\right\}$ for
$i=1,2,...,n$ entering in $M$ can be chosen in such a way that
other solutions can be obtained as their linear combinations with
non--negative coefficients}. The slice $\overrightarrow{m}$ has
the property of reflexivity provided that the vector
$\overrightarrow{1}$ is inner and all other points are on its
boundary. For our choice of vectors
$\left\{\overrightarrow{m}_{i}\right\}$ the slice will be
reflexive if $\overrightarrow{1}$ can be expanded as a linear
combination of $\left\{\overrightarrow{m}_{i}\right\}$
\begin{equation}
\overrightarrow{1}=\sum _{i=1}^nc_i\,\overrightarrow{m}_{i}
\label{cmi}
\end{equation}
{with positive coeficients $c_i$}. Note that according to eq.
(\ref{ml1}) these coefficients satisfy the constraints
\begin{equation}
\sum_{i=1}^n c_i=1\,,\,\,c_i>0
\label{ci}
\end{equation}
and therefore the vector $c_i$ is analogous to the vector $l_j$.
This implies that we can construct the dual slice
$\overrightarrow{\widetilde{m}}$ obeying the equation
\begin{equation}
\overrightarrow{\widetilde{m}}\overrightarrow{c}=1\,.
\end{equation}
Due to Eq. (\ref{cmi}) the vectors $\widetilde{m}_i^{(j)}=m_{ij}$
can be chosen as a basis of the dual lattice. In this case
relation (\ref{Ml1}) due to $l_j>0$ provides the Batyrev
reflexivity condition for the dual polyhedron. For each vector
$l_j$ we can construct several matrices $M$ satisfying condition
(\ref{cmi}) and therefore there is a recurrence procedure to
generate new reflexive vectors $c_i$ starting for example from the
simplest vector $l_j=1/n$.

To formulate the reflexivity requirement on $\overrightarrow{l}$
it is convenient to extend the matrix $M$ with the matrix elements
$M_{ij}$ ($i,j=1,2,...n$) adding to it one column and one row
composed from units
\begin{equation}
{\it M'}_{\mu \nu}=M_{ij}\,\delta _{\mu i}\,\delta _{\nu j}+\delta
_{\mu 0}+\delta _{\nu 0}-\delta _{\mu 0}\,\delta _{\nu 0} \,.
\label{defM'}
\end{equation}
For each matrix $M$  we obtain the following set of equations for
the reflexive weight vectors $l_{\nu}^{r}$ and $l_{\mu}^{s}$ ($\mu
,\nu =0,1,2,...n$) with integer positive components
\begin{equation}
{\it M'}_{\mu \nu}l_{\nu}^{r}={\it M'}_{\mu \nu}l_{\mu}^{s}=0\,.
\label{M'eq}
\end{equation}
For their self-consistency we should impose on the matrix elements
$M_{ij}$ the condition
\begin{equation}
\det {\it M'}=0.
\end{equation}
Clearly the existence of solutions of eqs (\ref{M'eq}) imposes
even more severe constraints on $M$ related to the positivity
conditions for $l_k$ ($k=1,2,...n$). Together with conditions
(\ref{det}) this means that the components $l_{\mu}$ for $\mu
=1,2,...n$ can be expressed in terms of $l_0=-1$ in such way that
equations (\ref{ls}) and (\ref{ml1}) are fulfilled.

 For a reflexive slice we can choose
the vectors $\overrightarrow{m}_{l}$ in such way that one of
their projections is zero. For example,
\begin{equation}
\overrightarrow{m}_{l}=(m_{l}^1,
m_{l}^2,...,m_l^{n-l}=0,...,m_{l}^n)\,. \label{boundar}
\end{equation}
Consequently the reflexivity condition imposes an additional constraint
on the weight vector $\overrightarrow{k}$: The existence of solutions
of the equation
\begin{equation}
\sum _{i \ne n-l} m^{i}_l\,k_i=d_k
\end{equation}
for all $l=1,2,...,n$.

Let us express $M$ as a product of 2 matrices
\begin{equation}
M=G\,\lambda\,,
\label{decM}
\end{equation}
where $\lambda$ is diagonal, $\lambda _{ij}=\delta _{ij}\,\lambda _i$,
and the matrix elements of $G$ satisfy the requirement
\begin{equation}
\sum _{j=1}^nG_{ij}=1_i\,,
\end{equation}
which means that $\overrightarrow{1}$ is an eigenvector of $G$ with its
eigenvalue equal to unity.
The {above--mentioned} decomposition of $M$ is unique for any
non--degenerate matrix $M$. Indeed, we have the relation
\begin{equation}
G_{ij}=M_{ij}/\lambda _j \,,
\end{equation}
where $1/\lambda _i$ can be found from the system of $n$ linear equations
\begin{equation}
\sum _{j=1}^n M_{ij}\,(1/\lambda _j)=1_i\,.
\end{equation}

The hyperplane corresponding to the slice generated by
$\overrightarrow{m}_i$ crosses the
coordinate axes at the rational points $\lambda _i$.
By comparing with the constraint in Eq.~(\ref{Ml1}) we see
that the vector components $l_{i}$ correspond to the inverse of the diagonal
elements of the matrix $\lambda$
\begin{equation}
l_{i}=1/\lambda _{i}\,.
\end{equation}

Note that the relation {in Eq.~(\ref{Ml1})} considered as an equation for $M$, has solutions related by the transformation
\begin{equation}
M\rightarrow G^{\prime }M\,,
\end{equation}
where $G'$ has rational matrix elements satisfying the condition
\begin{equation}
\sum_{j=1}^{n}G_{ij}^{\prime }=1_{i}\,.
\end{equation}
The matrix $G'$ should be chosen in such a way that $M$ remains integer--valued
and non--negative.
All such matrices produce a group of transmutations corresponding to different
choices
of the vectors $\overrightarrow{m}_i$ on the slice.

Geometrically, the end point of the vector $\overrightarrow{l}$ lies on the
hyperplane
which passes through the points
$(1,0,0,...)$, $(0,1,0,...)$, $(0,0,1,...)$, $\dots ,
(0,0,0,...,1)$. The end point of the inversed vector
\begin{equation}
\overrightarrow{l}'=\overrightarrow{l}/|\overrightarrow{l}|^2
\end{equation}
lies on the sphere
\begin{equation}
\sum_{i=1}^{n}|l_{i}^{\prime }-\frac{1}{2}|^{2}=\frac{n}{4}\,.\,
\label{sphere}
\end{equation}
The hyperplane corresponding to the slice generated by those
vectors $\overrightarrow{m}$
satisfying the condition~(\ref{ml1}) is orthogonal to the vector
$\overrightarrow{l}'$ and passes through its end point. The intersection
of this hyperplane with the sphere (\ref{sphere}) is again a sphere of
a lower dimension built on the vector
$\overrightarrow{l}'-\overrightarrow{1}$ as on a diameter. On the contrary,
for each point $\overrightarrow{l}'$ we can construct the above low dimensional
sphere belonging to the slice $\overrightarrow{m}$ and therefore the developed
geometrical picture allows us to relate the reflexive
number $\overrightarrow{k}$ with the Batyrev polyhedron $\overrightarrow{m}$.

Among the
possible sets of vectors $\overrightarrow{m}$ for the reflexive polyhedron
it is convenient to choose the set of vectors
$\overrightarrow{m}_l$ ($l=1,2,...,n$) satisfying conditions (\ref{boundar}). The
number $n_l$ of such vectors being solutions of the equation
$\overrightarrow{m}_l\overrightarrow{k}=d$ for fixed $l$
is given by the integral
\begin{equation}
n_l=\int _0^{2\pi}\frac{d\phi}{2\pi}\,
\frac{e^{-id\phi}\,(1-e^{ik_{n-l}\phi})}{\prod
_{r=1}^n(1-e^{ik_r\phi})}= \int _L\frac{dx}{2\pi ix^{1+d}}\,
\frac{1-x^{k_{n-l}}}{\prod _{r=1}^n(1-x^{k_r})}\,, \label{intrefl}
\end{equation}
where the integration contour $L$ is a small circle around
the point $x=0$ drawn anticlockwise. To obtain the
above expression for $n_l$
we used the well known representation for the Kronecker symbol
\begin{equation}
\delta _{d,\,\overrightarrow{m}\overrightarrow{k}}=
\int _0^{2\pi}\frac{d\phi}{2\pi}\,e^{-id\phi}\,
e^{i\overrightarrow{m}\overrightarrow{k}\,\phi }\,.
\end{equation}

The necessary condition for the reflexivity
of the weight vector $\overrightarrow{k}$ can be
formulated as the set of inequalities
\begin{equation}
n_l\ge 1\, \label{condrefl}
\end{equation}
valid for all $l=1,2,...,n$. Of course, it should be combined with
the non-degeneracy requirement (\ref{det}) for the matrix
$m_{ij}$. But even after it the slice could be non-reflexive. The
sufficient condition for the reflexivity includes the condition of
the existence of such matrix $M'$ (\ref{defM'}) constructed from
points of the slice which has an eigenvector $c_i$ in the dual
space (see (\ref{M'eq})).
\begin{equation}
N(\overrightarrow{k})=\sum _{k'_1=1}^{\infty}...\sum
_{k'_n=1}^{\infty}\prod _{t'=1}^n\int _L\frac{dy _{t'}}{2\pi
iy_{t'}^{1+d'}}\prod _{t=1}^n\int _L\frac{dx _t}{2\pi ix_t^{1+d}}
\, \prod _{s=1}^{n}\prod
_{r=1}^n\frac{1}{1-x_s^{k_r}y_r^{k'_s}}>0. \label{N(k)}
\end{equation}

The reflexivity of the slice $\overrightarrow{m}$ can be verified
also by extracting from it the basis $\overrightarrow{m}_i$
($i=1,2,...,n$) having the maximal value of the determinant
\begin{equation}
T=\max _{\{\overrightarrow{m}_i\}}|\det M|
\end{equation}
and checking the positivity of the coefficients $c_k$ in the
expansion (\ref{cmi}) of the vector $\overrightarrow{1}$. This
basis is similar to the basis of simple roots among all roots in
the Lie algebras. The vectors $\overrightarrow{m}_i$ in this basis
are directed almost along the coordinate axes. Below we
investigate the so-called simple-laced and quasi-simple laced
numbers for which $\overrightarrow{m}_i$ have this property.

The total number of solutions of the equation $\overrightarrow{m}
\cdot \overrightarrow{k}=d$, which is equal to the number of
{moduli} $c_{\overrightarrow{m}}$ for the Calabi-Yau space, is
given by the formula
\begin{equation}
n_{d}(\overrightarrow{k})=\int _0^{2\pi}\frac{d\phi}{2\pi}\,
\frac{e^{-i d \phi}}{\prod _{r=1}^n(1-e^{ik_r\phi})}=
\int _L\frac{dx}{2\pi i\,x^{1+d}}\,
\frac{1}{\prod _{r=1}^n(1-x^{k_r})}\,.
\label{n_d}
\end{equation}
All these points except $\overrightarrow{m}=\overrightarrow{1}$
belong to the polyhedron boundaries $\overrightarrow{m}_l$.

In the case of the higher level Calabi-Yau spaces one should introduce
several projective vectors $\overrightarrow{k}^{t}$ ($t=1,2,..,r$) and
construct the intersection of the corresponding slices
\begin{equation}
\overrightarrow{m}\overrightarrow{k}^t=d^t\,.
\end{equation}
In the above geometrical construction it would lead to several
points $l^{\prime t}$ lying on the sphere (\ref{sphere}). The
hyperplane $\overrightarrow{m}$ corresponding to the polyhedron
will go through the ends of all vectors
$\overrightarrow{l}^{\prime t}$ and will be orthogonal to them.
Its intersection with this sphere will be again a sphere with the
dimension  $n-r$. The obtained polyhedron should satisfy the
Batyrev reflexivity property. The number of {moduli}
$c_{\overrightarrow{m}}$ for this general case can be obtained
from the integral
\begin{equation}
n(\overrightarrow{k}^1,...,\overrightarrow{k}^r)= \int _L \prod
_{t=1}^r \frac{dx _t }{2\pi i\,x_t^{1+d^t}}\, \frac{1}{\prod
_{r=1}^n(1-\prod _{s=1}^rx_s^{k_r^s})}\,. \label{n_d}
\end{equation}


The number of the reflexive vectors $\overrightarrow{k}$ at the level 1
for the given dimension $n$ and degree $d$ can be expressed in terms
of $N(\overrightarrow{k})$ (\ref{N(k)})
\begin{equation}
N_n(d)=\sum _{k_1=1}^{\infty}\,\sum _{k_2=1}^{\infty}...\sum _{k_n=1}^{\infty}\,
\delta _{d,\,\sum _{r=1}^nk_r}\,\theta (N(\overrightarrow{k}))\,.
\label{N(d)}
\end{equation}

In the next subsection we develop a physical
model based on these relations for Calabi--Yau manifolds.

\subsection{Physical interpretation of the Batyrev polynomial}

Let us write down Eq.~(\ref{Batyr}) for the Calabi-Yau spaces as the
condition
\begin{equation}
\Psi (a_1^+,a_2^+,...,a_n^+)=0
\end{equation}
for zeroes of the Schr\"{o}dinger wave function
\begin{equation}
\Psi (a_1^+,a_2^+,...,a_n^+)=\exp (-iEt)\,\sum
_{\overrightarrow{m}}\, c_{\overrightarrow{m}} \prod_{r=1}^n
(a^+_r)^{m_r}\,\Psi _0\,, \label{Psi}
\end{equation}
written for {a} string--like mechanical system of $n$ harmonic
oscillators. Here instead of variables $x_r$ in (\ref{Batyr}) we
introduced the {creation} operators $a_r^+$ for the oscillators
with frequencies $\omega _r=k_r$ and $\Psi _0$ is the vacuum state
in the diagonal representation for $a_r^+$. The integer components
{$m_r$} of the vectors $\overrightarrow{m}$ coincide with the
occupation numbers for these oscillators. The hamiltonian of this
quantum mechanical system is
\begin{equation}
H=\sum _{r=1}^n\,k_r\,a_r^+a^-_r\,.
\end{equation}
Here $n$ plays the r{\^ o}le of the dimension $D$ in string theory. The
degree $d_k=\sum _{r=1}^n k_r$ {corresponds to} the total energy
of the state
\begin{equation}
d_k =E=\sum _{r=1}^n m_r k_r\,.
\end{equation}
In coordinate representation the {creation and annihilation
operators read}
\begin{eqnarray}
a_r^+ &=&(x_r-\partial _r)/\sqrt{2},\,\,\,a^-_r~=~(x_r+\partial
_r)/\sqrt{2}
\end{eqnarray}
respectively, and, therefore, 
the vacuum state is $\Psi _0=\exp (-\overrightarrow{x}^2/2)$
with $\overrightarrow{x}^2=\sum _{i=1}^n x_i^2$.

Among the degenerate states $\overrightarrow{m}$ the simplest one
is $\overrightarrow{1}$. Other states with the same energy  $d_k$
should have at least one occupation number $m_l$ equal to zero,
because {otherwise} $\overrightarrow{m}\cdot
\overrightarrow{k}>d_k$. The reflexivity constraint restricts the
number of excited states for the considered harmonic oscilator
model. In string theory a similar restriction follows from
the Virasoro group. {In principle the condition of reflexivity
could arise dynamically in a more general theory where oscillator
interactions or unharmonic corrections were included in the
potential}. A similar effect takes place for the fractional
quantum Hall effect, where the ground state wave function has a 
special property: It vanishes as an odd power of small
relative distances $x_i-x_j$ \cite{QHE}. In our
 model the additional interactions could destroy the degeneracy
of the energy levels of the harmonic oscillators leading to the
reflexivity constraint for the lowest energy states. Such
possibility seems to be related to the fact that the reflexive
weights are very special, namely, the number of solutions for the
equation $\overrightarrow{m} \cdot \overrightarrow{k}=d_k$ for
reflexive $\overrightarrow{k}$ is finite for fixed $n$ in
comparison with an infinite number of other vectors. Moreover, the
special properties of the reflexive numbers could be connected to
a hidden symmetry of the Calabi--Yau spaces appearing in the
existence of the Berger graphs which are discussed below.

On the other hand, it would be also interesting to investigate the
statistical properties of the above quantum mechanical model for
the Batyrev polyhedrons. In particular one can define the
micro--canonical ensemble for physical states in which the
probability $P(\overrightarrow{k})$ to find the mechanical system
in the state characterized by the reflexive vector
$\overrightarrow{k}$ with fixed energy $d_k$ is given by the
expression
\begin{equation}
P(\overrightarrow{k})=\frac{1}{n_{d_k}(\overrightarrow{k})}\,,
\end{equation}
where $n_{d_k}(\overrightarrow{k})$ is defined in (\ref{n_d}). As
usual, we can introduce the thermodynamic potentials for this
system and study their physical properties. {The investigation of
the statistical properties of the physical model related to the
Calabi-Yau spaces will be performed in future publications. In the
following we focus on a subclass of reflexive Calabi--Yau spaces
corresponding to the cases in which the vectors
$\overrightarrow{m}_{k}$ are almost collinear with the coordinate
axes. Examples of such configurations for simply and
quasi--simply--laced polyhedrons are considered below. In
particular, it is possible to use recurrent relations for the
construction of polyhedrons with increasing dimension $n$, using
the algebraic construction of Ref.~\cite{AENV1}. In the next
subsection we combine this approach with some geometrical
concepts}.

\subsection{Geometric Relations for Polyhedrons and UCYA}

If we assume that all weight vectors $\overrightarrow{k}$ for
the reflexive slices $\overrightarrow{m}$ generated by the vectors
$\overrightarrow{m}_r$
for all dimensions less than a given number are known,
then the Universal Calabi--Yau Algebra (UCYA)~\cite{AENV1, AENV2, AENV3}
provides the possibility to calculate similar weight vectors in higher
dimensions. We shall illustrate here this method using a simple example of
the construction of the $(n+1)$--dimensional
reflexive polyhedrons containing inside them an $n$--dimensional reflexive
polyhedron.

We introduce the $n$--dimensional weight vector $\overrightarrow{k}$ with
dimension $d_k=\sum _{i=1}^n\,k_i$
and the corresponding polyhedron generated by $n$ integer--valued vectors
$\overrightarrow{m}_r$ which satisfy the relations
\begin{equation}
\overrightarrow{m}_r \cdot \overrightarrow{k}=d_k\,.
\end{equation}
According to  the reflexivity {condition} the vector
$\overrightarrow{m}=\overrightarrow{1}$ is assumed to be inner, {\it i.e.}
\begin{equation}
\overrightarrow{1}=\sum _{r=1}^n a_r\,\overrightarrow{m}_r\,,\,\,a_r>0\,.
\end{equation}

To generalize this polyhedron to the $(n+1)$--dimensional space we can add to
the components of the vector $\overrightarrow{m}_r$ a
$(n+1)$--th component, $t$, equal to unity
\begin{equation}
\overrightarrow{M}_r=(m_{r_1},m_{r_2},...,m_{r_n},1)\,.
\end{equation}
In this case the constructed polyhedron in the dimension $n+1$ will
automatically include inside itself the point
$\overrightarrow{M}=\overrightarrow{1}$ and its reflexivity will follow
if in the extended slice there are points $\overrightarrow{M}$ with $t=0$
and $t>1$.

A weight vector $\overrightarrow{K}$ for the polyhedron in the
$(n+1)$--dimensional space should satisfy the equation
\begin{equation}
\overrightarrow{M}_r \cdot \overrightarrow{K} = d_{K}
\label{skal}
\end{equation}
for $r=1,2,...,n+1$, where $\overrightarrow{M}_{n+1}$ is a new basis vector.
It is natural to choose this vector such that all its components
except $t$ are equal to zero, {\it i.e.}
\begin{equation}
\overrightarrow{M}_{n+1}=(0,0,0,...,0,\lambda _{n+1})\,,
\end{equation}
where $\lambda _{n+1} \ge 2$ is an integer number.

Because the vectors $\overrightarrow{M}_i$ ($i=1,2,...,n$) are known,
the first $n$ components of the vectors $\overrightarrow{k}$ and \overrightarrow{K}
should coincide (up to a common factor which can be put
equal to unity without loss of generality)
\begin{equation}
k_i=K_i\,\,\,(i=1,2,...,n)\,.
\end{equation}
Hence, from the condition that the vector $\overrightarrow{1}$ belongs to
the constructed slice in the $(n+1)$--dimensional space, we obtain
\begin{equation}
d_k+K_{n+1}=d_{K}\,,
\label{39}
\end{equation}
where $K_{n+1}=1,2,...$ is an integer number. This integer number is
restricted from above
\begin{equation}
K_{n+1}\le d_ {k}\,.
\label{40}
\end{equation}
Indeed, from Eq.~(\ref{skal}) we have that
\begin{equation}
\lambda _{n+1}\,K_{n+1}=d_{K}\,
\end{equation}
and therefore to obtain the reflexivity for the polyhedron in the
($n+1$)--dimensional  space one should impose the constraint
\begin{equation}
d_K\ge 2\,K_{n+1}\,,
\end{equation}
because $\lambda _{n+1}\ge 2$, which leads to the inequality $d_k \ge K_{n+1}$
due to Eq.~(\ref{39}).

The above construction is a particular case of UCYA for arity 2. Indeed,
in the framework
of this method we can take two low--dimensional weight vectors and extend them
to the ($n+1$)--dimensional space in our example as follows
\begin{equation}
\overrightarrow{k}^{a}=(k_1,k_2,...,k_n,0)\,,\,\,
\overrightarrow{k}^{b}=(0,0,0,...,0,1)\,.
\end{equation}
It is possible to verify that the intersection of the
corresponding slices $\overrightarrow{m}^a$ and $\overrightarrow{m}^b$ has the
property of reflexivity, the reflexive weight vector in the
$(n+1)$--dimensional space can be constructed by taking the following linear
combination of $\overrightarrow{k}^{a}$ and $\overrightarrow{k}^{b}$:
\begin{equation}
\overrightarrow{k}=\overrightarrow{k}^{a}+s\,\overrightarrow{k}^{b}\,,
\end{equation}
where $s=1,2,...$ is an integer number, restricted from above (cf. (\ref{40})):
\begin{equation}
s\le d_{k^{a}}\,.
\end{equation}
In particular, using the UCYA approach, we obtain two weight vectors for $n=3$,
\begin{eqnarray}
{\rm a)}\quad (1,1,1)&=&(1,1,0)+(0,0,1), \nonumber\\
{\rm b)}\quad (1,1,2)&=&(1,1,0)+2(0,0,1)\,.
\end{eqnarray}
The third vector can be constructed by adding two other extended weight vectors,
\begin{eqnarray}
{\rm c)}\quad (1,2,3)&=&(1,0,1)+2(0,1,1)\,.
\label{six}
\end{eqnarray}
Here the intersection of the two slices $\overrightarrow{m}^a$ and
$\overrightarrow{m}^b$ consists of the three points $(0,0,2),
(1,1,1)$ and $(2,2,0)$ having the reflexivity property.

In the $n=4$ case starting from the above--mentioned three $n=3$ vectors
we obtain the following weight vectors
\begin{eqnarray}
(1,1,1,1) &=& a+I\,,\,\,(1,1,1,2)~=~a+2I \,,\,\,(1,1,1,3)~=~a+3I, \nonumber\\
(1,1,2,1)&=&b+I\,,\,\,(1,1,2,2)~=~b+2I
\,,\,\,(1,1,2,3)~=~b+3I\,,\,\,(1,1,2,4)~=~b+4I, \nonumber\\
(1,2,3,1)&=&c+I\,,\,\,(1,2,3,2)~=~c+2I
\,,\,\,(1,2,3,3)~=~c+3I\,, \nonumber\\
(1,2,3,4)&=&c+4I\,,\,\,(1,2,3,5)~=~c+5I\,,\,\,(1,2,3,6)~=~c+6I,
\label{n4weight}
\end{eqnarray}
with
\begin{eqnarray}
I&=&(0,0,0,1), \, a~=~(1,1,1,0), \, b~=~(1,1,2,0), \, c~=~(1,2,3,0).
\end{eqnarray}
Some of these vectors coincide with others
after a transmutation of their components.

It is well--known that using the general UCYA construction for arity 2
we can obtain 90 out of 95 weight vectors for dimension $n=4$. Other weight
vectors can be found using arity 3 and 4. Generally, before adding two or
several extended vectors $\overrightarrow{k}^t$ ($t=1,2,...$) 
with integer coefficients
$\overrightarrow{k}=\sum _tn_t\overrightarrow{k}^t$, one should
verify the reflexivity of the slice produced by the solutions of the set
of equations $\overrightarrow{m} \cdot \overrightarrow{k}^t=0$.

So far we have reviewed the concept of reflexivity in the language of
algebra and geometry. In the next
sub--section we concentrate on arithmetic properties of these numbers
making use of some methods common in the study of Feynman diagrams in Quantum
Field Theory.

\subsection{Simply--Laced Numbers}

{We will first} consider the simplest case with the monomial points
$\left\{\overrightarrow{m}_i\right\}$ satisfying the equation
\begin{equation}
\sum_{r=1}^{n}\,\frac{m_{r}}{s_{r}}=1\,,
\label{msimple}
\end{equation}
where $s_{r}$ are integer numbers obeying the constraint
\begin{equation}
\sum_{r=1}^{n}\,\frac{1}{s_{r}}=1\, .
\label{ssimple}
\end{equation}
Geometrically $s_{r}$ is the value at
which the hyperplane generated by $\overrightarrow{m}_i$ crosses the axis $r$.
It is obvious that in the case {under consideration}
the vector $\overrightarrow{1}$ is inner in the corresponding polyhedron.
The vectors
$(1/s_1,1/s_2,...,1/s_n)$ satisfying the above constraint are known as the
``Egyptian fractions''~\cite{Guy}. For the cases $n=2$ and $n=3$ all
reflexive weight vectors are the ``Egyptian fractions'':
\begin{equation}
1=\frac{1}{2}+\frac{1}{2}
\end{equation}
and
\begin{equation}
1=\frac{1}{3}+\frac{1}{3}+\frac{1}{3}=\frac{1}{2}+\frac{1}{4}+\frac{1}{4}=
\frac{1}{2}+\frac{1}{3}+\frac{1}{6}\,.
\end{equation}
Concerning these decompositions, we note that the plane can be covered by
triangles with angles $\alpha =2\pi /s_1, \beta =2\pi /s_2 , \gamma
=2\pi /s_3$ without {mutual overlapping or empty spaces} only in these
four cases and by strips $a<x<a+\Delta$ {in} the degenerate case
$s_1=1$, $s_2=s_3=\infty$.

There are solutions of Eq.~(\ref{ssimple})  with all $s_{r}$ different.
They  lead to the polyhedron $\left\{\overrightarrow{m}_k\right\}$
without any symmetry under
the transmutation of $m_{k}$. An interesting example of such ``Egyptian
fractions'' is the following decomposition of unity
\begin{equation}
1=\sum_{k=1}^{r-1}\frac{1}{2^{k}}+\sum_{k=0}^{r-1}\frac{1}{2^{k}(2^{r}-1)}\,.
\end{equation}
In particular, provided that
\begin{equation}
M_{r}=2^{r}-1\,
\end{equation}
is a prime number $M_{r}=3,7,31,127,...$ (which can only be for primes
$r=2,3,5,7,...$),
in the above decomposition $1=\sum_{k}1/s_{k}$ the integers
$s_{k}$ are all divisors (except of $1$) of the degree $d$ being
the so-called perfect number
\begin{equation}
d=M_{r}(M_{r}+1)/2\,.
\end{equation}
In this case $M_{r}$ are called ``Mercenna numbers'' and, according to
Euclid and Euler, all even perfect numbers, being the sum of all their
divisors $d/s_{k}$
\begin{equation}
d=\sum_{k}\frac{d}{s_{k}},
\end{equation}
can be expressed in terms of Mercenna numbers. Examples of such
decomposition are 6=1+2+3 and 28=1+2+4+7+14.

In Eq.~(\ref{n4weight}) we found 6 reflexive polyhedrons of the type
$\frac{x}{6}+\frac{y}{3}+\frac{z}{2}+\frac{t-1}{\kappa }=1$
with $\kappa =1,\,6/5\,,\,3/2\,,\,2\,,\,3\,,\,6$
starting from the 3--dimensional polyhedron corresponding to the perfect
number $d=6$ (see (\ref{six})). In a similar way one can construct $d$ different
($n+1$)--dimensional polyhedrons containing inside them the polyhedrons
corresponding to the weight vectors being  decompositions of
other perfect numbers $d=28, 496, ...$ in the sum of all their $n$ divisors.
If $M_{r}$ is not a prime number not
all of its divisors enter in the decomposition of $d$. Note that
odd perfect numbers are not known.

The number of ``Egyptian fractions'' grows rapidly with $n$ (see \cite{Guy}).
To find a recurrence relation for this number one can generalize the
decomposition of $1$ in Eq.~(\ref{ssimple}) for a general
rational number $x$
\begin{equation}
\sum_{r=1}^{n}\,\frac{1}{s_{r}}=x\,.
\label{decompx}
\end{equation}
We denote the number of such decompositions by $N_{n}(x)$. For example
one can calculate the number of decompositions of unity $N_n(1)$
for several values of $n$ (see \cite{Guy})
$$
N_1(1)=1\,,\,\,N_2(1)=1\,,\,\,N_3(1)=3\,,\,\,N_4(1)=14\,,\,\,N_5(1)=147\,,
$$
\begin{equation}
N_6(1)=3462\,,\,\,N_7(1)=294314\,,\,\,N_8(1)=159330691\,.
\end{equation}

Let us introduce the
symbol $N_{n}^{(\Lambda )}(x)$ for the number of decompositions of $x$ in the
sum of $n$ unit ratios $1/s_{r}$ (\ref{decompx}) satisfying the relations
\begin{equation}
s_{r}\leq \Lambda\,\,,
\end{equation}
where $\Lambda$ is a large integer number.
One can derive the following recurrent relations for $N_{n}^{\Lambda}(x)$:
\begin{equation}
N_{n}^{(\Lambda )}(x)=\sum_{t=0}^{\infty}
N_{n-t}^{(\Lambda -1 )}(x-t/\Lambda)\,
\label{recur}
\end{equation}
with the following initial conditions
\begin{equation}
N_n^{(0)}(x)=\delta _{x,0}\,,\,\,N_n^{(-k)}(x)=0\,\,(k=1,2,...)\,.
\end{equation}

The generating function for $N_{n}^{\Lambda}$ is given below
\begin{equation}
F_{\lambda }^{(\Lambda )}(y)=\sum_{n=0}^{\infty }\sum_{x}\,\lambda
^{n}\,y^{x}\,N_{n}^{(\Lambda )}(x)=\prod_{t=1}^{\Lambda }\,\left( 1-\lambda
y^{1/t}\right) ^{-1}.\,
\end{equation}
The recurrent relation in (\ref{recur}) corresponds to the following
equation for this
function
\begin{equation}
F_{\lambda }^{(\Lambda )}(y)=\left( 1-\lambda y^{1/\Lambda}\right)
^{-1}F_{\lambda }^{(\Lambda -1)}(y)\,.
\end{equation}
Note that $F_{\lambda }^{(\Lambda )}(y)$ grows rapidly for
$\Lambda \rightarrow \infty$, but $N_{n}^{(\Lambda )}(x)$ tends to
a finite limit. For the regularized case  the equation
$z=F_{\lambda}^{(\Lambda )}(y)$ defines a Riemann surface with a
finite genus.

The inverse relation reads
\begin{equation}
\Phi _n^{(\Lambda )}(y)=\sum _{x}\,y^x\,N_n^{(\Lambda )}(x) =
\frac{1}{2\pi i}\int _L\,
\frac{d\,\lambda}{\lambda ^{n+1}}\,F_{\lambda}^{(\Lambda)}(y),
\end{equation}
{where a small closed contour of integration, $L$, is taken anticlockwise
around the point $\lambda = 0$.}

To express $N_n^{(\Lambda )}(x)$ in terms of $F_{\lambda}^{(\Lambda )}(y)$
we should perform the
additional integration
\begin{equation}
N_n^{(\Lambda )}(x)=\frac{1}{2\pi i\,m}
\int _{l_{m}}\,\frac{d\,y}{y^{1+x}}\,\Phi_n ^{(\Lambda )}(y)\,.
\end{equation}
Here the closed contour of integration $l_m$ goes $m$--times around the
point
$y=0$ moving through
other sheets of the Riemann surface $z=y^{-x}\,\Phi _n^{\Lambda}(y)$.
The integer number $m$ is  chosen from the condition that
the point $y=0$ on the surface becomes regular
in  the new coordinate $u=y^{1/m}$.

Let us calculate the asymptotic behavior of $\Phi _n^{(\Lambda )}(y)$ for
large $n$ and $\Lambda$ using the saddle
point method. For this purpose we present the generating function in the
form
\begin{equation}
\ln F_{\lambda}^{(\Lambda)}(y)=\ln F_{\lambda}^{(\Lambda)}(1)+
\Delta \ln F_{\lambda}^{(\Lambda)}(y)\,,
\end{equation}
where
\begin{equation}
\ln F_{\lambda}^{(\Lambda )}(1)=-\Lambda \,\ln (1-\lambda)\,
\end{equation}
and
\begin{equation}
\Delta \ln F_{\lambda}^{(\Lambda)}(y)=\frac{\lambda}{1-\lambda}\,\ln y \,
\sum _{t=1}^{n}\frac{1}{t}+f_{\lambda}(y)\,.
\end{equation}
Here
\begin{equation}
f_{\lambda}(y)=\sum _{t=1}^{\infty}
\left(\ln
\frac{1-\lambda}{1-\lambda\,y^{1/t}}-
\frac{1}{t}\,\frac{\lambda}{1-\lambda}\,\ln y \right) \,.
\end{equation}
In the last expression for $f_{\lambda}(y)$ we pushed $\Lambda $ to
infinity, because the sum over $t$ is convergent.

Now we apply the saddle point method to the calculation of the integral
over $\lambda$, considering the extremum of the function
\begin{equation}
J_n^{(\Lambda)}(\lambda)=\ln F_{\lambda}^{(\Lambda)}(1)-n\,\ln \lambda =
\Lambda \,\ln \frac{1}{1-\lambda}-n\,\ln \lambda
\,.
\end{equation}

From the stationarity condition $\delta J_n^{(\Lambda)}(\lambda)=0$ one
can find the saddle point
\begin{equation}
\tilde{\lambda}=\frac{n}{n+\Lambda} \ll 1
\end{equation}
and therefore, with quadratic accuracy in
$\delta \lambda =\lambda -\tilde{\lambda}$, we obtain for this function
\begin{equation}
J_n^{(\Lambda)}(\lambda)=\Lambda \,
\ln \frac{n+\Lambda}{\Lambda}+n\,\ln
\frac{n+\Lambda}{n}+\frac{(\delta \lambda )^2}{2}\,
\frac{(n+\Lambda)^3}{n\,\Lambda}+...\,\,.
\end{equation}

It is obvious that the contour of integration over $\delta \lambda$ goes
through the saddle point in a correct direction
parallel to the imaginary axis. Thus, we obtain for $\Phi _n(y)$
the following expression at large $n$
after calculating the Gaussian integral over $\delta \lambda$
\begin{equation}
\Phi
_n^{(\Lambda )}
(y)=\left(\frac{n+\Lambda}{\Lambda}\right)^{\Lambda}
\left(\frac{n+\Lambda}{n}\right)^{n}
\sqrt{\frac{\Lambda}{2\pi \,n\,(n+\Lambda)}}\,e^{\Delta \ln
F_{\tilde{\lambda}}^{(\Lambda)}(y)}\,.
\end{equation}
We substituted $\lambda$ by its saddle point value $\tilde{\lambda}$ in the
slowly changing function $\Delta \ln F_{\lambda}^{(\Lambda)}(y)$.

{
The most interesting case, when $\tilde{\lambda}<<1$, is considered in the
following}. In this limit $f_{\tilde{\lambda}}(y)=0$ and the result is
significantly simplified
\begin{equation}
\Phi
_n^{(\Lambda )}(y)
=\left(\frac{n+\Lambda}{\Lambda}\right)^{\Lambda}
\left(\frac{n+\Lambda}{n}\right)^{n}
\sqrt{\frac{\Lambda}{2\pi \,n\,(n+\Lambda)}}\,y^x\,,
\end{equation}
where $x=\frac{\tilde{\lambda}}{1-\tilde{\lambda}}\,\sum_{t=1}^{\Lambda}\frac{1}{t}\,.$ As
$\sum_{t=1}^{\Lambda}\frac{1}{t}\approx \ln \Lambda -\Psi (1)+
\frac{1}{2\Lambda}+...\,$,
where $\gamma =-\Psi (1)$ is the Euler constant, we obtain that
$N_n(x)$ has a maximum
\begin{equation}
N_n^{(\Lambda )}(x_m)\approx \frac{(n+\Lambda)^n}{n!}
\end{equation}
at
\begin{equation}
x_m=\frac{n}{\Lambda}\,\left(\ln \Lambda -\Psi (1)+
\frac{1}{2\Lambda}\right)\,.
\end{equation}
For larger $x$ the above saddle point method should be modified.

The simple-laced numbers in dimension $n=5$ are constructed in
Appendix~\ref{QSLRWVn5}. We investigate also their structure in
terms of the UCYA construction.

It turns out that when going beyond dimension $n=3$ not all of the
reflexive weight numbers are simply--laced. A large number of them
has a different structure in the sense that some of their
components $k_r$ are not divisors of the degree $d_k$. These are
what we called non--simply--laced numbers. The simplest case is
when each component $k_r$ can be converted in  a divisor of the
difference of the degree $d_k$ and another component $k_{r'}$. The
next subsection is devoted to the study of this class of numbers
called ``quasi--simply--laced''.

\subsection{Classification of Quasi--Simply--Laced Numbers}

Quasi--simply--laced numbers are important generalizations of the
simply--laced ones. For example, in dimension
$n=4$ all 95 polyhedra with a single reflexive vector belong to
this class (among them 14 are obtained from the
simply--laced numbers). A simple example is the vector
$\overrightarrow{k} = (1,2,3,5)[11]$, corresponding to the
following decomposition of unity in the sum of ratios $l_i$
\[
1=\frac{1}{11}+\frac{2}{11}+\frac{3}{11}+\frac{5}{11}\,.
\]
For this vector
$d_k = 11$ and $d_k/k_1 = 11,\,(d_k - k_1)/k_2 = 5, \,(d_k - k_2)/k_3 = 3$
and $(d_k - k_1)/k_4 = 2$.

Hence we can generalize the diagonal ansatz for the
matrix $M_{ij}$ in the case of the simply--laced weight vectors, assuming
that the vector $\overrightarrow{l}$ satisfies the
set of equations
\begin{equation}
s_{i}l_{i}+l_{i^{\prime }}=1
\label{qsleq}
\end{equation}
for $i=1,2,...,n$ and $i^{\prime }=i^{\prime }(i)$ is also one of these
numbers. Here $s_{i}$ are positive integer parameters which will be
later chosen from
the condition that the vector $\overrightarrow{1}$ belongs  to the slice of
$\overrightarrow{m}$:
\begin{equation}
\sum_{i=1}^{n}l_{i}=1\,
\label{syst}
\end{equation}
and the corresponding point in the slice is  inner in accordance with the property of
reflexivity.
In this section we will classify
all sets of equations for quasi--simply--laced numbers
in such a way that the sets obtained by a
transmutation of indices are considered as belonging to the same class.

For this purpose we introduce a diagrammatic representation where the indices
$i$ and $i^{\prime }$ appearing in Eq.~(\ref{qsleq})
are connected by a line with an arrow directed from $i$ to $i^{\prime }$.
For each different class of sets of equations there is only
one ``Feynman diagram'' related to the function $i'(i)$ in Eq.~(\ref{qsleq}).
These Feynman diagrams can be obtained from the
``functional'' integral $Z(\lambda )$ with the ``action'' $L$
\begin{equation}
Z(\lambda )=\int \frac{d\,x\,dy}{\pi }\,e^{-L}\,,\,\,L=|z|^{2}-\lambda
\,z^{\ast }e^{z}\,,\,\,\,z=x+iy
\end{equation}
by expanding it in the ``coupling constant'' $\lambda $:
\begin{equation}
Z(\lambda )=\sum_{n=0}^{\infty }\lambda ^{n}\,Z_{n}\,\,,\,\,Z_{n}=\sum_{r}%
\frac{1}{G_{r}}\,\,.
\end{equation}
Here $r$ enumerates different Feynman diagrams
in the $n^{{th}}$-order of perturbation theory (corresponding to different
classes of sets of the equations shown above) and $G_{r}$ is the number of
group elements of the symmetry for the diagram $r$
under {permutations} of the index $i$. In agreement with Eq.~(\ref{qsleq})
these diagrams contain all
possible vertices $V_r$
($r=0,1,2,...$) in which $r$ particles are absorbed
by the field $z$ and only one particle
is emitted by the field $z^*$.

Using the above expression for $Z(\lambda )$ we obtain in the $n$-th order
\begin{equation}
Z_{n}=\int \frac{d\,x\,dy}{\pi }\,e^{-|z|^{2}}\frac{z^{\ast n}}{n!}%
e\,^{nz}\,.
\end{equation}
Therefore the number of diagrams of order $n$ weighted with the
symmetry factors $1/G_{r}$ equals
\begin{equation}
Z_{n}=\sum_{r}\frac{1}{G_{r}}=\frac{n^{n}}{n!}\,.
\end{equation}

\begin{figure}[tbp]
\begin{picture}(100,80)
\thicklines
\put(110,30){\circle{30}}
\put(117,-5){$G=2$}
\put(150,30){\circle{30}}
\put(110,14){\circle*{4}}
\put(150,14){\circle*{4}}
\put(240,30){\circle{30}}
\put(240,-5){$G=1$}
\put(270,14){\circle*{4}}
\put(240,14){\circle*{4}}
\put(240,14){\line(1,0){20}}
\put(270,14){\vector(-1,0){20}}
\put(370,30){\circle{30}}
\put(356,-5){$G=2$}
\put(354,30){\circle*{4}}
\put(386,30){\circle*{4}}
\put(106,45.8){\vector(-1,0){0.05}}
\put(146,45.8){\vector(-1,0){0.05}}
\put(236,45.8){\vector(-1,0){0.05}}
\put(366,45.8){\vector(-1,0){0.05}}
\put(374,14.2){\vector(1,0){0.05}}
\end{picture}
\caption{Diagrams for $n=2$}
\label{dia2}
\end{figure}
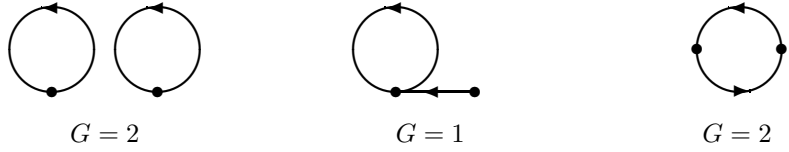

It is natural to expect that at large $n$ the saddle point configuration
for the Feynman diagrams corresponds to an almost constant averaged
symmetry factor $\frac{1}{\overline{G}(n)}$
for a subgroup of the permutation group
\begin{equation}
\frac{1}{\overline{G}(n)}=\frac{\sum_{r}1/G_{r}}{\sum_{r}\!1}
\,.
\end{equation}
In this case the number of different classes of solutions does not grow
very rapidly at large $n$
\begin{equation}
\sum_{r}1\approx \overline{G}(n)\,\frac{e^{n}}{\sqrt{2\pi n}}\,
\end{equation}
in comparison with the
total number of Calabi---Yau spaces. Let us consider the
Feynman diagrams for $n=2,3,4,5,...$. For $n=2$ there is one disconnected
and two connected diagrams (see Fig.~\ref{dia2}, where the symmetry weights $G_{r}$
are also indicated).

The corresponding sets of equations are
\begin{eqnarray}
&a)&s_{1}\,l_{1}+l_{1}=1\,,\qquad s_{2}\,l_{2}+l_{2}=1\,,\\
&b)&s_{1}\,l_{1}+l_{1}=1\,,\qquad s_{2}\,l_{2}+l_{1}=1\,,\\
&c)&s_{1}\,l_{1}+l_{2}=1\,,\qquad s_{2}\,l_{2}+l_{1}=1\,.
\end{eqnarray}
One can verify in this case the fulfillment of the relation
\[
Z_{2}=\frac{1}{2!}+1+\frac{1}{2!}=\frac{2^{2}}{2!}=2\,.
\]
For $n=3$ we have $7$ different Feynman diagrams and $7$ different sets of
equations, respectively
\begin{eqnarray}
&1)&s_{1}\,l_{1}+l_{1}=1\,,\quad s_{2}\,l_{2}+l_{2}=1\,,\quad s_{3}%
\,l_{3}+l_{3}=1;\,G=3!; \\
&2)&s_{1}\,l_{1}+l_{1}=1\,,\quad s_{2}\,l_{2}+l_{2}=1\,,\quad s_{3}%
\,l_{3}+l_{2}=1;\,G=1;\\
&3)&s_{1}\,l_{1}+l_{1}=1\,,\quad s_{2}\,l_{2}+l_{1}=1\,,\quad s_{3}%
\,l_{3}+l_{2}=1;\,G=1; \\
&4)&s_{1}\,l_{1}+l_{1}=1\,,\quad s_{2}\,l_{2}+l_{1}=1\,,\quad s_{3}%
\,l_{3}+l_{1}=1;\,G=2!;\\
&5)&s_{1}\,l_{1}+l_{1}=1\,,\quad s_{2}\,l_{2}+l_{3}=1\,,\quad s_{3}%
\,l_{3}+l_{2}=1;\,\,G=2!; \\
&6)&s_{1}\,l_{1}+l_{3}=1\,,\quad s_{2}\,l_{2}+l_{1}=1\,,\quad s_{3}%
\,l_{3}+l_{2}=1;\,G=3;\\
&7)&s_{1}\,l_{1}+l_{2}=1\,,\quad s_{2}\,l_{2}+l_{3}=1\,,\quad s_{3}%
\,l_{3}+l_{2}=1;\,G=1\,.
\end{eqnarray}
The number of diagrams weighted with their symmetry factors is
\[
Z_{3}=\frac{1}{6}+1+1+\frac{1}{2!}+\frac{1}{2!}+\frac{1}{3}+1=\frac{3^{3}}{3!%
}=\frac{9}{2}\,,
\]
which agrees with the above relation for $Z_n$.

For $n=4$ there are $19$ different Feynman diagrams and
\[
Z_{4}=\frac{4^{4}}{4!}=\frac{32}{3}\,.
\]

For $n=5,\,6$ and $7$  there are respectively $47,\,130$ and $342$ different
Feynman
diagrams with the corresponding values of $Z_n$
\[
Z_{5}=\frac{5^{5}}{5!}=\frac{625}{24}\,,\,\,Z_{6}=\frac{6^{6}}{6!}
=\frac{324}{5}\,,\,\,Z_{7}=\frac{7^{7}}{7!}=\frac{117649}{720}\,.
\]
It is possible to calculate the number of the corresponding diagrams
for larger values of $n$ (see~\cite{BLL}). It turns out that the
averaged symmetry $\overline{G}$ of the Feynman diagrams grows approximately linearly
from $\overline{G}=1$ for $n=1$ up to $\overline{G}(n) \approx 10,7$ for $n=27$.

Looking at these Feynman diagrams we can see that their connected parts
contain only one loop. In this way the quasi--classical approximation for the
``functional'' integral should be exact:
\begin{equation}
Z(\lambda )=\int \frac{d\,x\,dy}{\pi }\,e^{-L}\,=\frac{1}{1-z(\lambda )}%
\,,\,\,\,
\end{equation}
where $z(\lambda )$ is the solution of the classical equation $\delta L=0$:
\begin{equation}
z(\lambda )=\lambda \,e^{z(\lambda )}\,.\,\,
\end{equation}
Indeed, solving this equation with the use of perturbation theory we obtain
\begin{equation}
\,Z(\lambda )=\sum_{n=0}^{\infty }\lambda ^{n}\frac{n^{n}}{n!}\,,\,\,
\end{equation}
corresponding to
\begin{equation}
\,z(\lambda )=1-\frac{1}{\sum\limits_{n=0}^{\infty }\lambda ^{n}{\displaystyle{\frac{n^{n}}{n!}}}}%
=\lambda +\lambda ^{2}+\frac{3}{2}\lambda ^{3}+...\,.\,\,
\end{equation}

One can obtain a more detailed description of the Feynman diagrams in
terms of
the number of vertices $V_{r}$ with a different number $r+1$ of lines.
For this case we should consider the more general action
\begin{equation}
\,L=|z|^{2}-\,z^{\ast }\sum_{r=0}^{\infty }\,g_{r}\frac{z^{r}}{r!}\,,\,\,
\end{equation}
where $g_{r}$ are corresponding coupling constants.
Here we also obtain that the quasi--classical result is exact
\begin{equation}
Z=\int \frac{d\,x\,dy}{\pi }\,e^{-L}\,=\frac{1}{1-a}\,,\,\,\,
\end{equation}
where $a$ is the solution of the classical equation
\begin{equation}
a=\sum_{r=1}^{\infty }\,g_{r}\frac{a^{r-1}}{(r-1)!}\,\,\,
\end{equation}
and the perturbative expansion for $Z$ reads
\begin{equation}
Z=\sum_{r_{0}=0}^{\infty }\,\frac{g_{0}^{r_{0}}}{r_{0}!}\sum_{r_{1}=0}^{%
\infty }\frac{g_{1}^{r_{1}}}{r_{1}!(1!)^{r_{1}}}...\sum_{r_{\infty
}=0}^{\infty }\frac{g_{\infty }^{r_{\infty }}\left(\sum_{k=0}^{\infty }r_{k}\right)!}{%
r_{\infty }!(\infty !)^{r_{\infty }}}\,\,\,\delta \left(\sum_{k=0}^{\infty
}(k\,-1)r_{k}\right).\,\,\,
\end{equation}
The coefficient in front of the product of $g_{k}^{r_{k}}$ coincides with
the number of Feynman diagrams (with symmetry factors) having $r_{k}$
vertices with $k+1$ lines for each $k=0,1,2,...$. At large orders
$n=\sum_{k=0}^{\infty }r_{k} \gg 1$ of perturbation theory there exists
a saddle point
\begin{equation}
\,\,\tilde{r}_{k}=n\,\frac{e^{-1}}{k!}\,.\,
\end{equation}
in the sums over $r_{k}$.

In Appendix~\ref{QSLRWVn4}  we illustrate this classification of
the quasi--simply--laced reflexive weight vectors in the case
$n=4$. Here among 19 types of such numbers in 2 cases the
reflexivity condition for the polyhedrons is not fulfilled.
Moreover, using the freedom to extract from the slice
$\overrightarrow{m}$ different sets of the vectors
$\overrightarrow{m}_i$ in the slice $\overrightarrow{m}$ we can
restrict ourselves to a smaller number of possibilities $k<17$
corresponding to the above Feynman diagrams containing only closed loops.

After the above analysis of reflexive vectors in particular cases
of the Egyptian and quasi-Egyptian fractions with the use of the
methods of the Quantum Field Theory we turn now to a more
algebraic approach by associating to them some matrices similar to
those appearing in  the case of the Cartan-Lie algebras.

%% file: SecondPart.tex
\section{Simply--Laced Numbers as Generators of Berger Graphs}

In this Section we investigate
the Berger graphs for the simply-laced numbers
considered initially in Refs.~\cite{V, TV, ETV}.
These graphs are related to the CY$_d$ spaces of the first level described
by a single reflexive weight vector. In the case of small $d$ they coincide
with the Dynkin diagrams for the roots of the
Cartan--Lie algebras $A_r$, $D_r$, $E_6$, $E_7$ and $E_8$. In the Dynkin 
diagrams  for such root systems the nodes are connected by 
single lines. Consequently, their Cartan matrices turn out to be 
symmetric. The Berger matrices for the simply--laced case are also symmetric.

Similar to the Cartan--Lie case the Berger graph for a simply--laced
reflexive vector $\overrightarrow{k}$ is built by assigning 
the degree $d_k$ to the central node
as its Coxeter label.  This will be the maximal Coxeter
label in the graph. The number of legs attached to this central node
coincides with the dimension $n$ of the vector. In each leg the number of
nodes is $d_k/k_i-1$, with $k_i$ being the corresponding component of
$\overrightarrow{k}$. The Coxeter labels of these nodes are
$d_k-k_i\,,\,\,d_k-2k_i\,,...,k_i$. They decrease along the leg starting from 
the central node. For example in the primary graph of $(1,1,2)[4]$ for 
dimension 3 there are three legs, the central node has the Coxeter label 4 and
in the first and second legs there are additional nodes with 
the Coxeter labels $3,2,1$. In the third leg the Coxeter label of the 
additional node is 2.

The associated Berger matrix $B_{ij}$ for $\overrightarrow{k}$ is built from
scalar products $(\alpha_i,\alpha _j)$ of the root vectors
$\overrightarrow{\alpha}_l$ assigned to each node. The scalar product of two
vectors of the nodes connected by a line is $-1$. Disconnected nodes correspond
to orthogonal vectors. The diagonal matrix element $B_{ii}$ is the square 
of the vector assigned to the corresponding node. For all nodes $B_{ii}$ is 2 
with the exception of the central node, where the root square is equal to
$n-1$. The determinant of such a matrix is zero, which is a generalization of 
the similar result for the affine simply--laced Cartan--Lie matrices.
The Coxeter labels assigned to the nodes in a Berger graph coincide with
the cofficient $c_l$ in front of the corresponding root 
in the linear combination of the eigenvector $\sum _lc_l\alpha _l$ 
corresponding to a vanishing eigenvalue of the Berger matrix. As an example,
the primary graph for $\overrightarrow{k}=(1,1)[2]$ has the central
node with its Coxeter label equal 2 and $B_{ii}=1$. Further, each of its two 
legs has one node with its Coxeter label equal to 1 and $B_{ii}=2$.

The maximal Coxeter labels for the reflexive simply--laced vectors
$(1)$, $(1,1)$, $(1,1,1)$, $(1,1,2)$, $(1,2,3)$ are $1,2,3,4,6,$ respectively,
which coincides with the maximal Coxeter labels for the corresponding
simply--laced Lie algebras, $A_r$, $D_r$, $E_6$, $E_7$ and $E_8$. In more
detail, we have respectively for these primary graphs: one node of $A$--type 
with Coxeter label (1), 
three nodes of
$D$--type with Coxeter labels $(1,2,1)$ (one chain), seven nodes of $E_6$--type
with Coxeter labels $(1,2;1,2;1,2;3)$ (three chains), eight nodes of $E_7$--type
with Coxeter labels $(1,2,3;1,2,3;2;4)$ (three chains) and nine nodes of
$E_8$--type with Coxeter labels $(1,2,3,4,5;2,4;3;6)$ (three chains).

It is important to note that the primary graphs for the above cases are
generators of generalized Berger graphs in CY$_d$ polyhedra.
Namely, such Berger graphs can be built from one or several blocks of
the primary graphs. These blocks are connected by the lines 
appearing in the Cartan graphs for the $A_i$--series. Namely, 
on each line there are several nodes with the same Coxeter label
equal to the Coxeter label of two nodes to which these lines
are attached. Furthermore, the Coxeter labels for all nodes and 
the matrix elements of the Berger matrix $B_{ij}$ inside
each block are universal and coincide with those for the corresponding
elementary Cartan graph. Only  the square of the root corresponding
to the node with the attached lines is changed by adding to it the
number $l$ of these lines, i.e., $B_{ii}=2 \rightarrow B_{ii}=2+l$~\cite{V}.

Each of the reflexive vectors $\vec{k}$ with $n$ components can be extended
to extra dimensions $n\rightarrow p+n$ by adding to it several
vanishing components:
$\vec k^{\rm ext}_{p+n}=(0,0,...,0;k_1,...,k_n)$. These extended vectors
participate in the UCYA $r$--arity construction leading to new reflexive
vectors in higher dimensions.
The structure of the Berger graphs for the polyhedrons obtained 
by this method  depends on the number
$p$ of zero components for the corresponding extended vectors.

The UCYA $r$--arity construction can be used to build
new Calabi--Yau polyhedra (at level one) as it was discussed in the previous
section. In 
particular, to go
from the vectors $(1)$, $(1,1)$, $(1,1,1)$, $(1,1,2)$, $(1,2,3)$
to $n=4$ dimensions one should take the extended vectors
(0,0,0,1), (0,0,1,1), (0,1,1,1), (0,1,1,2), (0,1,2,3)
and those obtained by permutations of their components. Then in the framework 
of the 2--arity
approach we can compose the linear
combinations of two of these numbers with integer coefficients. For
each pair one should verify that the intersection of two
polyhedra corresponding to two extended vectors has the
reflexivity property. This condition is fulfilled in the case of the
polyhedron corresponding to the
eldest vector $\vec{k}_1+\vec{k}_2$. In the
$K_3$ reflexive polyhedron for each eldest vector one can
find the primary graphs of the $A^{(1)}_r$, $D^{(1)}_r$, $E^{(1)}_6$,
$E^{(1)}_7$ and $E^{(1)}_8$ types. They are situated at two opposite sides
of the polyhedron
divided by the above intersection. A similar
situation takes place with the reflexive $K_3$ polyhedron corresponding
to the sum of the vectors $\vec{k}_1$ and $\vec{k}_2$ with  integer
coefficients. For each case in the constructed polyhedron one can find
generalized graphs corresponding to the primary graphs.

\begin{figure}[!ht]
\vspace{-0.4cm}
\hspace{1cm}
\centerline{\epsfig{file=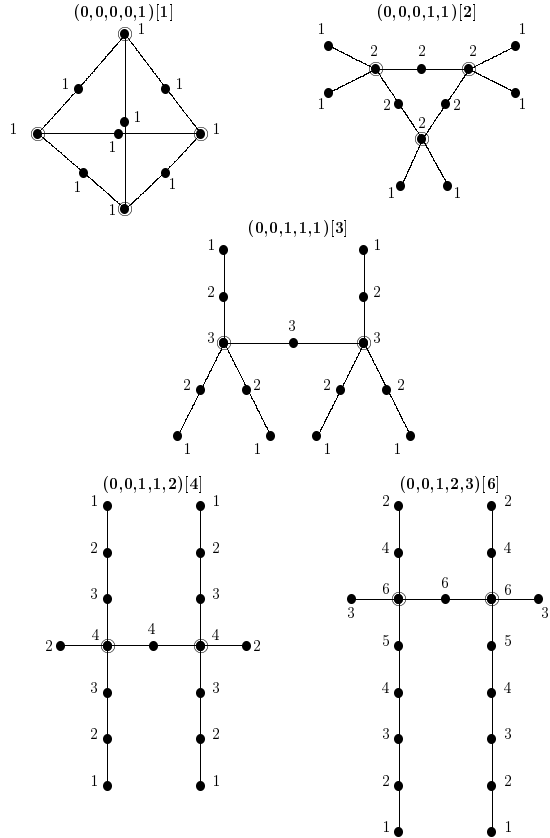,width=14cm}}
\vspace{-0.8cm}
\caption{Berger diagrams from a 2--arity construction in the $n=5$ dimensional
case, with the corresponding Coxeter labels at the nodes. They generate five
infinite series.}
\label{5graphs}
\end{figure}

When the 2--arity construction is used for the $n=5$
dimensional case (CY$_3$) then the structure of the corresponding graphs
becomes more complicated although they are
also built from the previous order
graphs. The difference is that some of the  Berger matrices corresponding
to the generalized Dynkin graphs can have  $B_{ii}=3$ instead of the 
usual $B_{ii}=2$. This is 
seen in Fig.~\ref{5graphs},
where the links between the previous Berger subgraphs belong to
the $A_l$ type with
modified Coxeter labels (1, 2, 3, \dots instead of 1).
It is remarkable that these five $n=5$ Berger graphs 
shown on Fig.~\ref{5graphs} generate
five infinite series because their structure holds for any $l$ in $A_l$.
This procedure can be generalized by linking with $A_l$--lines ($B_{ii}=2$)
not only a pair of triple nodes from corresponding primary graphs but also
other nodes sharing the same Coxeter label. When this happens the diagonal
element $A_{ii} = 2$ in the Cartan matrices is substituted by the matrix element
$B_{ii}=3$. The  determinants of the non--affine Berger matrices (0,0,1,1,1)[3],
(0,0,1,1,2)[4] and (0,0,1,2,3)[6] are equal to $3^4$, $4^3$ and $6^2$
independently of the number of nodes along the internal line connecting two
primary graphs (see Fig.~\ref{5graphs} and table~\ref{BERDET}).
Note that the labels of all the nodes along this line coincide
with those of the central nodes of two connected primary graphs, i.e.,
they are equal to $3, 4$ and 6 respectively. Thus, these graphs produce three
infinite series analogous to the graphs of the $D_r$--series generated by the
extended reflexive number $(0,0,1,1)$ in the $K3$ polyhedrons. In addition
to the infinite series of the Dynkin graphs of $A_r$ and $D_r$ with
maximal Coxeter numbers 1 and 2 there appear three new series with maximal
Coxeter numbers $3, 4$ and $6$. This construction of the infinite series of the
Berger graphs could lead to a possible generalization 
of the notion of the direct product of Lie algebras, {\it e.g.}, $E_8 \times E_8$ in the  heterotic string.

Probably the
various Berger graphs obtained by the UCYA construction  
should be placed in the same class. All new graphs contain several initial
diagrams joined by different numbers of the $A_i$-lines. The result
depends on the arity $2,3...$ in UCYA and on the dimension of
the constructed polyhedron.
The reflexive polyhedra allow us to build large classes of
graphs, among which the set of usual Dynkin diagrams, related to
the binary operations, is only a small one, because the Dynkin
diagrams and their affine generalizations are in one-to-one
correspondence with the well known  Cartan--Lie algebras and
infinite--dimensional Kac--Moody algebras, respectively. It is then natural
to think that the Berger graphs could lead
to algebras beyond the Cartan--Lie / Kac--Moody construction
and, in particular, could be related to ternary, quaternary, ...
generalizations of binary algebras.

To extend the class of Dynkin diagrams we  generalize
the rules for affine Cartan matrices~\cite{V}. Namely the Berger matrices
satisfy the following rules:
\begin{eqnarray}
{\mathbb B}_{ii}&=&2,\, \, {\rm or}\,\,  3, {\rm or} \,\, 4, ..., \,\,
{\mathbb B}_{ij}~\leq~ 0, \,\,
{\mathbb B}_{ij}=0 ~\mapsto~ {\mathbb B}_{ji}=0, \nonumber\\
{\mathbb B}_{ij} &\in& {\mathbb Z} ,\,\,
{\rm Det} \,\,{\mathbb B} ~=~ 0,\,\,
{\rm Det} \,\,{\mathbb B}_{\{(i)\}} ~>~ 0.
\end{eqnarray}
The constraint of a vanishing determinant is a generalization of the
``affine condition'' for Kac--Moody algebras. In the above new rules we relax 
the restriction on the diagonal element
${\mathbb B}_{ii}=2$, {\it i.e.}, to satisfy the affine condition
we allow for ${\mathbb B}_{ii}$ to be larger: ${\mathbb B}_{ii}=3,\,4,\,...$.
These large values can appear  in the lattice of reflexive polyhedra starting
from CY$_3$, CY$_4$, ...
 Below we shall check the coincidence of the
graph's labels indicated on figures with
the Coxeter labels obtained from the eigenvalues of
the Berger matrices. The proposed prescriptions for the Coxeter
labels are universal for all Berger graphs independently from their
dimension and arity construction. These prescriptions generalize
the Cartan and Kac--Moody rules in a natural way.
Note that the value of the diagonal term ${\mathbb A}_{ii}$
of the Cartan matrix  can be
related to one of the Casimir invariants of the simple
Lie algebras.  The number of these invariants is equal to the
algebra rank.  All Cartan--Lie algebras contain the Casimir
operator of degree 2, but there are other invariant operators.
For example, the exceptional $E_6$, $E_7$, $E_8$
algebras have the following degrees  of  Casimir invariants:
\begin{eqnarray}
E_6:  \{ \,2,\, 5,\, 6,  \,  8, \,  9,\, 12 \,           \},\,\,
E_7:  \{ \,2,\, 6,\, 8,  \, 10, \, 12,\, 14,\, 18 \,     \},\,\,
E_8:  \{ \,2,\, 8,\, 12, \, 14, \, 18,\, 20,\, 24,\, 30\,\}.
\end{eqnarray}
By subtracting $1$ from the Casimir operator degree we obtain the so--called
``Coxeter exponents'' of the corresponding Lie algebra.
One can see that the largest degree of the Casimir invariants  in this
list, $12$, $18$ and $30$,  is equal to the Coxeter number of the
$E_6$, $E_7$ and $E_8$ algebras, respectively. We note that the
diagonal element of the
Cartan matrices corresponding to each node on the (extended) Dynkin
diagram is always equal to 2 and can be calculated through the
Coxeter labels surrounding this node. Let us consider, for example,
the  $A_r$ series.
For each internal node  $N_i$ of the corresponding Dynkin diagram
the value of the
diagonal Cartan element $A_{ii}$ satisfies the  relation:
$A_{ii}=(C_{i-1}+C_{i+1})/C_{i}=2$, where $C_{...}=1$ are
Coxeter labels of the nodes $N_{i-1}$, $N_i$, $N_{i+1}$. To generalize
this relation  to boundary nodes one should consider the extended
Dynkin graph of the affine $A_r^{(1)}$ algebra. In this case all
nodes are linked  by lines with two neighbours. In the Dynkin graphs
of the $D_r^{(1)}$ series apart from several nodes with two lines
there are two nodes with three lines. The above
relation between the Coxeter labels and the diagonal Cartan element
can be easily checked for nodes with two
and three lines. In particular,
for a triple node $N_i$ one obtains
$A_{ii}=(C_{i+}+C_{i-1}+C_{i-})C_{i}=(1+2+1)/2=2$. For this rule to
be also valid for a boundary node one can formally add an additional  node
with a vanishing Coxeter label. The above two examples show that
the Coxeter labels 1 or 2 allow one to construct an infinite series
of Dynkin diagrams. The larger values $A_{ii}\geq 3$ for triple
nodes are allowed only for some special values of the algebra
rank, as we can see in the cases of $E_6$, $E_7$, $E_8$
algebras.

The extended reflexive number $(0,0,0,0,1)$ is the origin of the
infinite series of the Berger
graphs with a multi--cycle topology. One can compare them with the
Kac--Moody case of
the $A_r^{(1)}$ infinite series where the graphs have
only  one cycle.
The simplest example of multi--cycle topology corresponds to the
tetrahedron Berger graph
having 4 closed cycles with the corresponding  $4 \times 4 $
matrix:
{\small \begin{eqnarray}
B_3^{(1)}(00001)= \left (
\begin{array}{cccc}
 3& -1& -1& -1 \\
-1&  3& -1& -1 \\
-1& -1&  3& -1 \\
-1& -1& -1&  3 \\
\end{array}
\right )
\nonumber
\end{eqnarray}}
This matrix has the eigenvalues $ \{0, 4, 4, 4\}$ for the corresponding
eigenvectors $(1,1,1,1)$, $(-1,0,0,1)$, $(-1,0,1,0)$, $(-1,1,0,0)$. The Coxeter
labels are given by the zero eigenvector
$\{\,1,\, 1, \,1, \,1\, \}$, which provides the well--known relation
for the highest root
(affine condition):
$1 \cdot \alpha_0+1 \cdot \alpha_1+1 \cdot \alpha_2+1 \cdot \alpha_3=0$,
where $\alpha_i$, $i=1,2,3$ are the simple roots and $-\alpha_0$ is equal to
the highest root $\alpha_h $.
For the non--affine case one should remove one node from the Berger graph.
Thus the relation
between the affine and non--affine Berger graphs is similar to the relation
between the Cartan--Lie and Kac--Moody graphs.
In the last case the Cartan--Lie algebra produces the so--called horizontal
subalgebra of the Kac--Moody algebra where the highest root participates
in the construction of the  additional simple root,
more exactly,  $-\alpha_h$. Such non--affine Berger matrix is positively defined
and   Det $B_3^{(1)}(00001)=16$. The affine condition and
the positive definiteness of the non--affine
Berger matrix do not depend on the number of internal nodes with $B_{ii}=2$.
For instance, the non--affine Berger matrix after its inversion offers the
possibility to construct three ${\mathbb Z}_3$--symmetric fundamental weights:
\begin{eqnarray}
\Lambda_1 &=&\{\, \frac{1}{2}, \, \frac{1}{4},\, \frac{1}{4}\,\},\,\, \,
\Lambda_2 ~=~\{\, \frac{1}{4}, \, \frac{1}{2},\, \frac{1}{4}\,\},\,\,\,
\Lambda_3 ~=~\{\, \frac{1}{4}, \, \frac{1}{4},\, \frac{1}{2}\,\}.
\end{eqnarray}
Note that  in the Cartan--Lie $A_r$ algebras there are two
elementary fundamental representations, but in the Berger case
one already has three elementary fundamental ``representations''. In our 
example we
obtain the ${\mathbb Z}_3$ symmetry acting in the space of the fundamental
representations.  One can believe that this new invariance could
have some  applications to the solution of the family generation problem
in the electro-weak theory.

Another example is related to the generalization of the $D_r$--infinite series
of the Cartan--Lie  Dynkin graphs.
From our point of view the B(011) graph is exceptional. The graphs B(0011) in the K3
2--arity polyhedra produce an infinite series of $D_r$--Cartan--Lie Dynkin graphs.
They are built from two $(1,1)[2]$ blocks connected by a segment with 
$l$--internal nodes
having $B_{ii}=2$ and Coxeter label equal to 2.
The Berger graphs B(00011) could have three blocks $(1,1)[2]$. Each two of them
~are connected by a line with the nodes having $B_{ii}=2$ and the Coxeter labels
equal to 2. We illustrate this by the following  Berger matrix:
{\small \begin{eqnarray}
B_8^{(1)}(00011)= \left (
\begin{array}{ccccccccc}
  2&  0&-1& 0& 0& 0& 0& 0& 0\\
  0&  2&-1& 0& 0& 0& 0& 0& 0\\
 -1& -1& 3& 0& 0&-1& 0& 0&-1\\
  0&  0& 0& 2& 0&-1& 0& 0& 0\\
  0&  0& 0& 0& 2&-1& 0& 0& 0\\
  0&  0&-1&-1&-1& 3& 0& 0&-1\\
  0&  0& 0& 0& 0& 0& 2& 0&-1\\
  0&  0& 0& 0& 0& 0& 0& 2&-1\\
  0&  0&-1& 0& 0&-1&-1&-1& 3\\
\end{array}
\right )
\nonumber
\end{eqnarray}}\\
For simplicity we did not include any of the internal nodes.
The eigenvalues of this matrix are $\{0, 2, 2, 2, 3, 3 - \sqrt{3},
3 -  \sqrt{3}, 3+\sqrt{3},3 +\sqrt{3} \}$ and the
eigenvector with the vanishing eigenvalue is $\{1, 1, 2, 1, 1, 2, 1, 1, 2\}$. To
discuss the infinite $B(00011)$ series let us recall the usual 
$D_r^{(1)}$ Dynkin
graphs which are described by the $B(0011)$ graphs in the $K3$ polyhedra.
For example we consider the diagram $D_8^{(1)}=B_8^{(1)}(0011)$ with three
internal nodes:
{\small \begin{eqnarray}
B_8^{(1)}(0011)= \left (
\begin{array}{ccccccccc}
  2&  0&-1& 0& 0& 0& 0& 0& 0\\
  0&  2&-1& 0& 0& 0& 0& 0& 0\\
 -1& -1& 2&-1& 0& 0& 0& 0& 0\\
  0&  0&-1& 2&-1& 0& 0& 0& 0\\
  0&  0& 0&-1& 2&-1& 0& 0& 0\\
  0&  0& 0& 0&-1& 2&-1& 0& 0\\
  0&  0& 0& 0& 0&-1& 2&-1&-1\\
  0&  0& 0& 0& 0& 0&-1& 2& 0\\
  0&  0& 0& 0& 0& 0&-1& 0& 2\\
\end{array}
\right )
\nonumber
\end{eqnarray}}\\
The determinant of this matrix is equal to zero. The affine condition reads
$1 \cdot \alpha_0+ 1\cdot  \alpha_1+2\cdot  \alpha_2+
  2 \cdot \sum_{i=3}^{i=5}  \alpha_i +2\cdot  \alpha_6+
1 \cdot \alpha_7+ 1\cdot  \alpha_8 ~=~ 0$, with  $-\alpha_0 $ being the
highest root of the $D_8$ Cartan--Lie algebra. This condition relates the
$D_8^{(1)}$  Kac--Moody algebra to the non--affine case of the Cartan--Lie
$D_8$ algebra. One can construct the Cartan matrix corresponding to this algebra
{\small \begin{eqnarray}
B_8(0011)= \left (
\begin{array}{ccccccccc}
  2&-1& 0& 0& 0& 0& 0& 0\\
 -1& 2&-1& 0& 0& 0& 0& 0\\
  0&-1& 2&-1& 0& 0& 0& 0\\
  0& 0&-1& 2&-1& 0& 0& 0\\
  0& 0& 0&-1& 2&-1& 0& 0\\
  0& 0& 0& 0&-1& 2&-1&-1\\
  0& 0& 0& 0& 0&-1& 2& 0\\
  0& 0& 0& 0& 0&-1& 0& 2\\
\end{array}
\right )
\nonumber
\end{eqnarray}}\\
The determinant of this matrix is equal to 4 independently of the
internal nodes. The determinant  of the
non--affine Berger matrix Det $B(00011)$ in the above example is equal
to 48 and depends on the internal nodes. In this case one can also find the
fundamental nodes:
{\small {\begin{eqnarray}
G_8(00011)=
 \left (
\begin{array}{c||cc|ccc|ccc}
 F.W.      &\alpha_{a1}&\alpha_{a2} & \alpha_{a3}&\alpha_{a4}
           &\alpha_{a5}&\alpha_{a6}&\alpha_{a7} & \alpha_{a8}
\\\hline
\Lambda_{a1}& 1   & 1& 1/2 & 1/2 & 1  & 1/2 & 1/2 & 1  \\
\Lambda_{a2}& 1   & 2& 1   & 1   & 2  & 1   & 1   & 2  \\
\Lambda_{a3}& 1/2 & 1& 7/6 & 2/3 & 4/3& 7/12& 7/12& 7/6\\
\Lambda_{a4}& 1/2 & 1& 2/3 & 7/6 & 4/3& 7/12& 7/12& 7/6\\
\Lambda_{a5}& 1   & 2& 4/3 & 4/3 & 8/3& 7/6 & 7/6 & 7/3\\
\Lambda_{a6}& 1/2 & 1& 7/12& 7/12& 7/6& 7/6 & 2/3 & 4/3\\
\Lambda_{a7}& 1/2 & 1& 7/12& 7/12& 7/6& 2/3 & 7/6 & 4/3\\
\Lambda_{a8}& 1   & 2& 7/6 & 7/6 & 7/3& 4/3 & 4/3 & 8/3\\
\end{array}
\right )
\nonumber
\end{eqnarray}}}\\
The $B(0011)$ graphs help us to clarify the structure of the Berger graphs 
determined by the matrices
$B(00111)$, $B(00112)$, $B(00123)$ in CY$_3$ reflexive polyhedra of 2--arity.
In the case of $K3$ the $B(0111)$ graph generates  one exceptional
$E_6^{(1)}$ graph.
In the higher dimension $n=5$ we obtain the graph from an infinite series,
constructed from two
$E_6^{(1)}$ blocks in which we should change two nodes
$A_{ii}=2 \rightarrow B_{ii}=3$.

Now we consider the case of the $B(00111)$ Berger graph with one
internal node $b$ having the Coxeter label 3 placed  between two generalized
forms of $E_6^{(1)}$ exceptional graphs having the central nodes $B_{ii}=a$:
{\small \begin{eqnarray}
B_{14}^{(1)}(00111)= \left (
\begin{array}{ccccccc|c|ccccccc}
    2&-1& 0& 0& 0& 0& 0& 0& 0& 0& 0& 0& 0& 0& 0 \\
   -1& 2& 0& 0& 0& 0&-1& 0& 0& 0& 0& 0& 0& 0& 0 \\
    0& 0& 2&-1& 0& 0& 0& 0& 0& 0& 0& 0& 0& 0& 0 \\
    0& 0&-1& 2& 0& 0&-1& 0& 0& 0& 0& 0& 0& 0& 0 \\
    0& 0& 0& 0& 2&-1& 0& 0& 0& 0& 0& 0& 0& 0& 0 \\
    0& 0& 0& 0&-1& 2&-1& 0& 0& 0& 0& 0& 0& 0& 0 \\
    0&-1& 0&-1& 0&-1& a&-1& 0& 0& 0& 0& 0& 0& 0 \\  \hline
    0& 0& 0& 0& 0& 0&-1& b&-1& 0& 0& 0& 0& 0& 0 \\  \hline
    0& 0& 0& 0& 0& 0& 0&-1& a&-1& 0&-1& 0&-1& 0 \\
    0& 0& 0& 0& 0& 0& 0& 0&-1& 2&-1& 0& 0& 0& 0 \\
    0& 0& 0& 0& 0& 0& 0& 0& 0&-1& 2& 0& 0& 0& 0 \\
    0& 0& 0& 0& 0& 0& 0& 0&-1& 0& 0& 2&-1& 0& 0 \\
    0& 0& 0& 0& 0& 0& 0& 0& 0& 0& 0&-1& 2& 0& 0 \\
    0& 0& 0& 0& 0& 0& 0& 0&-1& 0& 0& 0& 0& 2&-1 \\
    0& 0& 0& 0& 0& 0& 0& 0& 0& 0& 0& 0& 0&-1& 2 \\
\end{array}
\right )
\nonumber
\end{eqnarray}}\\
with the determinant ${\rm Det} \, B_{14}(00111)=27^2(a-2)[b(a-2)-2)]$. For the
choice of parameters $a=3$ and $b=2$, the eigenvalues of this Berger matrix read
$\{0,\, 1,\, 1,\, 1,\, 1,\, 2,\, 3,\, 3,\, 3,\, 3,\, 4,\,
2 - \sqrt{3-\sqrt {7}},\, 2 + \sqrt{3-\sqrt {7}},\,
2-\sqrt{3+\sqrt {7}} ,\, 2+\sqrt{3+\sqrt {7}}\,\}$. The eigenvector with zero
eigenvalue corresponds to the Coxeter labels:
$C_i=\{1,\, 2,\, 1,\, 2,\, 1,\, 2,\, 3,\,: 3,:\, 3,\, 2,
\,1, \,2,\, 1,\, 2,\, 1\,\}$,
which are similar to those for the $E_6^{(1)}$--diagrams.
The link between two parts of the graph can be extended by an arbitrary
number of internal nodes with Coxeter labels 3. This choice of labels is
supported by the construction of the corresponding reflexive polyhedron in
CY$_3$. A different selection of parameters: $a=2$, $b=2$, generates unusual
Coxeter labels:
$C_i=\{-1,\, -2,\, -1,\, -2,\, -1,\, -2,\, -3,\,: 0,: \,3,\, 2,\, 1,\, 2,\,
1,\, 2,\, 1\,\}$,
with Coxeter number equal to 0.
We would like to stress the fact that the solution $a=3$ and $b_l=2$ exists for any
$l=0,1,2,3,....$, and gives an infinite series of corresponding Berger graphs,
$B^{(1)}_{l}(00111)[3]$.

To obtain non--affine Berger graphs, {\it i.e.} the analog of the Cartan--Lie
case, one should remove one root with the Coxeter label equal to one. This means
that the simple roots $\alpha_i$ on the Berger graph
define the highest root $\alpha_h=-\alpha_0$ (the affine condition), {\it i.e.},
$\alpha_0+\sum_i C_i \cdot \alpha_i=0$, where $C_i$ are the Coxeter labels. In
this case one can check that the determinant of the
Berger matrix is equal to 81, a value which does not depend on the number $l$ of the
internal nodes. Also, all principal minors are positive--defined in a similar way to
the Cartan--Lie case.
In a complete analogy with the Cartan case the Berger non--affine graph also
defines the fundamental weights which read
{\small {\begin{eqnarray}
&&\hspace{-16cm}G_{15}(00111)= \nonumber\\
 \left (
\begin{array}{c||cccccc|c|ccccccc}
 F.W.      &\alpha_{a1}&\alpha_{a2}& \alpha_{a3}&\alpha_{a4}&
\alpha_{a5}&\alpha_{a6}&\alpha_{c} & \alpha_{b6}&
\alpha_{b5}&\alpha_{b4}&\alpha_{b3}& \alpha_{b2}&\alpha_{b1}&\alpha_{b0}
\\\hline\hline
\Lambda_{a1}& 2& 1  & 2   &1  & 2   & 3& 3   & 3   &2   &1    &2    & 1  & 2    & 1    \\
\Lambda_{a2}& 1& 4/3& 5/3 &2/3& 4/3 & 2& 2   &2    &4/3 &2/3  & 4/3 & 2/3 & 4/3 & 2/3  \\
\Lambda_{a3}& 2& 5/3& 10/3&4/3& 8/3 & 4& 4   &4    &8/3 &4/3  & 8/3 & 4/3 & 8/3 & 4/3  \\
\Lambda_{a4}& 1& 2/3& 4/3 &4/3& 5/3 & 2& 2   &2    &4/3 &2/3  & 4/3 & 2/3 & 4/3 & 2/3  \\
\Lambda_{a5}& 2& 4/3& 8/3 &5/3& 10/3& 4& 4   &4    &8/3 &4/3  & 8/3 & 4/3 & 8/3 & 4/3  \\
\Lambda_{a6}& 3& 2  &4    & 2 &4    & 6& 6   & 6   &4   &2    &4    & 2   & 4   & 2    \\
\hline
\Lambda_{c}&  3& 2  &4    & 2 & 4   &6 & 7   & 7   &14/3&7/3  &14/3 & 7/3 & 14/3& 7/3  \\
\hline
\Lambda_{b6}& 3& 2  &4    & 2 & 4   &6 & 7   & 8   &16/3&8/3  &16/3 & 8/3 & 16/3& 8/3  \\
\Lambda_{b5}& 2& 4/3& 8/3 &4/3& 8/3 & 4& 14/3&16/3 &38/9& 19/9& 32/9& 16/9& 32/9& 16/9 \\
\Lambda_{b4}& 1& 2/3& 4/3 &2/3& 4/3 & 2& 7/3 &8/3  &19/9& 14/9& 16/9& 8/9 & 16/9& 8/9  \\
\Lambda_{b3}& 2& 4/3& 8/3 &4/3& 8/3 & 4& 14/3&16/3 &32/9& 16/9& 38/9& 19/9& 32/9& 16/9 \\
\Lambda_{b2}& 1& 2/3& 4/3 &2/3& 4/3 & 2& 7/3 &8/3  &16/9& 8/9 & 19/9& 14/9& 16/9& 8/9  \\
\Lambda_{b1}& 2& 4/3& 8/3 &4/3& 8/3 & 4& 14/3&16/3 &32/9& 16/9& 32/9& 16/9& 38/9& 19/9 \\
\Lambda_{b0}& 1& 2/3& 4/3 &2/3& 4/3 & 2& 7/3 &8/3  &16/9& 8/9 & 16/9& 8/9 & 19/9& 14/9 \\
\end{array}
\right ) \nonumber
\end{eqnarray}}}
The Dynkin diagrams for the Cartan--Lie/Kac--Moody algebras can have
the nodes with the maximal number of edges equal to 3.  For instance, let us
take the $E_6^{(1)}$,$E_7^{(1)}$, $E_8^{(1)}$ graphs and consider the
vertex--nodes  having three edges and  the Coxeter labels equal to
3, 4 and 6, respectively. It is known that in the case of the Cartan--Lie
algebras the number of Casimir invariants coincides with the
algebra rank.  The $r$-degrees of these Casimir take values from 2 up
to the maximum equal to the Coxeter number.
The important  cases correspond to the degrees of invariants for the three
above--mentioned algebras equal to the following sums:
Cas$_{i}=C_{i-1}+C_{i+}+C_{i-}=2+2+2=6$, Cas$=3+3+2=8$, Cas$=5+4+3=12$, respectively.
The diagonal elements of the Cartan matrices for the nodes in these cases are
equal to $A_{ii}={\rm Cas}/C_{ii}=6/3=2$, $A_{ii}={\rm Cas}/C_{ii}=8/4=2$,
$A_{ii}={\rm Cas}/C_{ii}=12/6=2$. The relation between the Coxeter labels and
$B_{ii}$ is also valid for all nodes with 2, 3, 4,... edges. Here
the diagonal elements of the Berger matrix are  $B_{ii} = 3, 4, ...$.

The extension of the $A_r^{(1)}$ series for the $B_{ii}=3$ case
gives a new infinite series $B(00001)$ in which for all
triple nodes $N_i$ we have
$B_{ii}=(C_{i-1}+C_{i+}+C_{i-})/C_{i}=3$, where all $C_{...}=1$. Hence
one can obtain the infinite series of graphs both for the Cartan nodes
$A_{ii}=2$ and for the Berger nodes with $B_{ii}=3$. Note that
all  Coxeter labels for the  Berger graphs $B(0...01)$ are equal
to 1.

A similar extension of the $D_r^{(1)}$ series is $B(00011)$,
where apart from the Cartan nodes $A_{ii}=2$ there appear two nodes
$B_{ii}=3$ with 4 edges. Here one has $B_{ii}=
(C_{i-1}+C_{i+1}+C_{i+}+C_{i-})/C_{i}= (2+2+1+1)/2=3$. Note that
for the Berger graphs the maximal Coxeter label is equal
to 2. Therefore for all Berger graphs we have only two possibilities for
Coxeter labels, 1 or 2.  In the Cartan--Lie case one obtains just two infinite
series of simply--laced Dynkin diagrams, with the maximal Coxeter
labels equal to 1 and 2. For other examples of simply--laced Cartan--Lie
algebras their maximal values are 3, 4 and 6. This is related
to the fact that the corresponding  algebras are exceptional.

\begin{figure}
\vspace{-3cm}
\centerline{\hspace{2cm}\epsfig{file=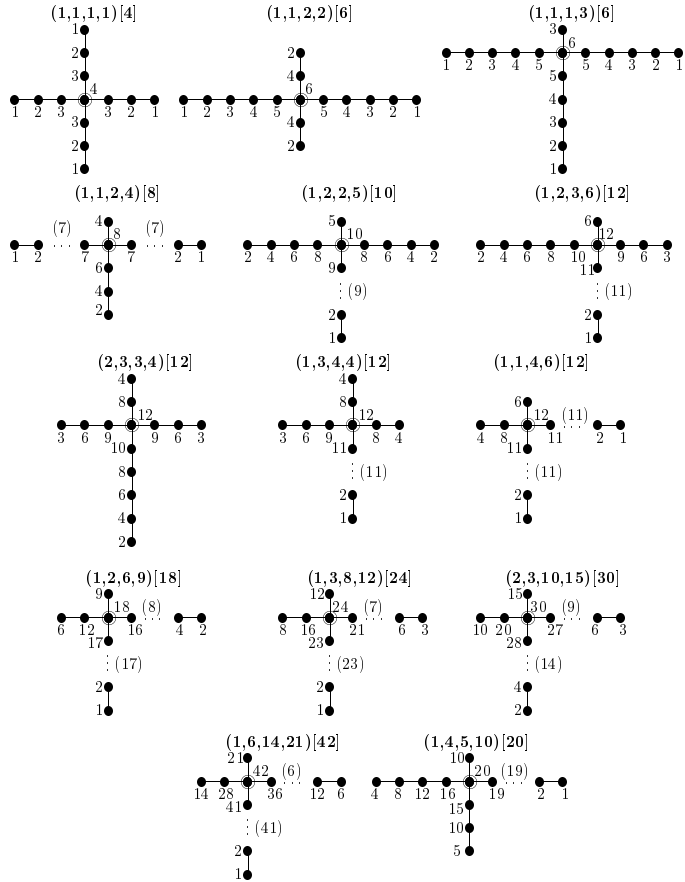,width=10cm}}
\vspace{-0.4cm}
\caption{14 exceptional new Berger
diagrams with the Coxeter labels at the nodes.}
\label{14All}
\end{figure}

Apart from the five types of infinite series, which can be
interpreted as generalizations of the corresponding Cartan--Lie
simply--laced graphs, we also found 14
exceptional completely new graphs (see Fig.~\ref{14All})
corresponding to 14 simply--laced numbers
inside the 95 $K_3$ reflexive numbers as shown
in Table~\ref{BERDET}.
As it was mentioned above, the affine graphs have symmetric Berger matrices
with their determinant equal to zero, which is similar to the 
Kac--Moody type of
infinite-dimensional algebras. Removing one node with a minimal Coxeter label
we can obtain the non--affine graph generalizing the
Cartan--Lie case. For the corresponding Berger matrix the determinant is
positively defined
and all principal minors are also positive, see Table~\ref{BERDET}.
{\small \begin{table}
\vspace{-3cm}
\hspace{2cm}\begin{tabular}{|l|c|c|c|c|} \hline
${\vec k}_{3,4}^{\rm ext}        $ & $ {\rm Rank} $&$h            $ & ${\rm Casimir} (B_{ii})
$ & $ {\rm Determinant}                                                   $  \\
\hline \hline
$(0,1,1,1)[3]      $ & $ 6 (E_6)  $ & $12         $ & $6          $ & $3  $  \\
$(0,1,1,2)[4]      $ & $ 7 (E_7)  $ & $18         $ & $8          $ & $2  $  \\
$(0,1,2,3)[6]      $ & $ 8 (E_8)  $ & $30         $ & $12         $ & $1  $  \\
$(0,0,1,1,1)[3]    $ & $ 2_3+10+l $ & $18+3(l+1)  $ & $9          $ & $3^4$  \\
$(0,0,1,1,2)[4]    $ & $ 2_3+13+l $ & $32+4(l+1)  $ & $12         $ & $4^3$  \\
$(0,0,1,2,3)[6]    $ & $ 2_3+15l  $ & $60+6(l-1)  $ & $18         $ & $6^2$  \\ \hline
$(0,1,1,1,1)[4]    $ & $ 1_3+11   $ & $28         $ & $12         $ & $16 $  \\
$(0,2,3,3,4)[12]   $ & $ 1_3+12   $ & $90         $ & $36         $ & $ 8 $  \\
$(0,1,1,2,2)[6]    $ & $ 1_3+13   $ & $48         $ & $18         $ & $ 9 $  \\
$(0,1,1,1,3)[6]    $ & $ 1_3+15   $ & $54         $ & $18         $ & $ 12$  \\
$(0,1,1,2,4)[8]    $ & $ 1_3+17   $ & $80         $ & $24         $ & $ 8 $  \\
$(0,1,2,2,5)[10]   $ & $ 1_3+17   $ & $100        $ & $30         $ & $ 5 $  \\
$(0,1,3,4,4)[12]   $ & $ 1_3+17   $ & $120        $ & $36         $ & $ 3 $  \\
$(0,1,2,3,6)[12]   $ & $ 1_3+19   $ & $132        $ & $36         $ & $ 6 $  \\
$(0,1,4,5,10)[20]  $ & $ 1_3+26   $ & $290        $ & $60         $ & $ 2 $  \\ \hline
$(0,1,1,4,6)[12]   $ & $ 1_3+24   $ & $162        $ & $36         $ & $ 6 $  \\
$(0,1,2,6,9)[18]   $ & $ 1_3+27   $ & $270        $ & $54         $ & $ 3 $  \\
$(0,1,3,8,12)[24]  $ & $ 1_3+32   $ & $420        $ & $72         $ & $ 2 $  \\
$(0,2,3,10,15)[30] $ & $ 1_3+25   $ & $420        $ & $90         $ & $ 4 $  \\
$(0,1,6,14,21)[42] $ & $ 1_3+49   $ & $1092       $ & $126        $ & $ 1 $  \\
\hline
\end{tabular}
\caption{Rank, Coxeter number $h$, Casimir depending on $B_{ii}$ and determinants
for the non--affine exceptional Berger graphs. The maximal Coxeter labels
coincide with the degree of the corresponding reflexive simply--laced vector.
The determinants in the last
column for the infinite series (0,0,1,1,1)[3], (0,0,1,1,2)[4] and (0,0,1,2,3)[6]
are independent from the number $l$ of internal binary $B_{ii}=2$ nodes.
The numbers $1_3$ and $2_3$ denote the number of nodes with $B_{ii}=3$.}
\label{BERDET}
\end{table}}

We see that for the Berger graphs one can build the infinite series
with the maximal Coxeter labels 3, 4, 6 due to the presence of new nodes
with $B_{ii}=3$. These new nodes lead to the appearance of 14 exceptional
simply-laced Berger graphs with their maximal Coxeter labels:
4, 6, ..., 42. When we introduce the new nodes $B_{ii} = 4$ these 14
exceptional examples produce new 14 infinite series of Berger graphs.
This could be a key point to understand the nature of exceptional cases in
Cartan--Lie  and  Berger algebras.

The Cartan graphs of dimension 1, 2, 3 are well--known and correspond to the
classical Cartan--Lie algebras. The main purpose of the above discussion was to
enlarge their list with graphs generated by the reflexive weight vectors of dimension
four (corresponding to CY$_3$). In this dimension, corresponding to
$K3$--sliced CY$_3$ spaces, we have singled out the
following fourteen reflexive weight vectors from the total of 95 $K3$--vectors
(1,1,1,1)[4], (1,1,2,2)[6], (1,1,1,3)[6], (1,1,2,4)[8], (2,3,3,4)[12],
(1,3,4,4)[12], (1,2,3,6)[12], (1,2,2,5)[10], (1,4,5,10)[20], (1,1,4,6)[12],
(1,2,6,9)[18], (1,3,8,12)[24], (2,3,10,15)[30] and (1,6,14,21)[42].
The Coxeter labels for the nodes in the Berger graphs were assigned
consistently from the geometrical and algebraic points of view. These Berger matrices
have simple properties: they are symmetric and affine. In addition, these graphs
and matrices are not  extendable, i.e. other graphs and Berger matrices cannot be
obtained from them by adding more nodes to any of the legs.
In this respect these graphs are ``exceptional''. Similarly  to
classical non--exceptional graphs, one can construct infinite series 
containing them.
Apparently these fourteen vectors are the only set of vectors
with symmetric Berger matrices among
the  total 95 reflexive vectors. We
investigate such Berger matrix in the simplest case 
in Appendix~\ref{CaseStudy}.

%% file: Conclusions.tex
\section{Conclusions, outlook and further possibilities}

In this work we have investigated different properties of the weight 
vectors and graphs related to the CY reflexive polyhedra.
The  important correspondence between
CY$_{3,4}$ spaces and reflexive polyhedra was discovered several years ago
by Batyrev~\cite{Bat}. In previous works~\cite{AENV1, V,TV, ETV}
an interesting relation between reflexive vectors  and
Dynkin and Berger graphs with dimensions $n=4,5,..$ was established.
It was shown with the use of the Universal Calabi--Yau Algebra (UCYA) and
the arity construction,
that the structure of the reflexive polyhedra can be 
described in terms of the Berger graphs.

In this paper
we studied the reflexivity properties of the Batyrev polyhedra and
the UCYA construction using some geometrical and algebraic ideas.
The reflexivity condition for a polyhedron corresponds to the existence of one 
or several dual polyhedra. This condition was formulated as a positivity
requirement for an expression depending on the corresponding weight vector.   
A physical interpretation of the Calabi-Yau spaces in terms of zeroes of 
the wave 
function for a string-like harmonic oscillator model was presented. 
The UCYA approach was illustrated in the problem of constructing the
reflexive polyhedra containing the given polyhedron inside it. We extracted
large classes of reflexive polyhedrons based 
on the simple-laced and quasi-simple-laced numbers and investigated their
properties  with the use of number theory, recurrence relations and 
functional methods of quantum field theory. In particular the simple-laced 
reflexive vectors were related to the so-called "Egyptian fractions"~\cite{Guy}.
We suggested a classification of quasi-simple-laced numbers. Our 
general approach was illustrated by numerous examples (see Appendices).  

We proceeded to study the relation between the reflexive vectors and the 
structure of the Berger graphs . It was demonstrated that the simply--laced 
reflexive numbers are generators of the n--dimensional Berger graphs. 
The well--known Dynkin diagrams for the Cartan--Lie algebras produce
a subclass of the full set of the Berger graphs. We assigned the Coxeter
labels and weights $B_{ii}$ for the nodes in these graphs.
The suggested  prescriptions for the 
corresponding Berger matrices for arbitrary $n$ in agreement
with the rules for the Cartan matrices $A_{ij}$. The
constraints for the Cartan case
are well known~\cite{CSM, FSS, Fuchs, Gil, Wyb}:
$$
A_{ii} = 2, \,\, A_{ij} \leq 0  (i \neq j), \,\, A_{ij}=0  
\leftrightarrow A_{ji}=0, \,\, 
A_{ij} \in {\mathbb Z}, \,\, {\rm Det} \,A  > 0.
$$
Thus, the rank of $A$ is equal to $r$. For the Kac--Moody case one
should modify only the last condition for the
determinant $A$. Indeed, by neglecting the positivity requirement
for Det $A$ one obtains  a 
class of the Kac--Moody algebras. Interesting subclasses
of Kac--Moody algebras can be constructed if this restriction on the
determinant is replaced as follows
\begin{eqnarray}
{\rm Det} \, A_{i} > 0, \qquad i=0,1,2,...,r,
\end{eqnarray}
where $A_{i}$ are the matrices in which the $i^{\rm th}$ row and the 
$i^{\rm th}$ column are removed.
The determinants of $A_{i}$  are called the principal minors. In
a general case the rank of $A$ can be arbitrary, but provided that the new
restriction is imposed its rank is $r$ or $r+1$.
For the Cartan--Lie algebras the rank is equal to $r+1$. For the
affine Lie case the
Cartan matrices have rank $r$ and all principal minors
are positive. In the Berger case we impose  the same restrictions
on the determinant and principal minors.
Then the Berger matrix turns out to be
a degenerated semi--definite matrix which is called the
affine matrix.
We modify the first condition for $A_{ij}$ and allow the
diagonal element $B_{ii}$ to be all positive integers,
i.e., 2, 3, 4, ....

Summing up the above discussion, we conclude that one can construct not
only diagrams for the Cartan (Det $B>0$) and
affine  (Det $B=0$, Det $B_{i}>0$) cases, but  
it is possible to obtain some information from the generalized Berger
graphs also
about  the roots and weights for extended algebraic structures.
In particular one can generalize  the
known simply--laced series $A_r$ and $D_r$ and exceptional
simply--laced algebras $E_{6,7,8}$ and $E_{6,7,8}^{(1)}$.  The
Berger matrices for the new simply--laced graphs in dimensions
$n = 4, 5, ...$ share a number of properties with the Cartan matrices
in dimensions $n=1,2,3$.

The interest in the construction of new algebraic structures beyond
the Lie algebras started  from the investigation of
$SU(2)$--conformal field theories~\cite{CIZ,FZ,Itz,Ito}
(see also~\cite{DWA}). One can
expect that geometrical concepts, in particular, algebraic geometry,
are a natural and promising way to discover new algebras. Historically the
marriage of algebra and geometry was useful for both 
branches of mathematics. In particular, to prove the mirror symmetry of the
Calabi--Yau spaces the powerful technique of
the Newton reflexive polyhedra was used.
Furthermore, the $ADE$--type singularities~\cite{Klein,DuVal, Lam, KoSha}
in $K3 ={\rm CY}_2$ spaces~\cite{Asp, Bersh, KV} and their resolution
were related to the
Dynkin diagrams for the Cartan--Lie algebra. One can
formulate these relations
in the following way: 
\begin{enumerate}
\item{The algebraic origin of the Cartan--Lie algebras is the Torus and
$SU(2)/U(1)$}.
\item{The geometrical origin of the Cartan--Lie algebras is
$S^1$ and $CP^1\cong S^2$.}
\end{enumerate}
It is possible to continue this correspondence to the Berger graphs
\begin{enumerate}
\item{The algebraic origin of the Berger graphs is the
Torus and $SU(3)/SU(2)\times U(1)$}.
\item{The geometrical origin of the Berger graphs is $S^1$ and $CP^2$}.
\end{enumerate}
Another possibility to understand the origin of the
Berger graphs is related to the resolution of the quotient singularities.
The Calabi--Yau spaces are defined by their holonomy 
groups~\cite{Berger,Beau,Bry}.
Typical quotient singularities $C^n/G$ are characterized by the list of
finite subgroups $G\subset H$ of the holonomy groups, $H=SU(2),SU(3),...$.
For the case of the $ SU(2)$ holonomy group there are five well--known
Klein--Du--Val singularities of $A$--$D$--$E$
type~\cite{Klein, Du Val, Lam, KoSha}.
And the crepent  resolution of $A$--$D$--$E$--types singularities in $K3$
gives us the corresponding Dynkin diagrams in the $K3$ polyhedra. So it is
natural to relate the Berger graphs in CY$_3$ polyhedra to the
resolution of the $C^3/G$ quotient singularities, where $G$ are
finite subgroups of $SU(3)$\cite{YY}. Although this relation is not 
established,
we note that the number of finite subgroups
of $SU(3)$ is $5+12$ (see~\cite{YY}). Further, the first five
finite subgroups  are isomorphic to the groups  $SU(2)$ and one can
guess that they could be the origin of our five Berger graphs
$B(00001), B(00011), B(00111), B(00112), B(00123)$ discussed in this paper.

As a final remark, one can assume that the Berger graphs could
be linked to new algebras which are realized in
quadratic or/and  cubic matrices~\cite{Kern}.  The main question
at this point is how to unify in one approach the Cartan--Lie algebras 
and new hypothetical (ternary) algebras \cite{Nambu, Takhtajan, Filippov,
Kern, CIZ, FZ, Traubenberg}.

%% file: Appendix.tex
\appendix

\section{Quasi--Simply--Laced Reflexive Weight Vectors for $n=4$}
\label{QSLRWVn4}

To find all possible weight vectors in the $n=4$ case we firstly
construct all expansions of the form
\begin{equation}\label{eq.1}
d=k_{1}+k_{2}+k_{3}+k_{4}
\end{equation}
which fulfill the simply--laced condition
\begin{equation}\label{eq.2}
\frac{d}{k_{1}}=s_{1}\,,\,\,
\frac{d}{k_{2}}=s_{2}\,,\,\,
\frac{d}{k_{3}}=s_{3}\,,\,\,
\frac{d}{k_{4}}=s_{4}\,,
\end{equation}
with $s_{1},\ s_{2},\ s_{3},\ s_{4}>1$ being integer numbers.
For this purpose one can use a {computer code} based on the recurrent
relation~(\ref{recur}) for
the number $N_n(x,1/s)$ of decompositions of the rational number $x>0$ in the
ratios $1/s_k$.

To search for all possible quasi--simply--laced numbers it is helpful to
apply the above classification
of the classes of sets of equations. Let us consider in the following
 the different cases where one, two, three or four numerators
of the above ratios are modified.

For one modified numerator there is only one possibility (from now on
$s_i \in {\mathbb Z}^+$ and $s_i> 1$)
\begin{equation}\label{eq.3}
\frac{d}{k_{1}}=s_{1}\,,\,\,
\frac{d-k_1}{k_{2}}=s_{2}\,,\,\,
\frac{d}{k_{3}}=s_{3}\,,\,\,\
\frac{d}{k_{4}}=s_{4}\,.
\end{equation}
For two modified numerators there are four classes
\begin{equation}\label{eq.4}
\frac{d}{k_{1}}=s_{1}\,,\,\,
\frac{d-k_{1}}{k_{2}}=s_{2}\,,\,\,
\frac{d-k_{1}}{k_{3}}=s_{3}\,,\,\,
\frac{d}{k_{4}}=s_{4}\,,
\end{equation}
\begin{equation}\label{eq.5}
\frac{d}{k_{1}}=s_{1}\,,\,\,
\frac{d-k_{1}}{k_{2}}=s_{2}\,,\,\,
\frac{d-k_{4}}{k_{3}}=s_{3}\,,\,\,
\frac{d}{k_{4}}=s_{4}\,,
\end{equation}
\begin{equation}\label{eq.6}
\frac{d}{k_{1}}=s_{1}\,,\,\,
\frac{d-k_{1}}{k_{2}}=s_{2}\,,\,\,
\frac{d-k_{2}}{k_{3}}=s_{3}\,,\,\,
\frac{d}{k_{4}}=s_{4}\,,
\end{equation}
\begin{equation}\label{eq.7}
\frac{d}{k_{1}}=s_{1}\,,\,\,
\frac{d-k_{3}}{k_{2}}=s_{2}\,,\,\,
\frac{d-k_{2}}{k_{3}}=s_{3}\,,\,\,
\frac{d}{k_{4}}=s_{4}\,.
\end{equation}
In the case of three modified numerators seven classes exist
\begin{equation}\label{eq.8}
\frac{d}{k_{1}}=s_{1}\,,\,\,
\frac{d-k_{1}}{k_{2}}=s_{2}\,,\,\,
\frac{d-k_{1}}{k_{3}}=s_{3}\,,\,\,
\frac{d-k_{1}}{k_{4}}=s_{4}\,,
\end{equation}
\begin{equation}\label{eq.9}
\frac{d}{k_{1}}=s_{1}\,,\,\,
\frac{d-k_{3}}{k_{2}}=s_{2}\,,\,\,
\frac{d-k_{1}}{k_{3}}=s_{3}\,,\,\,
\frac{d-k_{1}}{k_{4}}=s_{4}\,,
\end{equation}
\begin{equation}\label{eq.10}
\frac{d}{k_{1}}=s_{1}\,,\,\,
\frac{d-k_{3}}{k_{2}}=s_{2}\,,\,\,
\frac{d-k_{2}}{k_{3}}=s_{3}\,,\,\,
\frac{d-k_{1}}{k_{4}}=s_{4}\,,
\end{equation}
\begin{equation}\label{eq.11}
\frac{d}{k_{1}}=s_{1}\,,\,\,
\frac{d-k_{3}}{k_{2}}=s_{2}\,,\,\,
\frac{d-k_{4}}{k_{3}}=s_{3}\,,\,\,
\frac{d-k_{1}}{k_{4}}=s_{4}\,,
\end{equation}
\begin{equation}\label{eq.12}
\frac{d}{k_{1}}=s_{1}\,,\,\,
\frac{d-k_{1}}{k_{2}}=s_{2}\,,\,\,
\frac{d-k_{2}}{k_{3}}=s_{3}\,,\,\,
\frac{d-k_{2}}{k_{4}}=s_{4}\,,
\end{equation}
\begin{equation}\label{eq.13}
\frac{d}{k_{1}}=s_{1}\,,\,\,
\frac{d-k_{3}}{k_{2}}=s_{2}\,,\,\,
\frac{d-k_{2}}{k_{3}}=s_{3}\,,\,\,
\frac{d-k_{2}}{k_{4}}=s_{4}\,,
\end{equation}
\begin{equation}\label{eq.14}
\frac{d}{k_{1}}=s_{1}\,,\,\,
\frac{d-k_{3}}{k_{2}}=s_{2}\,,\,\,
\frac{d-k_{4}}{k_{3}}=s_{3}\,,\,\,
\frac{d-k_{2}}{k_{4}}=s_{4}\,.
\end{equation}
And, finally, for four numerators six classes should be considered
\begin{equation}\label{eq.15}
\frac{d-k_{2}}{k_{1}}=s_{1}\,,\,\,
\frac{d-k_{1}}{k_{2}}=s_{2}\,,\,\,
\frac{d-k_{1}}{k_{3}}=s_{3}\,,\,\,
\frac{d-k_{1}}{k_{4}}=s_{4}\,,
\end{equation}
\begin{equation}\label{eq.16}
\frac{d-k_{2}}{k_{1}}=s_{1}\,,\,\,
\frac{d-k_{3}}{k_{2}}=s_{2}\,,\,\,
\frac{d-k_{1}}{k_{3}}=s_{3}\,,\,\,
\frac{d-k_{1}}{k_{4}}=s_{4}\,,
\end{equation}
\begin{equation}\label{eq.17}
\frac{d-k_{3}}{k_{1}}=s_{1}\,,\,\,
\frac{d-k_{3}}{k_{2}}=s_{2}\,,\,\,
\frac{d-k_{1}}{k_{3}}=s_{3}\,,\,\,
\frac{d-k_{1}}{k_{4}}=s_{4}\,,
\end{equation}
\begin{equation}\label{eq.18}
\frac{d-k_{4}}{k_{1}}=s_{1}\,,\,\,
\frac{d-k_{3}}{k_{2}}=s_{2}\,,\,\,
\frac{d-k_{1}}{k_{3}}=s_{3}\,,\,\,
\frac{d-k_{1}}{k_{4}}=s_{4}\,,
\end{equation}
\begin{equation}\label{eq.19}
\frac{d-k_{4}}{k_{1}}=s_{1}\,,\,\,
\frac{d-k_{3}}{k_{2}}=s_{2}\,,\,\,
\frac{d-k_{2}}{k_{3}}=s_{3}\,,\,\,
\frac{d-k_{1}}{k_{4}}=s_{4}\,,
\end{equation}
\begin{equation}\label{eq.20}
\frac{d-k_{2}}{k_{1}}=s_{1}\,,\,\,
\frac{d-k_{3}}{k_{2}}=s_{2}\,,\,\,
\frac{d-k_{4}}{k_{3}}=s_{3}\,,\,\,
\frac{d-k_{1}}{k_{4}}=s_{4}\,.
\end{equation}
For the simply--laced case in Eq.~(\ref{eq.2}) we have 14 solutions
\begin{eqnarray}
1\!\!\!&=&\ \frac{1}{4}+\frac{1}{4}+\frac{1}{4}+\frac{1}{4}\,,\quad
1\!\!\!~=~\ \frac{1}{2}+\frac{1}{6}+\frac{1}{6}+\frac{1}{6}\,,\quad
1\!\!\!~=~\ \frac{1}{3}+\frac{1}{3}+\frac{1}{6}+\frac{1}{6}\,,\nonumber\\
1\!\!\!&=&\ \frac{1}{2}+\frac{1}{4}+\frac{1}{8}+\frac{1}{8}\,,\quad
1\!\!\!~=~\ \frac{1}{2}+\frac{1}{3}+\frac{1}{12}+\frac{1}{12}\,,\quad
1\!\!\!~=~\ \frac{1}{2}+\frac{1}{5}+\frac{1}{5}+\frac{1}{10}\,,\nonumber\\
1\!\!\!&=&\ \frac{1}{2}+\frac{1}{4}+\frac{1}{6}+\frac{1}{12}\,,\quad
1\!\!\!~=~\ \frac{1}{2}+\frac{1}{3}+\frac{1}{9}+\frac{1}{18}\,,\quad
1\!\!\!~=~\ \frac{1}{3}+\frac{1}{3}+\frac{1}{4}+\frac{1}{12}\,,\nonumber\\
1\!\!\!&=&\ \frac{1}{2}+\frac{1}{3}+\frac{1}{8}+\frac{1}{24}\,,\quad
1\!\!\!~=~\ \frac{1}{2}+\frac{1}{4}+\frac{1}{5}+\frac{1}{20}\,,\quad
1\!\!\!~=~\ \frac{1}{2}+\frac{1}{3}+\frac{1}{7}+\frac{1}{42}\,,\nonumber\\
1\!\!\!&=&\ \frac{1}{3}+\frac{1}{4}+\frac{1}{4}+\frac{1}{6}\,,\quad
1\!\!\!~=~\ \frac{1}{2}+\frac{1}{3}+\frac{1}{10}+\frac{1}{15}\,.
\end{eqnarray}
In the class corresponding to Eq.~(\ref{eq.3}) there are 37 new solutions:
\begin{eqnarray}
1\!\!\!&=&  \frac{1}{3}+\frac{2}{9}+\frac{1}{3}+\frac{1}{9}\,,\quad
1\!\!\!~=~  \frac{1}{3}+\frac{2}{15}+\frac{1}{2}+\frac{1}{30}\,,\quad
1\!\!\!~=~  \frac{1}{3}+\frac{2}{15}+\frac{1}{3}+\frac{1}{5}\,,\nonumber\\
1\!\!\!&=&  \frac{1}{3}+\frac{2}{21}+\frac{1}{2}+\frac{1}{14}\,,\quad
1\!\!\!~=~  \frac{1}{4}+\frac{3}{8}+\frac{1}{3}+\frac{1}{24}\,,\quad
1\!\!\!~=~  \frac{1}{4}+\frac{3}{8}+\frac{1}{4}+\frac{1}{8}\,,\nonumber\\
1\!\!\!&=&  \frac{1}{4}+\frac{3}{16}+\frac{1}{2}+\frac{1}{16}\,,\quad
1\!\!\!~=~  \frac{1}{4}+\frac{3}{20}+\frac{1}{2}+\frac{1}{10}\,,\quad
1\!\!\!~=~  \frac{1}{4}+\frac{3}{28}+\frac{1}{2}+\frac{1}{7}\,,\nonumber\\
1\!\!\!&=&  \frac{1}{5}+\frac{2}{5}+\frac{1}{3}+\frac{1}{15}\,,\quad
1\!\!\!~=~  \frac{1}{5}+\frac{2}{5}+\frac{1}{5}+\frac{1}{5}\,,\quad
1\!\!\!~=~  \frac{1}{5}+\frac{4}{15}+\frac{1}{2}+\frac{1}{30}\,,\nonumber\\
1\!\!\!&=&  \frac{1}{5}+\frac{4}{15}+\frac{1}{3}+\frac{1}{5}\,,\quad
1\!\!\!~=~  \frac{1}{5}+\frac{2}{15}+\frac{1}{2}+\frac{1}{6}\,,\quad
1\!\!\!~=~  \frac{1}{10}+\frac{9}{20}+\frac{1}{5}+\frac{1}{4}\,,\nonumber\\
1\!\!\!&=&  \frac{1}{6}+\frac{5}{12}+\frac{1}{3}+\frac{1}{12}\,,\quad
1\!\!\!~=~  \frac{1}{6}+\frac{5}{12}+\frac{1}{4}+\frac{1}{6}\,,\quad
1\!\!\!~=~  \frac{1}{6}+\frac{5}{18}+\frac{1}{2}+\frac{1}{18}\,,\nonumber\\
1\!\!\!&=&  \frac{1}{6}+\frac{5}{24}+\frac{1}{2}+\frac{1}{8}\,,\quad
1\!\!\!~=~  \frac{1}{7}+\frac{2}{7}+\frac{1}{2}+\frac{1}{14}\,,\quad
1\!\!\!~=~  \frac{1}{7}+\frac{3}{14}+\frac{1}{2}+\frac{1}{7}\,,\nonumber\\
1\!\!\!&=&  \frac{1}{8}+\frac{7}{24}+\frac{1}{2}+\frac{1}{12}\,,\quad
1\!\!\!~=~  \frac{1}{8}+\frac{7}{24}+\frac{1}{3}+\frac{1}{4}\,,\quad
1\!\!\!~=~  \frac{1}{8}+\frac{7}{40}+\frac{1}{2}+\frac{1}{5}\,,\nonumber\\
1\!\!\!&=&  \frac{1}{9}+\frac{4}{9}+\frac{1}{3}+\frac{1}{9}\,,\quad
1\!\!\!~=~  \frac{1}{9}+\frac{2}{9}+\frac{1}{2}+\frac{1}{6}\,,\quad
1\!\!\!~=~  \frac{1}{10}+\frac{3}{10}+\frac{1}{2}+\frac{1}{10}\,,\nonumber\\
1\!\!\!&=&  \frac{1}{11}+\frac{5}{66}+\frac{1}{2}+\frac{1}{3}\,,\quad
1\!\!\!~=~  \frac{1}{12}+\frac{11}{24}+\frac{1}{3}+\frac{1}{8}\,,\quad
1\!\!\!~=~  \frac{1}{12}+\frac{11}{36}+\frac{1}{2}+\frac{1}{9}\,,\nonumber\\
1\!\!\!&=&  \frac{1}{15}+\frac{7}{30}+\frac{1}{2}+\frac{1}{5}\,,\quad
1\!\!\!~=~  \frac{1}{16}+\frac{5}{16}+\frac{1}{2}+\frac{1}{8}\,,\quad
1\!\!\!~=~  \frac{1}{16}+\frac{5}{48}+\frac{1}{2}+\frac{1}{3}\,,\nonumber\\
1\!\!\!&=&  \frac{1}{21}+\frac{10}{21}+\frac{1}{3}+\frac{1}{7}\,,\quad
1\!\!\!~=~  \frac{1}{21}+\frac{5}{42}+\frac{1}{2}+\frac{1}{3}\,,\quad
1\!\!\!~=~  \frac{1}{28}+\frac{9}{28}+\frac{1}{2}+\frac{1}{7}\,,\nonumber\\
1\!\!\!&=&  \frac{1}{36}+\frac{5}{36}+\frac{1}{2}+\frac{1}{3}\,.
\end{eqnarray}
In the classes  shown in Eqs.~(\ref{eq.4})--(\ref{eq.7}) it is possible
to find 32 additional solutions
\begin{eqnarray}
1\!\!\!&=&\ \frac{1}{7}+\frac{1}{7}+\frac{2}{7}+\frac{3}{7}\,,\quad
1\!\!\!~=~\ \frac{1}{10}+\frac{1}{5}+\frac{3}{10}+\frac{2}{5}\,,\quad
1\!\!\!~=~\ \frac{1}{15}+\frac{2}{15}+\frac{1}{3}+\frac{7}{15}\,,\nonumber\\
1\!\!\!&=&\ \frac{1}{15}+\frac{1}{5}+\frac{4}{15}+\frac{7}{15}\,,\quad
1\!\!\!~=~\ \frac{1}{22}+\frac{3}{22}+\frac{7}{22}+\frac{1}{2}\,,\quad
1\!\!\!~=~\ \frac{1}{16}+\frac{1}{4}+\frac{5}{16}+\frac{3}{8}\,,\nonumber\\
1\!\!\!&=&\ \frac{1}{18}+\frac{2}{9}+\frac{1}{3}+\frac{7}{18}\,,\quad
1\!\!\!~=~\ \frac{1}{22}+\frac{2}{11}+\frac{3}{11}+\frac{1}{2}\,,\quad
1\!\!\!~=~\ \frac{1}{21}+\frac{5}{21}+\frac{1}{3}+\frac{8}{21}\,,\nonumber\\
1\!\!\!&=&\ \frac{1}{26}+\frac{5}{26}+\frac{7}{26}+\frac{1}{2}\,,\quad
1\!\!\!~=~\ \frac{1}{8}+\frac{3}{16}+\frac{1}{4}+\frac{7}{16}\,,\quad
1\!\!\!~=~\ \frac{1}{9}+\frac{1}{6}+\frac{5}{18}+\frac{4}{9}\,,\nonumber\\
1\!\!\!&=&\ \frac{2}{21}+\frac{1}{7}+\frac{1}{3}+\frac{3}{7}\,,\quad
1\!\!\!~=~\ \frac{1}{13}+\frac{3}{26}+\frac{4}{13}+\frac{1}{2}\,,\quad
1\!\!\!~=~\ \frac{1}{11}+\frac{2}{11}+\frac{5}{22}+\frac{1}{2}\,,\nonumber\\
1\!\!\!&=&\ \frac{1}{10}+\frac{1}{4}+\frac{3}{10}+\frac{7}{20}\,,\quad
1\!\!\!~=~\ \frac{1}{13}+\frac{5}{26}+\frac{3}{13}+\frac{1}{2}\,,\quad
1\!\!\!~=~\ \frac{1}{16}+\frac{5}{32}+\frac{9}{32}+\frac{1}{2}\,,\nonumber\\
1\!\!\!&=&\ \frac{1}{6}+\frac{2}{9}+\frac{5}{18}+\frac{1}{3}\,,\quad
1\!\!\!~=~\ \frac{3}{20}+\frac{1}{5}+\frac{1}{4}+\frac{2}{5}\,,\quad
1\!\!\!~=~\ \frac{1}{8}+\frac{1}{6}+\frac{7}{24}+\frac{5}{12}\,,\nonumber\\
1\!\!\!&=&\ \frac{1}{10}+\frac{2}{15}+\frac{1}{3}+\frac{13}{30}\,,\quad
1\!\!\!~=~\ \frac{1}{7}+\frac{5}{21}+\frac{2}{7}+\frac{1}{3}\,,\quad
1\!\!\!~=~\ \frac{1}{11}+\frac{5}{33}+\frac{1}{3}+\frac{14}{33}\,,\nonumber\\
1\!\!\!&=&\ \frac{1}{6}+\frac{5}{24}+\frac{1}{4}+\frac{3}{8}\,,\quad
1\!\!\!~=~\ \frac{1}{8}+\frac{5}{32}+\frac{7}{32}+\frac{1}{2}\,,\quad
1\!\!\!~=~\ \frac{1}{11}+\frac{5}{44}+\frac{13}{44}+\frac{1}{2}\,,\nonumber\\
1\!\!\!&=&\ \frac{2}{27}+\frac{5}{54}+\frac{1}{3}+\frac{1}{2}\,,\quad
1\!\!\!~=~\ \frac{1}{7}+\frac{3}{14}+\frac{1}{4}+\frac{11}{28}\,,\quad
1\!\!\!~=~\ \frac{1}{6}+\frac{1}{5}+\frac{4}{15}+\frac{11}{30}\,,\nonumber\\
1\!\!\!&=&\ \frac{7}{50}+\frac{4}{25}+\frac{1}{5}+\frac{1}{2}\,,\quad
1\!\!\!~=~\ \frac{7}{36}+\frac{2}{9}+\frac{1}{4}+\frac{1}{3}\,.
\end{eqnarray}
In the sets of Eqs.~(\ref{eq.9})--(\ref{eq.14}) there are 10 new solutions,
namely
\begin{eqnarray}
1\!\!\!&=&\ \frac{1}{11}+\frac{2}{11}+\frac{3}{11}+\frac{5}{11}\,,\quad
1\!\!\!~=~\ \frac{1}{13}+\frac{3}{13}+\frac{4}{13}+\frac{5}{13}\,,\quad
1\!\!\!~=~\ \frac{1}{7}+\frac{3}{14}+\frac{2}{7}+\frac{5}{14}\,,\nonumber\\
1\!\!\!&=&\ \frac{2}{27}+\frac{5}{27}+\frac{1}{3}+\frac{11}{27}\,,\quad
1\!\!\!~=~\ \frac{3}{34}+\frac{2}{17}+\frac{5}{17}+\frac{1}{2}\,,\quad
1\!\!\!~=~\ \frac{3}{38}+\frac{5}{38}+\frac{11}{38}+\frac{1}{2}\,,\nonumber\\
1\!\!\!&=&\ \frac{4}{25}+\frac{1}{5}+\frac{7}{25}+\frac{9}{25}\,,\quad
1\!\!\!~=~\ \frac{2}{17}+\frac{3}{17}+\frac{7}{34}+\frac{1}{2}\,,\quad
1\!\!\!~=~\ \frac{5}{27}+\frac{2}{9}+\frac{7}{27}+\frac{1}{3}\,,\nonumber\\
1\!\!\!&=&\ \frac{5}{38}+\frac{3}{19}+\frac{4}{19}+\frac{1}{2}\,.
\end{eqnarray}
For the last case in Eqs.~(\ref{eq.15})--(\ref{eq.20}) we only have 2
new solutions
\begin{equation}
1=\frac{2}{17}+\frac{3}{17}+\frac{5}{17}+\frac{7}{17}\,,\quad
1=\frac{3}{19}+\frac{4}{19}+\frac{5}{19}+\frac{7}{19}\,.
\end{equation}
The total number of quasi--simply--laced numbers is therefore 95, which is
in agreement with the known number of Calabi--Yau spaces with one
reflexive vector.

\begin{table}
\begin{center}
\begin{tabular}{|ccc|c||ccc|c|}
\hline
&&&&&&&\\[-3mm]
&Matrix & &  Equation && Matrix & & Equation\\
&&&&&&&\\[-3mm]
\hline
&&&&&&&\\[-3mm]
&\scriptsize
\begin{tabular}{|cccc|}
  $\tilde s_1$ & 0 & 0 & 0\\
  1 & $s_2$ & 0 & 0\\
  1 & 0 & $s_3$ & 0\\
  1 & 0 & 0 & $s_4$
\end{tabular}
&&
{\bf{(\ref{eq.8})}}
&
&\scriptsize
\begin{tabular}{|cccc|}
  $\tilde s_1$ & 0 & 0 & 0 \\
  0 & $\tilde s_2$ & 0 & 0 \\
  0 & 0 & $s_3$ & 1 \\
  0 & 0 & 1 & $s_4$
\end{tabular}
&&
{\bf{(\ref{eq.7})}}
\\[8mm]
&\scriptsize
\begin{tabular}{|cccc|}
  $\tilde s_1$ & 0 & 0 & 0\\
  1 & $s_2$ & 0 & 0\\
  1 & 0 & $s_3$ & 0\\
  0 & 1 & 0 & $s_4$
\end{tabular}
&&
{\bf{(\ref{eq.9})}}
&
&\scriptsize
\begin{tabular}{|cccc|}
 $\tilde s_1$ & 0 & 0 & 0 \\
 0 & $s_2$ & 1 & 0 \\
 0 & 1 & $s_3$ & 0 \\
 0 & 1 & 0 & $s_4$
\end{tabular}
&&
{\bf{(\ref{eq.13})}}
\\[8mm]
&\scriptsize
\begin{tabular}{|cccc|}
 $\tilde s_1$ & 0 & 0 & 0 \\
 1 & $s_2$ & 0 & 0 \\
 1 & 0 & $s_3$ & 0 \\
 0 & 0 & 0 & $\tilde s_4$
\end{tabular}
&&
{\bf{(\ref{eq.4})}}
&
&\scriptsize
\begin{tabular}{|cccc|}
 $\tilde s_1$ & 0 & 0 & 0 \\
 0 & $s_2$ & 1 & 0 \\
 0 & 0 & $s_3$ & 1 \\
 0 & 1 & 0 & $s_4$
\end{tabular}
&&
{\bf{(\ref{eq.14})}}
\\[8mm]
&\scriptsize
\begin{tabular}{|cccc|}
 $\tilde s_1$ & 0 & 0 & 0 \\
 1 & $s_2$ & 0 & 0 \\
 0 & 1 & $s_3$ & 0 \\
 0 & 1 & 0 & $s_4$
\end{tabular}
&&
{\bf{(\ref{eq.12})}}&
&\scriptsize
\begin{tabular}{|cccc|}
 $s_1$ & 1 & 0 & 0 \\
 1 & $s_2$ & 0 & 0 \\
 1 & 0 & $s_3$ & 0 \\
 1 & 0 & 0 & $s_4$
\end{tabular}
&&
{\bf{(\ref{eq.15})}}\\[8mm]
&\scriptsize
\begin{tabular}{|cccc|}
 $\tilde s_1$ & 0 & 0 & 0 \\
 1 & $s_2$ & 0 & 0 \\
 0 & 1 & $s_3$ & 0 \\
 0 & 0 & 1 & $s_4$
\end{tabular}
&&
{\bf{(\ref{eq.11})}}
&
&\scriptsize
\begin{tabular}{|cccc|}
 $s_1$ & 1 & 0 & 0 \\
 1 & $s_2$ & 0 & 0 \\
 1 & 0 & $s_3$ & 0 \\
 0 & 1 & 0 & $s_4$
\end{tabular}
&&
{\bf{(\ref{eq.17})}}
\\[8mm]
&\scriptsize
\begin{tabular}{|cccc|}
 $\tilde s_1$ & 0 & 0 & 0 \\
 1 & $s_2$ & 0 & 0 \\
 0 & 1 & $s_3$ & 0 \\
 0 & 0 & 0 & $\tilde s_4$
\end{tabular}
&&
{\bf{(\ref{eq.6})}}
&
&\scriptsize
\begin{tabular}{|cccc|}
 $s_1$ & 1 & 0 & 0 \\
 1 & $s_2$ & 0 & 0 \\
 1 & 0 & $s_3$ & 0 \\
 0 & 0 & 1 & $s_4$
\end{tabular}
&&
{\bf{(\ref{eq.18})}}
\\[8mm]
&\scriptsize
\begin{tabular}{|cccc|}
 $\tilde s_1$ & 0 & 0 & 0 \\
 1 & $s_2$ & 0 & 0 \\
 0 & 0 & $\tilde s_3$ & 0 \\
 0 & 0 & 1 & $s_4$
\end{tabular}
&&
{\bf{(\ref{eq.5})}}
&
&\scriptsize
\begin{tabular}{|cccc|}
 $s_1$ & 1 & 0 & 0 \\
 1 & $s_2$ & 0 & 0 \\
 0 & 0 & $s_3$ & 1 \\
 0 & 0 & 1 & $s_4$
\end{tabular}
&&
{\bf{(\ref{eq.19})}}
\\[8mm]
&\scriptsize
\begin{tabular}{|cccc|}
 $\tilde s_1$ & 0 & 0 & 0 \\
 1 & $s_2$ & 0 & 0 \\
 0 & 0 & $\tilde s_3$ & 0 \\
 0 & 0 & 0 & $\tilde s_4$
\end{tabular}
&&
{\bf{(\ref{eq.3})}}
&
&\scriptsize
\begin{tabular}{|cccc|}
 $s_1$ & 1 & 0 & 0 \\
 0 & $s_2$ & 1 & 0 \\
 1 & 0 & $s_3$ & 0 \\
 1 & 0 & 0 & $s_4$
\end{tabular}
&&
{\bf{(\ref{eq.16})}}
\\[8mm]
&\scriptsize
\begin{tabular}{|cccc|}
 $\tilde s_1$ & 0 & 0 & 0 \\
 1 & $s_2$ & 0 & 0 \\
 0 & 0 & $s_3$ & 1 \\
 0 & 0 & 1 & $s_4$
\end{tabular}
&&
{\bf{(\ref{eq.10})}}
&
&\scriptsize
\begin{tabular}{|cccc|}
 $s_1$ & 1 & 0 & 0 \\
 1 & $s_2$ & 1 & 0 \\
 0 & 1 & $s_3$ & 1 \\
 1 & 0 & 0 & $s_4$
\end{tabular}
&&
{\bf{(\ref{eq.20})}}
\\[8mm]
&\scriptsize
\begin{tabular}{|cccc|}
 $\tilde s_1$ & 0 & 0 & 0 \\
 0 & $\tilde s_2$ & 0 & 0 \\
 0 & 0 & $\tilde s_3$ & 0 \\
 0 & 0 & 0 & $\tilde s_4$
\end{tabular}
&&
{\bf{(\ref{eq.2})}}
&
&\scriptsize
&&
\scriptsize
\\[8mm]
\hline
\end{tabular}
\caption{Matrix and number of the corresponding equation}
\label{tab:matrix}
\end{center}
\end{table}

The simplest way to find the above--mentioned decompositions is to solve the
following system of linear equations
\begin{eqnarray}\label{LE}
  l_{r'} + s_r l_r &=& 1\,, \qquad\qquad r,r'=1,...,n \\
  \sum_{r=1}^{n}l_r &=& 1
\end{eqnarray}
for all integer $s_r$. This gives $n^n$ sets of linear equations,
which have non--trivial solutions only if the determinant
of the extended $(n+1)\times (n+1)$ matrix equals zero
\begin{equation}
J=\begin{array}{|ccccc|}
  s_1    & 0      & \ldots & 0      & -1 \\[2mm]
  0      & s_2    & \ldots & 0      & -1 \\[2mm]
  \vdots & \vdots & \ddots & \vdots & \vdots \\[2mm]
  0      & 0      & \ldots & s_n    & -1 \\[2mm]
  1      & 1      & \ldots & 1      & - 1
\end{array}\ \equiv\ 0\,.
\label{ExMat}
\end{equation}
Here in every row $i$ the zero in the place $i'$ is substituted by
unity. Some of the
$n^n$ sets of linear equations are equivalent and are obtained by
transmutations of indices $i$. For example, in the $n=4$ case there are
only 19 non--equivalent sets of equations according to the
classification of the quasi--simply--laced numbers given in the previous
subsection. The
corresponding matrices and numbers of solutions for each matrix
are listed in Table \ref{tab:matrix}, where $\tilde s_i=s_i+1$ and
one should add to each matrix
the row with the numbers ``1'' and the column with numbers ``-1''.

Note that for the ansatz in Eq.~(\ref{eq.8}) the above determinant $J$
is factorized, which means that the corresponding set of equations
has an infinite number of solutions. The condition of vanishing
for this determinant can be written as follows
\begin{equation}\label{RelEq}
s_1(s_3 s_4 + s_2 s_4 + s_2 s_3 - s_2 s_3 s_4 )=0\ \Leftrightarrow \
\frac{1}{s_2} + \frac{1}{s_3} + \frac{1}{s_4} - 1 =0
\end{equation}
and in fact, apart from many (non--reflexive) solutions with $s_1=0$,
we obtain those solutions corresponding to the decompositions of
unity for $n=3$. The same is true for Eq.~(\ref{eq.15}). We should
neglect these two classes of sets of equations due to their degeneracy.

\section{Five--Dimensional Simply--Laced Numbers}
\label{QSLRWVn5}

As it was discussed above, the arity {approach} offers the possibility
to construct all reflexive vectors for an arbitrary dimension $n$, but
the difficulty for its application is that at each step of the calculations
we should verify the reflexivity of the polyhedron
obtained as an intersection of the slices corresponding to the extended
reflexive vectors in lower dimensions. The simply--laced numbers are
a particular case of the reflexive numbers and they
can be found by the same method. {In Table \ref{Table147a} we
list all the 147 simply--laced numbers for $n=5$ with their corresponding
expansion in linear combinations of the
extended simply--laced vectors of Table~\ref{Table100}}.
Note that almost all simply--laced numbers have the 2--arity expansion.
Only one number $(1,15,24,40,40)\,[120]$ cannot be obtained as a sum of
two extended simply--laced vectors
and it is constructed according to the 3--arity approach:
\begin{equation}
(1,15,24,40,40)=(1,0,0,1,1)+15\,(0,1,0,1,1)+24\,(0,0,1,1,1)\,.
\end{equation}

\begin{table}
\tiny
\vspace{.05in}
\begin{center}
\begin{tabular}{|l|l||l|l|}
\hline
$1   $&$  (1, 1, 1, 1, 1           ) = 1  (1, 1, 1, 0, 0)   + 1
(0, 0, 0, 1, 1)   %
$&$2 $&$  (1, 1, 1, 1, 2           ) = 1  (1, 1, 0, 0, 1)   + 1
(0, 0, 1, 1, 1)   $ \\ \hline
$3   $&$  (1, 1, 1, 1, 4           ) = 1  (1, 1, 0, 0, 2)   + 1
(0, 0, 1, 1, 2)   %
$&$4 $&$  (1, 1, 1, 2, 5           ) = 1  (1, 1, 0, 0, 2)   + 1
(0, 0, 1, 2, 3)   $ \\ \hline
$5   $&$  (1, 1, 1, 3, 3           ) = 1  (1, 1, 0, 1, 0)   + 1
(0, 0, 1, 2, 3)   %
$&$6 $&$  (1, 1, 1, 3, 6           ) = 1  (1, 0, 0, 1, 2)   + 1
(0, 1, 1, 2, 4)   $ \\ \hline
$7   $&$  (1, 1, 1, 6, 9           ) = 1  (1, 0, 1, 4, 6)   + 1
(0, 1, 0, 2, 3)   %
$&$8 $&$  (1, 1, 2, 2, 2           ) = 1  (1, 0, 1, 0, 0)   + 1
(0, 1, 1, 2, 2)   $ \\ \hline
$9   $&$  (1, 1, 2, 2, 6           ) = 1  (1, 0, 0, 1, 2)   + 1
(0, 1, 2, 1, 4)   %
$&$10$&$  (1, 1, 2, 4, 4           ) = 1  (0, 1, 1, 0, 2)   + 1
(1, 0, 1, 4, 2)   $ \\ \hline
$11  $&$  (1, 1, 2, 4, 8           ) = 1  (1, 0, 1, 0, 2)   + 1
(0, 1, 1, 4, 6)   %
$&$12$&$  (1, 1, 2, 8, 12          ) = 1  (1, 0, 1, 4, 6)   + 1
(0, 1, 1, 4, 6)   $ \\ \hline
$13  $&$  (1, 1, 3, 3, 4           ) = 1  (1, 0, 1, 2, 0)   + 1
(0, 1, 2, 1, 4)   %
$&$14$&$  (1, 1, 3, 5, 5           ) = 1  (1, 0, 0, 1, 1)   + 1
(0, 1, 3, 4, 4)   $ \\ \hline
$15  $&$  (1, 1, 3, 10, 15         ) = 1  (1, 0, 0, 2, 3)   + 1
(0, 1, 3, 8, 12)  %
$&$16$&$  (1, 1, 4, 4, 10          ) = 1  (1, 0, 3, 2, 6)   + 1
(0, 1, 1, 2, 4)   $ \\ \hline
$17  $&$  (1, 1, 4, 6, 12          ) = 1  (1, 0, 2, 0, 3)   + 1
(0, 1, 2, 6, 9)   %
$&$18$&$  (1, 1, 4, 12, 18         ) = 1  (1, 0, 2, 6, 9)   + 1
(0, 1, 2, 6, 9)   $ \\ \hline
$19  $&$  (1, 1, 6, 8, 8           ) = 1  (1, 0, 3, 6, 2)   + 1
(0, 1, 3, 2, 6)   %
$&$20$&$  (1, 1, 6, 16, 24         ) = 1  (0, 1, 0, 2, 3)   + 1
(1, 0, 6, 14, 21) $ \\ \hline
$21  $&$  (1, 1, 8, 10, 20         ) = 1  (0, 1, 4, 5, 10)  + 1
(1, 0, 4, 5, 10)  %
$&$22$&$  (1, 1, 12, 28, 42        ) = 1  (0, 1, 6, 14, 21) + 1
(1, 0, 6, 14, 21) $ \\ \hline
$23  $&$  (1, 2, 2, 2, 7           ) = 1  (0, 0, 2, 1, 3)   + 1
(1, 2, 0, 1, 4)   %
$&$24$&$  (1, 2, 2, 3, 4           ) = 1  (1, 1, 0, 2, 0)   + 1
(0, 1, 2, 1, 4)   $ \\ \hline
$25  $&$  (1, 2, 2, 5, 10          ) = 1  (0, 1, 2, 3, 6)   + 1
(1, 1, 0, 2, 4)   %
$&$26$&$  (1, 2, 2, 10, 15         ) = 1  (0, 1, 2, 6, 9)   + 1
(1, 1, 0, 4, 6)   $ \\ \hline
$27  $&$  (1, 2, 3, 3, 3           ) = 1  (1, 0, 2, 3, 0)   + 1
(0, 2, 1, 0, 3)   %
$&$28$&$  (1, 2, 3, 3, 9           ) = 1  (1, 0, 2, 0, 3)   + 1
(0, 2, 1, 3, 6)   $ \\ \hline
$29  $&$  (1, 2, 3, 6, 6           ) = 1  (1, 0, 2, 3, 0)   + 1
(0, 2, 1, 3, 6)   %
$&$30$&$  (1, 2, 3, 6, 12          ) = 1  (1, 0, 2, 0, 3)   + 1
(0, 2, 1, 6, 9)   $ \\ \hline
$31  $&$  (1, 2, 3, 12, 18         ) = 1  (1, 0, 0, 2, 3)   + 1
(0, 2, 3, 10, 15) %
$&$32$&$  (1, 2, 4, 7, 14          ) = 1  (1, 0, 2, 6, 9)   + 1
(0, 2, 2, 1, 5)   $ \\ \hline
$33  $&$  (1, 2, 6, 6, 15          ) = 1  (0, 1, 2, 6, 9)   + 1
(1, 1, 4, 0, 6)   %
$&$34$&$  (1, 2, 6, 9, 18          ) = 1  (1, 0, 3, 8, 12)  + 1
(0, 2, 3, 1, 6)   $ \\ \hline
$35  $&$  (1, 2, 6, 18, 27         ) = 1  (0, 1, 6, 14, 21) + 1
(1, 1, 0, 4, 6)   %
$&$36$&$  (1, 2, 9, 12, 12         ) = 1  (0, 1, 3, 12, 8)  + 1
(1, 1, 6, 0, 4)   $ \\ \hline
$37  $&$  (1, 2, 9, 24, 36         ) = 1  (1, 0, 6, 14, 21) + 1
(0, 2, 3, 10, 15) %
$&$38$&$  (1, 2, 12, 15, 30        ) = 1  (0, 2, 6, 1, 9)   + 1
(1, 0, 6, 14, 21) $ \\ \hline
$39  $&$  (1, 2, 18, 42, 63        ) = 1  (1, 0, 6, 14, 21) + 2
(0, 1, 6, 14, 21) %
$&$40$&$  (1, 3, 3, 3, 5           ) = 1  (1, 1, 0, 0, 1)   + 1
(0, 2, 3, 3, 4)   $ \\ \hline
$41  $&$  (1, 3, 3, 7, 7           ) = 1  (1, 3, 0, 4, 4)   + 3
(0, 0, 1, 1, 1)   %
$&$42$&$  (1, 3, 3, 14, 21         ) = 1  (0, 1, 3, 8, 12)  + 1
(1, 2, 0, 6, 9)   $ \\ \hline
$43  $&$  (1, 3, 4, 4, 12          ) = 1  (0, 1, 1, 4, 6)   + 1
(1, 2, 3, 0, 6)   %
$&$44$&$  (1, 3, 4, 8, 8           ) = 1  (1, 0, 3, 2, 6)   + 1
(0, 3, 1, 6, 2)   $ \\ \hline
$45  $&$  (1, 3, 4, 16, 24         ) = 1  (1, 0, 3, 8, 12)  + 1
(0, 3, 1, 8, 12)  %
$&$46$&$  (1, 3, 5, 6, 15          ) = 1  (0, 1, 2, 6, 9)   + 1
(1, 2, 3, 0, 6)   $ \\ \hline
$47  $&$  (1, 3, 6, 6, 8           ) = 1  (1, 2, 3, 0, 6)   + 1
(0, 1, 3, 6, 2)   %
$&$48$&$  (1, 3, 6, 10, 10         ) = 1  (1, 2, 0, 6, 9)   + 1
(0, 1, 6, 4, 1)   $ \\ \hline
$49  $&$  (1, 3, 6, 20, 30         ) = 1  (0, 1, 6, 14, 21) + 1
(1, 2, 0, 6, 9)   %
$&$50$&$  (1, 3, 8, 12, 24         ) = 1  (1, 0, 6, 2, 9)   + 1
(0, 3, 2, 10, 15) $ \\ \hline
$51  $&$  (1, 3, 8, 24, 36         ) = 1  (1, 0, 6, 14, 21) + 1
(0, 3, 2, 10, 15) %
$&$52$&$  (1, 3, 12, 16, 16        ) = 1  (1, 0, 3, 4, 4)   + 3
(0, 1, 3, 4, 4)   $ \\ \hline
$53  $&$  (1, 3, 12, 32, 48        ) = 1  (1, 1, 0, 4, 6)   + 2
(0, 1, 6, 14, 21) %
$&$54$&$  (1, 3, 24, 56, 84        ) = 1  (1, 0, 6, 14, 21) + 3
(0, 1, 6, 14, 21) $ \\ \hline
$55  $&$  (1, 4, 4, 9, 18          ) = 1  (1, 0, 3, 8, 12)  + 1
(0, 4, 1, 1, 6)   %
$&$56$&$  (1, 4, 5, 5, 5           ) = 1  (1, 0, 2, 4, 1)   + 1
(0, 4, 3, 1, 4)   $ \\ \hline
$57  $&$  (1, 4, 5, 10, 20         ) = 1  (0, 2, 3, 10, 15) + 1
(1, 2, 2, 0, 5)   %
$&$58$&$  (1, 4, 5, 20, 30         ) = 1  (1, 0, 3, 8, 12)  + 2
(0, 2, 1, 6, 9)   $ \\ \hline
$59  $&$  (1, 4, 10, 15, 30        ) = 1  (1, 0, 8, 3, 12)  + 2
(0, 2, 1, 6, 9)   %
$&$60$&$  (1, 4, 15, 20, 20        ) = 1  (1, 0, 3, 4, 4)   + 4
(0, 1, 3, 4, 4)   $ \\ \hline
$61  $&$  (1, 4, 15, 40, 60        ) = 1  (1, 0, 3, 8, 12)  + 4
(0, 1, 3, 8, 12)  %
$&$62$&$  (1, 4, 20, 25, 50        ) = 1  (1, 0, 4, 5, 10)  + 4
(0, 1, 4, 5, 10)  $ \\ \hline
$63  $&$2(1, 5, 9, 15, 15         ) = 1  (2, 0, 3, 15, 10) + 5
(0, 2, 3, 3, 4))  %
$&$64$&$  (1, 5, 9, 30, 45         ) = 1  (1, 2, 0, 6, 9)   + 3
(0, 1, 3, 8, 12)  $ \\ \hline
$65  $&$  (1, 5, 12, 12, 30        ) = 1  (1, 3, 8, 0, 12)  + 2
(0, 1, 2, 6, 9)   %
$&$66$&$  (1, 5, 24, 30, 60        ) = 1  (1, 0, 4, 5, 10)  + 5
(0, 1, 4, 5, 10)  $ \\ \hline
$67  $&$  (1, 6, 6, 26, 39         ) = 1  (1, 0, 2, 6, 9)   + 2
(0, 3, 2, 10, 15) %
$&$68$&$  (1, 6, 7, 7, 21          ) = 1  (1, 0, 4, 1, 6)   + 3
(0, 2, 1, 2, 5)   $ \\ \hline
$69  $&$  (1, 6, 7, 14, 14         ) = 1  (1, 0, 1, 2, 2)   + 6
(0, 1, 1, 2, 2)   %
$&$70$&$  (1, 6, 7, 28, 42         ) = 1  (1, 0, 3, 8, 12)  + 2
(0, 3, 2, 10, 15) $ \\ \hline
$71  $&$  (1, 6, 14, 21, 42        ) = 1  (1, 0, 2, 3, 6)   + 6
(0, 1, 2, 3, 6)   %
$&$72$&$  (1, 6, 14, 42, 63        ) = 1  (1, 0, 2, 6, 9)   + 6
(0, 1, 2, 6, 9)   $ \\ \hline
$73  $&$  (1, 6, 21, 28, 28        ) = 1  (1, 0, 3, 4, 4)   + 6
(0, 1, 3, 4, 4)   %
$&$74$&$  (1, 6, 21, 56, 84        ) = 1  (1, 0, 3, 8, 12)  + 6
(0, 1, 3, 8, 12)  $ \\ \hline
$75  $&$  (1, 6, 42, 98, 147       ) = 1  (1, 0, 6, 14, 21) + 6
(0, 1, 6, 14, 21) %
$&$76 $&$ (1, 7, 48, 112, 168      ) = 1  (1, 0, 6, 14, 21) + 7
(0, 1, 6, 14, 21) $ \\ \hline
$77   $&$ (1, 8, 9, 18, 36         ) = 1  (1, 2, 6, 0, 9)   + 3
(0, 2, 1, 6, 9)   %
$&$78 $&$ (1, 8, 27, 72, 108       ) = 1  (1, 0, 3, 8, 12)  + 8
(0, 1, 3, 8, 12)  $ \\ \hline
$79   $&$ (1, 9, 20, 60, 90        ) = 1  (1, 0, 2, 6, 9)   + 9
(0, 1, 2, 6, 9)   %
$&$80 $&$ (1, 10, 10, 14, 35       ) = 1  (1, 0, 5, 4, 10)  + 5
(0, 2, 1, 2, 5)   $ \\ \hline
$81   $&$ (1, 10, 22, 22, 55       ) = 1  (1, 0, 2, 2, 5)   + 10
(0, 1, 2, 2, 5)   %
$&$82 $&$ (1, 10, 44, 55, 110      ) = 1  (1, 0, 4, 5, 10)  + 10
(0, 1, 4, 5, 10)  $ \\ \hline
$83   $&$ (1, 12, 12, 15, 20       ) = 1  (1, 4, 10, 5, 0)  + 2
(0, 4, 1, 5, 10)  %
$&$84 $&$ (1, 12, 13, 52, 78       ) = 1  (1, 0, 1, 4, 6)   + 12
(0, 1, 1, 4, 6)   $ \\ \hline
$85   $&$ (1, 12, 26, 39, 78       ) = 1  (1, 0, 2, 3, 6)   + 12
(0, 1, 2, 3, 6)   %
$&$86 $&$ (1, 12, 39, 52, 52       ) = 1  (1, 0, 3, 4, 4)   + 12
(0, 1, 3, 4, 4)   $ \\ \hline
$87   $&$ (1, 12, 39, 104, 156     ) = 1  (1, 0, 3, 8, 12)  + 12
(0, 1, 3, 8, 12)  %
$&$88 $&$ (1, 14, 20, 35, 70       ) = 1  (1, 4, 0, 5, 10)  + 10
(0, 1, 2, 3, 6)   $ \\ \hline
$89   $&$ (1, 14, 90, 210, 315     ) = 1  (1, 0, 6, 14, 21) + 14
(0, 1, 6, 14, 21) %
$&$90 $&$ (1, 15, 20, 24, 60       ) = 1  (1, 5, 0, 4, 10)  + 10
(0, 1, 2, 2, 5)   $ \\ \hline
$91   $&$ (1, 15, 24, 40, 40       ) = 1  (1, 0, 0, 1, 1)   + 15
(0, 1, 0, 1, 1) + 24 (0, 0, 1, 1, 1)
$&$92 $&$ (1, 15, 24, 80, 120      ) = 1  (1, 3, 0, 8, 12)  + 12
(0, 1, 2, 6, 9)   $ \\ \hline
$93   $&$ (1, 18, 38, 114, 171     ) = 1  (1, 0, 2, 6, 9)   + 18
(0, 1, 2, 6, 9)   %
$&$94 $&$ (1, 20, 84, 105, 210     ) = 1  (1, 0, 4, 5, 10)  + 20
(0, 1, 4, 5, 10)  $ \\ \hline
$95   $&$ (1, 21, 132, 308, 462    ) = 1  (1, 0, 6, 14, 21) + 21
(0, 1, 6, 14, 21) %
$&$96 $&$ (1, 24, 75, 200, 300     ) = 1  (1, 0, 3, 8, 12)  + 24
(0, 1, 3, 8, 12)  $ \\ \hline
$97   $&$ (1, 42, 258, 602, 903    ) = 1  (1, 0, 6, 14, 21) + 42
(0, 1, 6, 14, 21) %
$&$98 $&$ (2, 2, 2, 3, 3           ) = 1  (1, 2, 0, 3, 0)   + 1
(1, 0, 2, 0, 3)   $ \\ \hline
$99   $&$ (2, 2, 2, 3, 9           ) = 1  (1, 2, 0, 0, 3)   + 1
(1, 0, 2, 3, 6)   %
$&$100$&$ (2, 2, 3, 14, 21         ) = 1  (1, 0, 3, 8, 12)  + 1
(1, 2, 0, 6, 9)   $ \\ \hline
$101  $&$ (2, 2, 5, 6, 15          ) = 1  (1, 0, 2, 6, 9)   + 1
(1, 2, 3, 0, 6)   %
$&$102$&$ (2, 3, 3, 4, 12          ) = 1  (1, 0, 1, 4, 6)   + 1
(1, 3, 2, 0, 6)   $ \\ \hline
$103  $&$ (2, 3, 3, 8, 8           ) = 1  (1, 3, 0, 2, 6)   + 1
(1, 0, 3, 6, 2)   %
$&$104$&$ (2, 3, 3, 16, 24         ) = 1  (1, 3, 0, 8, 12)  + 1
(1, 0, 3, 8, 12)  $ \\ \hline
$105  $&$ (2, 3, 4, 9, 18          ) = 1  (1, 3, 0, 8, 12)  + 1
(1, 0, 4, 1, 6)   %
$&$106$&$ (2, 3, 5, 5, 15          ) = 1  (0, 1, 4, 5, 10)  + 1
(2, 2, 1, 0, 5)   $ \\ \hline
$107  $&$ (2, 3, 5, 10, 10         ) = 1  (0, 1, 4, 5, 10)  + 1
(2, 2, 1, 5, 0)   %
$&$108$&$ (2, 3, 5, 20, 30         ) = 1  (2, 0, 3, 10, 15) + 1
(0, 3, 2, 10, 15) $ \\ \hline
$109  $&$ (2, 3, 6, 22, 33         ) = 1  (1, 0, 6, 14, 21) + 1
(1, 3, 0, 8, 12)  %
$&$110$&$ (2, 3, 10, 15, 30        ) = 1  (2, 3, 0, 10, 15) + 5
(0, 0, 2, 1, 3)   $ \\ \hline
$111  $&$ (2, 3, 10, 30, 45        ) = 1  (2, 0, 1, 6, 9)   + 3
(0, 1, 3, 8, 12)  %
$&$112$&$ (2, 3, 15, 20, 20        ) = 1  (2, 3, 0, 15, 10) + 5
(0, 0, 3, 1, 2)   $ \\ \hline
$113  $&$ (2, 3, 15, 40, 60        ) = 2  (1, 0, 3, 8, 12)  + 3
(0, 1, 3, 8, 12)  %
$&$114$&$ (2, 3, 30, 70, 105       ) = 2  (1, 0, 6, 14, 21) + 3
(0, 1, 6, 14, 21) $ \\ \hline
$115  $&$ (2, 4, 4, 5, 5           ) = 1  (2, 0, 1, 4, 1)   + 1
(0, 4, 3, 1, 4)   %
$&$116$&$ (2, 4, 9, 9, 12          ) = 1  (1, 0, 3, 8, 12)  + 1
(1, 4, 6, 1, 0)   $ \\ \hline
$117  $&$ (2, 5, 5, 8, 20          ) = 1  (1, 0, 5, 4, 10)  + 1
(1, 5, 0, 4, 10)  %
$&$118$&$ (2, 5, 14, 14, 35        ) = 2  (1, 0, 2, 2, 5)   + 5
(0, 1, 2, 2, 5)   $ \\ \hline
$119  $&$ (2, 5, 28, 35, 70        ) = 2  (1, 0, 4, 5, 10)  + 5
(0, 1, 4, 5, 10)  %
$&$120$&$ (2, 6, 6, 7, 21          ) = 1  (0, 6, 2, 1, 9)   + 2
(1, 0, 2, 3, 6)   $ \\ \hline
$121  $&$ (2, 7, 12, 21, 42        ) = 1  (2, 1, 0, 3, 6)   + 6
(0, 1, 2, 3, 6)   %
$&$122$&$ (2, 7, 54, 126, 189      ) = 2  (1, 0, 6, 14, 21) + 7
(0, 1, 6, 14, 21) $ \\ \hline
$123  $&$ (2, 9, 22, 66, 99        ) = 2  (1, 0, 2, 6, 9)   + 9
(0, 1, 2, 6, 9)   %
$&$124$&$ (2, 10, 15, 18, 45       ) = 2  (1, 2, 6, 0, 9)   + 3
(0, 2, 1, 6, 9)   $ \\ \hline
$125  $&$ (2, 12, 21, 21, 28       ) = 1  (1, 6, 21, 0, 14) + 1
(1, 6, 0, 21, 14) %
$&$126$&$ (2, 21, 138, 322, 483    ) = 2  (1, 0, 6, 14, 21) + 21
(0, 1, 6, 14, 21) $ \\ \hline
$127  $&$ (2, 33, 42, 154, 231     ) = 2  (1, 6, 0, 14, 21) + 21
(0, 1, 2, 6, 9)   %
$&$128$&$ (3, 3, 4, 6, 8           ) = 1  (1, 0, 3, 6, 2)   + 1
(2, 3, 1, 0, 6)   $ \\ \hline
$129  $&$ (3, 3, 4, 20, 30         ) = 1  (3, 0, 2, 10, 15) + 1
(0, 3, 2, 10, 15) %
$&$130$&$ (3, 4, 7, 28, 42         ) = 1  (3, 0, 1, 8, 12)  + 2
(0, 2, 3, 10, 15) $ \\ \hline
$131  $&$ (3, 4, 14, 21, 42        ) = 1  (3, 0, 8, 1, 12)  + 2
(0, 2, 3, 10, 15) %
$&$132$&$ (3, 4, 21, 28, 28        ) = 3  (1, 0, 3, 4, 4)   + 4
(0, 1, 3, 4, 4)   $ \\ \hline
$133  $&$ (3, 4, 21, 56, 84        ) = 3  (1, 0, 3, 8, 12)  + 4
(0, 1, 3, 8, 12)  %
$&$134$&$ (3, 5, 6, 6, 10          ) = 1  (1, 2, 0, 6, 9)   + 1
(2, 3, 6, 0, 1)   $ \\ \hline
$135  $&$ (3, 5, 10, 12, 30        ) = 1  (1, 5, 0, 4, 10)  + 2
(1, 0, 5, 4, 10)  %
$&$136$&$ (3, 5, 12, 20, 20        ) = 1  (3, 0, 2, 15, 10) + 5
(0, 1, 2, 1, 2)   $ \\ \hline
$137  $&$ (3, 5, 12, 40, 60        ) = 2  (0, 1, 6, 14, 21) + 3
(1, 1, 0, 4, 6)   %
$&$138$&$ (3, 7, 60, 140, 210      ) = 3  (1, 0, 6, 14, 21) + 7
(0, 1, 6, 14, 21) $ \\ \hline
$139  $&$ (3, 8, 33, 88, 132       ) = 3  (1, 0, 3, 8, 12)  + 8
(0, 1, 3, 8, 12)  %
$&$140$&$ (3, 10, 12, 15, 20       ) = 1  (1, 10, 4, 5, 0)  + 2
(1, 0, 4, 5, 10)  $ \\ \hline
$141  $&$ (3, 14, 102, 238, 357    ) = 3  (1, 0, 6, 14, 21) + 14
(0, 1, 6, 14, 21) %
$&$142$&$ (3, 22, 30, 110, 165     ) = 1  (3, 2, 0, 10, 15) + 10
(0, 2, 3, 10, 15) $ \\ \hline
$143  $&$ (4, 5, 6, 15, 30         ) = 2  (2, 1, 0, 6, 9)   + 3
(0, 1, 2, 1, 4)   %
$&$144$&$ (4, 5, 36, 45, 90        ) = 4  (1, 0, 4, 5, 10)  + 5
(0, 1, 4, 5, 10)  $ \\ \hline
$145  $&$ (4, 6, 15, 15, 20        ) = 1  (2, 3, 15, 0, 10) + 1
(2, 3, 0, 15, 10) %
$&$146$&$ (6, 7, 78, 182, 273      ) = 6  (1, 0, 6, 14, 21) + 7
(0, 1, 6, 14, 21) $ \\ \hline
$147  $&$ (6, 14, 15, 70, 105      ) = 2  (3, 2, 0, 10, 15) + 5
(0, 2, 3, 10, 15) %
$&$ $&$$  \\
\hline
\end{tabular}
\end{center}
\caption{Expansion of the $n=5$ reflexive simply--laced numbers
on the $n = 4$ reflexive simply--laced numbers. 146 are expanded using
2--arity, the vector with number 91 is expanded using 3--arity.}
\label{Table147a}
\end{table}
To find these 147 simply--laced numbers  we can also use the recurrent
relations~(\ref{recur}). Moreover, using these relations,
also 3462 simply--laced numbers for $n=6$ were calculated. However, in the
following, we discuss a different method to generate these numbers in a
recurrent way. The idea of this recurrent procedure is to present unit
fractions in each step of the iteration as a sum of two (or several) unit
fractions.
We begin with the expansion of unity in the sum of two unit fractions:
\begin{equation}
1=\frac{1}{2}+\frac{1}{2}\,.
\end{equation}
If in the fraction $1/p$ the number $p$ is a prime number, we {only have
two} of such expansions
\begin{equation}
\frac{1}{p}=\frac{1}{2p}+\frac{1}{2p}=\frac{1}{p+1}+\frac{1}{p(p+1)}\,.
\end{equation}
If the number in its denominator is a product of two
(or several) prime numbers $p_1p_2...$,
the fraction has more unit expansions. For example,
\[
\frac{1}{p_1p_2}=\frac{1}{2p_1p_2}+\frac{1}{2p_1p_2}=\frac{1}{p_1p_2+1}+
\frac{1}{p_1p_2(p_1p_2+1)}=
\]
\begin{equation}
\frac{1}{p_2(p_1+1)}+\frac{1}{p_1p_2(p_1+1)}=
\frac{1}{p_1(p_2+1)}+\frac{1}{p_1p_2(p_2+1)}=
\frac{1}{p_2(p_1+p_2)}+\frac{1}{p_1(p_1+p_2)}\,.
\end{equation}
Using the above relations we obtain for $n=3$ from the unit expansion
for $n=2$:
\begin{equation}
1=\frac{1}{2}+\frac{1}{4}+\frac{1}{4}=\frac{1}{2}+\frac{1}{3}+\frac{1}{6}\,.
\end{equation}
The following expansion is obvious and easily generalized for an arbitrary $n$:
\begin{equation}
1=\frac{1}{3}+\frac{1}{3}+\frac{1}{3}\,.
\end{equation}
In the $n=4$ case the corresponding unity expansions obtained from the above 3
decompositions for $n=3$ are given below
\begin{eqnarray}
1&=&\frac{1}{4}+\frac{1}{4}+\frac{1}{4}+\frac{1}{4}~=~
\frac{1}{3}+\frac{1}{4}+\frac{1}{4}+\frac{1}{6}~=~
\frac{1}{2}+\frac{1}{4}+\frac{1}{5}+\frac{1}{20}~=~
\frac{1}{2}+\frac{1}{4}+\frac{1}{8}+\frac{1}{8}\nonumber\\
&=&\frac{1}{3}+\frac{1}{3}+\frac{1}{6}+\frac{1}{6}~=~
\frac{1}{3}+\frac{1}{3}+\frac{1}{6}+\frac{1}{6}~=~
\frac{1}{2}+\frac{1}{4}+\frac{1}{6}+\frac{1}{12}~=~
\frac{1}{2}+\frac{1}{6}+\frac{1}{6}+\frac{1}{6}\nonumber\\
&=&\frac{1}{2}+\frac{1}{3}+\frac{1}{12}+\frac{1}{12}~=~
\frac{1}{2}+\frac{1}{3}+\frac{1}{7}+\frac{1}{42}~=~
\frac{1}{2}+\frac{1}{3}+\frac{1}{9}+\frac{1}{18}~=~
\frac{1}{2}+\frac{1}{3}+\frac{1}{8}+\frac{1}{24}\nonumber\\
&=&\frac{1}{2}+\frac{1}{3}+\frac{1}{10}+\frac{1}{15}~=~
\frac{1}{2}+\frac{1}{5}+\frac{1}{5}+\frac{1}{10}.
\end{eqnarray}
Only the last unity decomposition cannot be obtained
by the use of the above relations. It is necessary
to generalize this procedure by using
the expansion of a unit fraction in the sum of
three unit fractions. Namely, we should apply the
relation
\begin{equation}
\frac{1}{p_1p_2p_3}=\frac{1}{p_1(p_1p_2+p_1p_3+p_2p_3)}+
\frac{1}{p_2(p_1p_2+p_1p_3+p_2p_3)}+
\frac{1}{p_3(p_1p_2+p_1p_3+p_2p_3)}\,.
\end{equation}

\begin{table}[!ht]
\scriptsize
\vspace{.05in}
\begin{center}
\begin{tabular}{|l|l||l|l||l|l|}
\hline
$1     $&$  (1/5, 1/5, 1/5, 1/5, 1/5)        %
$&$2   $&$  (1/6, 1/6, 1/6, 1/6, 1/3)        %
$&$3   $&$  (1/8, 1/8, 1/8, 1/8, 1/2)        $ \\ \hline
$4     $&$  (1/10, 1/10, 1/10, 1/5, 1/2)     %
$&$5   $&$  (1/9, 1/9, 1/9, 1/3, 1/3)        %
$&$6   $&$  (1/12, 1/12, 1/12, 1/4, 1/2)     $ \\ \hline
$7     $&$  (1/18, 1/18, 1/18, 1/3, 1/2)     %
$&$8   $&$  (1/8, 1/8, 1/4, 1/4, 1/4)        %
$&$9   $&$  (1/12, 1/12, 1/6, 1/6, 1/2)      $ \\ \hline
$10    $&$  (1/12, 1/12, 1/6, 1/3, 1/3)      %
$&$11  $&$  (1/16, 1/16, 1/8, 1/4, 1/2)      %
$&$12  $&$  (1/24, 1/24, 1/12, 1/3, 1/2)     $ \\ \hline
$13    $&$  (1/12, 1/12, 1/4, 1/4, 1/3)      %
$&$14  $&$  (1/15, 1/15, 1/5, 1/3, 1/3)      %
$&$15  $&$  (1/30, 1/30, 1/10, 1/3, 1/2)     $ \\ \hline
$16    $&$  (1/20, 1/20, 1/5, 1/5, 1/2)      %
$&$17  $&$  (1/24, 1/24, 1/6, 1/4, 1/2)      %
$&$18  $&$  (1/36, 1/36, 1/9, 1/3, 1/2)      $ \\ \hline
$19    $&$  (1/24, 1/24, 1/4, 1/3, 1/3)      %
$&$20  $&$  (1/48, 1/48, 1/8, 1/3, 1/2)      %
$&$21  $&$  (1/40, 1/40, 1/5, 1/4, 1/2)      $ \\ \hline
$22    $&$  (1/84, 1/84, 1/7, 1/3, 1/2)      %
$&$23  $&$  (1/14, 1/7, 1/7, 1/7, 1/2)       %
$&$24  $&$  (1/12, 1/6, 1/6, 1/4, 1/3)       $ \\ \hline
$25    $&$  (1/20, 1/10, 1/10, 1/4, 1/2)     %
$&$26  $&$  (1/30, 1/15, 1/15, 1/3, 1/2)     %
$&$27  $&$  (1/12, 1/6, 1/4, 1/4, 1/4)       $ \\ \hline
$28    $&$  (1/18, 1/9, 1/6, 1/6, 1/2)       %
$&$29  $&$  (1/18, 1/9, 1/6, 1/3, 1/3)       %
$&$30  $&$  (1/24, 1/12, 1/8, 1/4, 1/2)      $ \\ \hline
$31    $&$  (1/36, 1/18, 1/12, 1/3, 1/2)     %
$&$32  $&$  (1/28, 1/14, 1/7, 1/4, 1/2)      %
$&$33  $&$  (1/30, 1/15, 1/5, 1/5, 1/2)      $ \\ \hline
$34    $&$  (1/36, 1/18, 1/6, 1/4, 1/2)      %
$&$35  $&$  (1/54, 1/27, 1/9, 1/3, 1/2)      %
$&$36  $&$  (1/36, 1/18, 1/4, 1/3, 1/3)      $ \\ \hline
$37    $&$  (1/72, 1/36, 1/8, 1/3, 1/2)      %
$&$38  $&$  (1/60, 1/30, 1/5, 1/4, 1/2)      %
$&$39  $&$  (1/126, 1/63, 1/7, 1/3, 1/2)     $ \\ \hline
$40    $&$  (1/15, 1/5, 1/5, 1/5, 1/3)       %
$&$41  $&$  (1/21, 1/7, 1/7, 1/3, 1/3)       %
$&$42  $&$  (1/42, 1/14, 1/14, 1/3, 1/2)     $ \\ \hline
$43    $&$  (1/24, 1/8, 1/6, 1/6, 1/2)       %
$&$44  $&$  (1/24, 1/8, 1/6, 1/3, 1/3)       %
$&$45  $&$  (1/48, 1/16, 1/12, 1/3, 1/2)     $ \\ \hline
$46    $&$  (1/30, 1/10, 1/6, 1/5, 1/2)      %
$&$47  $&$  (1/24, 1/8, 1/4, 1/4, 1/3)       %
$&$48  $&$  (1/30, 1/10, 1/5, 1/3, 1/3)      $ \\ \hline
$49    $&$  (1/60, 1/20, 1/10, 1/3, 1/2)     %
$&$50  $&$  (1/48, 1/16, 1/6, 1/4, 1/2)      %
$&$51  $&$  (1/72, 1/24, 1/9, 1/3, 1/2)      $ \\ \hline
$52    $&$  (1/48, 1/16, 1/4, 1/3, 1/3)      %
$&$53  $&$  (1/96, 1/32, 1/8, 1/3, 1/2)      %
$&$54  $&$  (1/168, 1/56, 1/7, 1/3, 1/2)     $ \\ \hline
$55    $&$  (1/36, 1/9, 1/9, 1/4, 1/2)       %
$&$56  $&$  (1/20, 1/5, 1/4, 1/4, 1/4)       %
$&$57  $&$  (1/40, 1/10, 1/8, 1/4, 1/2)      $ \\ \hline
$58    $&$  (1/60, 1/15, 1/12, 1/3, 1/2)     %
$&$59  $&$  (1/60, 1/15, 1/6, 1/4, 1/2)      %
$&$60  $&$  (1/60, 1/15, 1/4, 1/3, 1/3)      $ \\ \hline
$61    $&$  (1/120, 1/30, 1/8, 1/3, 1/2)     %
$&$62  $&$  (1/100, 1/25, 1/5, 1/4, 1/2)     %
$&$63  $&$  (1/45, 1/9, 1/5, 1/3, 1/3)       $ \\ \hline
$64    $&$  (1/90, 1/18, 1/10, 1/3, 1/2)     %
$&$65  $&$  (1/60, 1/12, 1/5, 1/5, 1/2)      %
$&$66  $&$  (1/120, 1/24, 1/5, 1/4, 1/2)     $ \\ \hline
$67    $&$  (1/78, 1/13, 1/13, 1/3, 1/2)     %
$&$68  $&$  (1/42, 1/7, 1/6, 1/6, 1/2)       %
$&$69  $&$  (1/42, 1/7, 1/6, 1/3, 1/3)       $ \\ \hline
$70    $&$  (1/84, 1/14, 1/12, 1/3, 1/2)     %
$&$71  $&$  (1/84, 1/14, 1/6, 1/4, 1/2)      %
$&$72  $&$  (1/126, 1/21, 1/9, 1/3, 1/2)     $ \\ \hline
$73    $&$  (1/84, 1/14, 1/4, 1/3, 1/3)      %
$&$74  $&$  (1/168, 1/28, 1/8, 1/3, 1/2)     %
$&$75  $&$  (1/294, 1/49, 1/7, 1/3, 1/2)     $ \\ \hline
$76     $&$ (1/336, 1/48, 1/7, 1/3, 1/2)     %
$&$77   $&$ (1/72, 1/9, 1/8, 1/4, 1/2)       %
$&$78   $&$ (1/216, 1/27, 1/8, 1/3, 1/2)     $ \\ \hline
$79     $&$ (1/180, 1/20, 1/9, 1/3, 1/2)     %
$&$80   $&$ (1/70, 1/7, 1/7, 1/5, 1/2)       %
$&$81   $&$ (1/110, 1/11, 1/5, 1/5, 1/2)     $ \\ \hline
$82     $&$ (1/220, 1/22, 1/5, 1/4, 1/2)     %
$&$83   $&$ (1/60, 1/5, 1/5, 1/4, 1/3)       %
$&$84   $&$ (1/156, 1/13, 1/12, 1/3, 1/2)    $ \\ \hline
$85     $&$ (1/156, 1/13, 1/6, 1/4, 1/2)     %
$&$86   $&$ (1/156, 1/13, 1/4, 1/3, 1/3)     %
$&$87   $&$ (1/312, 1/26, 1/8, 1/3, 1/2)     $   \\ \hline
$88     $&$ (1/140, 1/10, 1/7, 1/4, 1/2)     %
$&$89   $&$ (1/630, 1/45, 1/7, 1/3, 1/2)     %
$&$90   $&$ (1/120, 1/8, 1/6, 1/5, 1/2)      $ \\ \hline
$91     $&$ (1/120, 1/8, 1/5, 1/3, 1/3)      %
$&$92   $&$ (1/240, 1/16, 1/10, 1/3, 1/2)    %
$&$93   $&$ (1/342, 1/19, 1/9, 1/3, 1/2)     $ \\ \hline
$94     $&$ (1/420, 1/21, 1/5, 1/4, 1/2)     %
$&$95   $&$ (1/924, 1/44, 1/7, 1/3, 1/2)     %
$&$96   $&$ (1/600, 1/25, 1/8, 1/3, 1/2)     $ \\ \hline
$97     $&$ (1/1806, 1/43, 1/7, 1/3, 1/2)    %
$&$98   $&$ (1/6, 1/6, 1/6, 1/4, 1/4)        %
$&$99   $&$ (1/9, 1/9, 1/9, 1/6, 1/2)        $ \\ \hline
$100    $&$ (1/21, 1/21, 1/14, 1/3, 1/2)     %
$&$101  $&$ (1/15, 1/15, 1/6, 1/5, 1/2)      %
$&$102  $&$ (1/12, 1/8, 1/8, 1/6, 1/2)       $ \\ \hline
$103    $&$ (1/12, 1/8, 1/8, 1/3, 1/3)       %
$&$104  $&$ (1/24, 1/16, 1/16, 1/3, 1/2)     %
$&$105  $&$ (1/18, 1/12, 1/9, 1/4, 1/2)      $ \\ \hline
$106    $&$ (1/15, 1/10, 1/6, 1/6, 1/2)      %
$&$107  $&$ (1/15, 1/10, 1/6, 1/3, 1/3)      %
$&$108  $&$ (1/30, 1/20, 1/12, 1/3, 1/2)     $ \\ \hline
$109    $&$ (1/33, 1/22, 1/11, 1/3, 1/2)     %
$&$110  $&$ (1/30, 1/20, 1/6, 1/4, 1/2)      %
$&$111  $&$ (1/45, 1/30, 1/9, 1/3, 1/2)      $ \\ \hline
$112    $&$ (1/30, 1/20, 1/4, 1/3, 1/3)      %
$&$113  $&$ (1/60, 1/40, 1/8, 1/3, 1/2)      %
$&$114  $&$ (1/105, 1/70, 1/7, 1/3, 1/2)     $ \\ \hline
$115    $&$ (1/10, 1/5, 1/5, 1/4, 1/4)       %
$&$116  $&$ (1/18, 1/9, 1/4, 1/4, 1/3)       %
$&$117  $&$ (1/20, 1/8, 1/8, 1/5, 1/2)       $ \\ \hline
$118    $&$ (1/35, 1/14, 1/5, 1/5, 1/2)      %
$&$119  $&$ (1/70, 1/28, 1/5, 1/4, 1/2)      %
$&$120  $&$ (1/21, 1/7, 1/7, 1/6, 1/2)       $ \\ \hline
$121    $&$ (1/42, 1/12, 1/7, 1/4, 1/2)      %
$&$122  $&$ (1/189, 1/54, 1/7, 1/3, 1/2)     %
$&$123  $&$ (1/99, 1/22, 1/9, 1/3, 1/2)      $ \\ \hline
$124    $&$ (1/45, 1/9, 1/6, 1/5, 1/2)       %
$&$125  $&$ (1/42, 1/7, 1/4, 1/4, 1/3)       %
$&$126  $&$ (1/483, 1/46, 1/7, 1/3, 1/2)     $ \\ \hline
$127    $&$ (1/231, 1/14, 1/11, 1/3, 1/2)    %
$&$128  $&$ (1/8, 1/8, 1/6, 1/4, 1/3)        %
$&$129  $&$ (1/20, 1/20, 1/15, 1/3, 1/2)     $ \\ \hline
$130    $&$ (1/28, 1/21, 1/12, 1/3, 1/2)     %
$&$131  $&$ (1/28, 1/21, 1/6, 1/4, 1/2)      %
$&$132  $&$ (1/28, 1/21, 1/4, 1/3, 1/3)      $ \\ \hline
$133    $&$ (1/56, 1/42, 1/8, 1/3, 1/2)      %
$&$134  $&$ (1/10, 1/6, 1/5, 1/5, 1/3)       %
$&$135  $&$ (1/20, 1/12, 1/6, 1/5, 1/2)      $ \\ \hline
$136    $&$ (1/20, 1/12, 1/5, 1/3, 1/3)      %
$&$137  $&$ (1/40, 1/24, 1/10, 1/3, 1/2)     %
$&$138  $&$ (1/140, 1/60, 1/7, 1/3, 1/2)     $ \\ \hline
$139    $&$ (1/88, 1/33, 1/8, 1/3, 1/2)      %
$&$140  $&$ (1/20, 1/6, 1/5, 1/4, 1/3)       %
$&$141  $&$ (1/238, 1/51, 1/7, 1/3, 1/2)     $ \\ \hline
$142    $&$ (1/110, 1/15, 1/11, 1/3, 1/2)    %
$&$143  $&$ (1/15, 1/12, 1/10, 1/4, 1/2)     %
$&$144  $&$ (1/45, 1/36, 1/5, 1/4, 1/2)      $ \\ \hline
$145    $&$ (1/15, 1/10, 1/4, 1/4, 1/3)      %
$&$146  $&$ (1/91, 1/78, 1/7, 1/3, 1/2)      %
$&$147  $&$ (1/35, 1/15, 1/14, 1/3, 1/2)     $  \\
\hline
\end{tabular}
\end{center}
\caption{The 147 simply--laced number for $n=5$ written in the form of
unit fractions.}
\label{Table147b}
\end{table}

The generalization to a higher number of terms
in the sum is trivial and one can verify that indeed the
simply--laced numbers for $n=5$ written in Table~\ref{Table147b} in the
form of the unit fractions can
be obtained by this method from the corresponding expansions for
$n=2,3,4$.

It is important to note that among all unit decompositions for each
$n$ there is a decomposition with the maximal denominator
corresponding to the maximal dimension $d(n)=\max d_k$.
This number satisfies the simple recurrence relation
\begin{equation}
d(n+1)=d(n)\,(d(n)+1)\,.
\end{equation}
It grows very rapidly
\begin{equation}
d(1)=1\,,\,\,d(2)=2\,,\,\,d(3)=6\,,\,\,d(4)=42\,,\,\,
d(5)=1806\,...\,.
\end{equation}
Note that the numbers $d(n)+1$ are not  prime
because $1807 = 13 \cdot 139$.

The above recurrence relation can be written as the finite difference
equation
\begin{equation}
d(n+1)-d(n)=(d(n))^2\,.
\end{equation}
Note, that it can not be substituted by the differential equation
$\widetilde{d}'=\widetilde{d}^2$ for
large $n$, because its solution in this case would have a pole
$\widetilde{d}=1/(n_0-n)$ absent in the solution of the recurrence
relation.

\section{A case study: The $B(01111)$ Berger Graph}
\label{CaseStudy}

\begin{figure}
\vspace{-2cm}
\hspace{1cm}
\centerline{\hspace{2cm}\epsfig{file=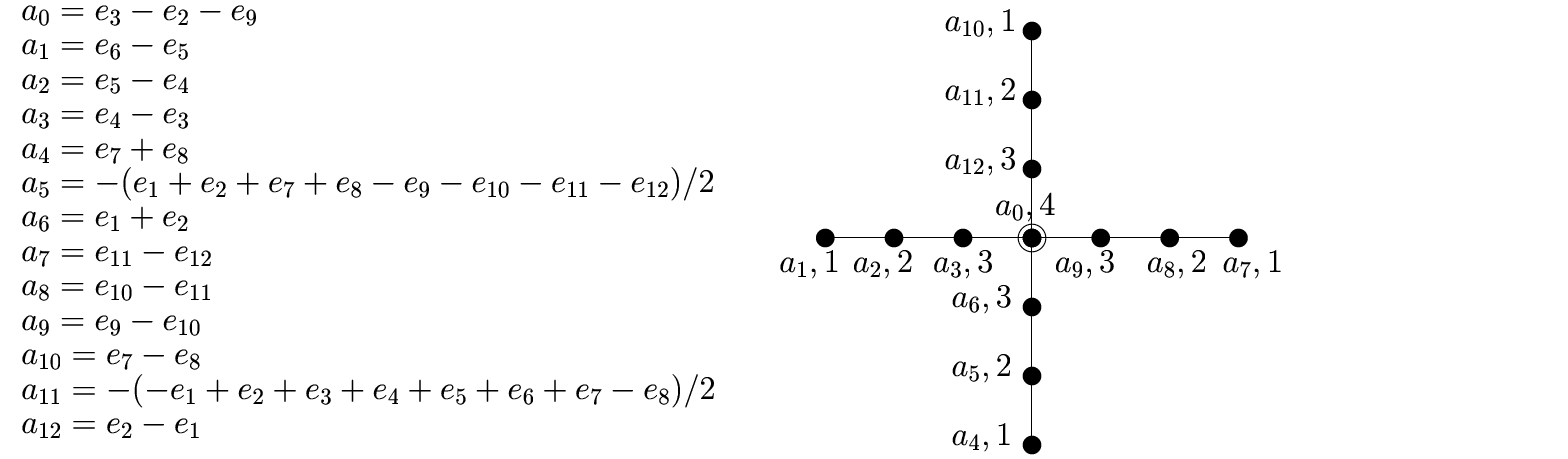,width=14cm}}
\vspace{-0.1cm}
\caption{Berger graph for the B(01111) vector with Coxeter labels and 
simple roots in orthogonal basis.}
\label{01111}
\end{figure}

For illustrative purposes we now consider the affine Berger graph $B(01111)$
as a case study among the 14 exceptional simply--laced graphs. This graph can 
be found  in the four dimensional
polyhedron determined by the reflexive number ${\vec k}_5=(1,1,1,1,1)[5]$,
constructed by 2--arity from the two simply--laced numbers
 ${\vec k}_4=(1,1,1,1)[4]$ and ${\vec k}_1=(1)[1]$.  In this case the
Berger graph $B(01111)$ coincides with the primary $B(1111)$ graph
(see Fig.~\ref{01111}).
This graph looks like a natural generalization of the extended Dynkin graph
for $E_6$. It is then possible to propose the next generalizations of
such graphs in dimensions $n = 6, 7, ...$. This proposal can be confirmed directly
by looking for these graphs in the polyhedra determined by the 2--arity
expansion:  ${\vec k}_{n+1}=(1,...,1)[n+1]=(1,...,1,0)_{n+1}+
(0,...,0,1)_{n+1}$.
Thus the reflexive simply--laced numbers $(1,1,...,1)[n]$ determine the set of
primary graphs, which have the form of stars with $n$--legs each of them
having $(n-1)$ nodes, with Coxeter labels $1, 2, ..., (n-1)$. The center of
the star has the maximal Coxeter label equals to $n$.

\begin{figure}
\vspace{-2cm}
\centerline{\hspace{2cm}\epsfig{file=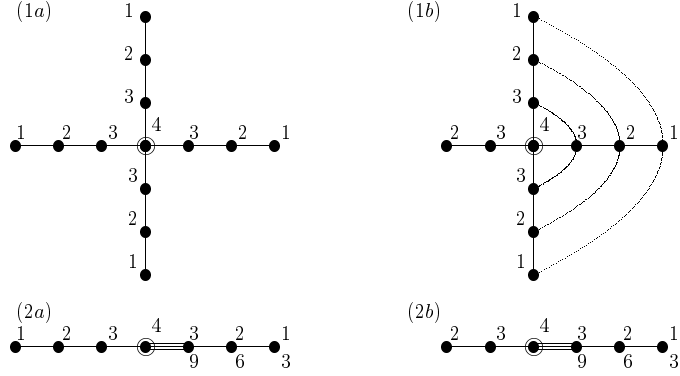,width=14cm}}
\caption{Relation between symmetric simply--laced  B(01111) and
non--simply--laced graphs.}
\label{01111SL}
\end{figure}

On such a case it is possible to extract the full information from
the affine graph, {\it i.e.}, Berger matrices for affine and non-affine cases, 
Coxeter labels, simple roots, fundamental weights, etc.
The corresponding affine Berger matrix can be constructed by the canonical way:
{\small \begin{eqnarray}
B{(01111)}= \left (
\begin{array}{ccc|ccc|ccc|ccc||c}
 2 &-1 & 0   & 0 & 0 & 0   & 0 & 0 & 0   & 0 & 0 & 0  &-1 \\
-1 & 2 &-1   & 0 & 0 & 0   & 0 & 0 & 0   & 0 & 0 & 0  & 0 \\
 0 & -1& 2   & 0 & 0 & 0   & 0 & 0 & 0   & 0 & 0 & 0  & 0 \\
\hline
 0 & 0 & 0  & 2 &-1 & 0    & 0 & 0 & 0   & 0 & 0 & 0  &-1 \\
 0 & 0 & 0  &-1 & 2 &-1    & 0 & 0 & 0   & 0 & 0 & 0  & 0 \\
 0 & 0 & 0  & 0 &-1 & 2    & 0 & 0 & 0   & 0 & 0 & 0  & 0 \\
\hline
 0 & 0 & 0  & 0 & 0 & 0    & 2 &-1 & 0   & 0 & 0 & 0  &-1 \\
 0 & 0 & 0  & 0 & 0 & 0    &-1 & 2 &-1   & 0 & 0 & 0  & 0 \\
 0 & 0 & 0  & 0 & 0 & 0    & 0 & 1 & 2   & 0 & 0 & 0  & 0 \\
\hline
 0 & 0 & 0  & 0 & 0 & 0    & 0 & 0 & 0   & 2 &-1 & 0  &-1 \\
 0 & 0 & 0  & 0 & 0 & 0    & 0 & 0 & 0   &-1 & 2 &-1  & 0 \\
 0 & 0 & 0  & 0 & 0 & 0    & 0 & 0 & 0   & 0 &-1 & 2  & 0 \\
\hline\hline
-1 & 0 & 0  &-1 & 0 & 0    &-1 & 0 & 0   &-1 & 0 & 0  & 3 \\
\end{array}
\right )\nonumber
\end{eqnarray}}\\
The generalization of this graph $B(01111)$ to $B((01...1)_n)$ is 
straightforward. Instead of the maximal diagonal element being $B_{ii}=3$ one
 can change it to $B_{ii}=n$. For all these cases the determinant cancels. 
The ``simple roots'' for this graph in an orthonormal basis 
$\{e_i\}, i=1,\ldots, 12$ are written in Fig~\ref{01111}.
They obey our restrictions:
\begin{eqnarray}
<\hat \alpha_{0} \cdot \hat \alpha_{0}> &=&3,  \nonumber\\
<\alpha_{i} \cdot \alpha_{i}> &=&2, \qquad i =1,2,3,\ldots 12,\nonumber\\
<\alpha_{i} \cdot \alpha_{j}> &=&-1, \qquad j=i\pm 1, \nonumber\\
<\alpha_{i} \cdot \alpha_{j}> &=&0, \qquad |j-i| >1.
\end{eqnarray}
The following linear combination of the ``roots'' satisfies  the
affine condition:
\begin{eqnarray}
4 \hat \alpha_{0}+ 3 \alpha_{a1}     + 2 \alpha_{a2}  +
\alpha_{a3}  + 3 \alpha_{b1}     + 2 \alpha_{b2}  +
\alpha_{b3}  + 3 \alpha_{c1}     + 2 \alpha_{c2}  +
\alpha_{c3}  + 3 \alpha_{d1}    + 2 \alpha_{d2} +    \alpha_{d3}
= 0. \nonumber
 \end{eqnarray}
The last affine link gives a possibility to find a non--affine  graph 
from an affine Berger graph. For this it is enough to remove one zero 
node with Coxeter label 1.
Correspondingly, removing, for example, the first row and column from
the affine Berger matrix one can obtain
a non--affine Berger matrix having the
determinant equal to 16. One can write this number as
$4\times 4$, where one coefficient $4$ could be related to the number
of nodes having Coxeter label one plus 1, and the other coefficient 4
could be
related to a ${\mathbb Z}_4$ symmetry of the affine Berger graph.
Let us recall that in the Cartan--Lie case the determinant of Cartan simply--laced
matrices is equal to the number of non--equivalent representations.
For example, in the  $A_r$, $D_r$, $E_6$, $E_7$, $E_8$ cases the number of
non--equivalent representations is $r+1$, $4$, $3$, $2$,$1$, respectively.
For any $n$ of the non--affine $B(01...1)$ matrix its determinant is equal
to $n^{n-2}$, $n>3$.
In the Cartan case from the inverse Cartan matrix one can  find the set of
$r$ fundamental weights.
Also, from the non--affine  Berger matrix one can obtain all fundamental weights:
{\small \begin{eqnarray}
G= \left (
\begin{array}{c||cc|ccc|ccc|ccc||c}
 F.W.     &\alpha_{a1}&\alpha_{a2}&\alpha_{b1}&\alpha_{b2}&
\alpha_{b3}&\alpha_{c1}&\alpha_{c2}&
\alpha_{c3}&\alpha_{d1}&\alpha_{d2}& \alpha_{d3}&\alpha_{0}
\\\hline\hline
\Lambda_{a1}&6 & 3 & 6    & 4   & 2   & 6    & 4   & 2   & 6    & 4   & 2   & 8 \\
\Lambda_{a2}&3 & 2 & 3    & 2   & 1   & 3    & 2   & 1   & 3    &
2   & 1   & 4 \\\hline
\Lambda_{b1}&6 & 3 & 15/2 & 5   & 5/2 & 27/4 & 9/2 & 9/4 & 27/4 & 9/2 & 9/4 & 9 \\
\Lambda_{b2}&4 & 2 & 5    & 4   & 2   & 9/2  & 3   & 3/2 & 9/2  & 3   & 3/2 & 6 \\
\Lambda_{b3}&2 & 1 & 5/2  & 2   & 3/2 & 9/4  & 3/2 & 3/4 & 9/4  &
3/2 & 3/4 & 3 \\\hline
\Lambda_{c1}&6 & 3 & 27/4 & 9/2 & 9/4 & 15/2 & 5   & 5/2 & 27/4 & 9/2 & 9/4 & 9\\
\Lambda_{c2}&4 & 2 & 9/2  & 3   & 3/2 & 5    & 4   & 2   & 9/2  & 3   & 3/2 & 6\\
\Lambda_{c3}&2 & 1 & 9/4  & 3/2 & 3/4 & 5/2  & 2   & 3/2 & 9/4  &
3/2 & 3/4 & 3 \\\hline
\Lambda_{d1}&6 & 3 & 27/4 & 9/2 & 9/4 & 27/4 & 9/2 & 9/4 & 15/2 & 5   & 5/2 & 9\\
\Lambda_{d2}&4 & 2 & 9/2  & 3   & 3/2 & 9/2  & 3   & 3/2 & 5    & 4   & 2   & 6\\
\Lambda_{d3}&2 & 1 & 9/4  & 3/2 & 3/4 & 9/4  & 3/2 & 3/4 & 5/2  &
2   & 3/2 & 3 \\\hline\hline
\Lambda_{0} &8 & 4 & 9    & 6   & 3   & 9    & 6   & 3   & 9    & 6   & 3   & 12\\
\end{array}
\right ) \nonumber
\end{eqnarray}}
Finally we briefly discuss the non--simply--laced case.
Apart from the calculation of the reflexive numbers
and the particular case of the quasi--simply--laced ones,
one can attempt to construct  the non--simply--laced graphs from the
above--investigated simply--laced graphs. For this purpose we shall return
again  to
the Cartan--Lie algebras and use the known method
to construct the root system of the non--simply--laced  $B_r-$, $C_r-$,
$F_4-$, $G_2-$ algebras from the simply--laced
$D_{r+1}$,$A_{2r-1}$, $E_6$,$D_4$ root systems,
respectively. The root systems of these simply--laced
algebras have the following diagram automorphisms $f$:
\begin{enumerate}
\item{$D_{r+1}$: $f(\alpha_i)=\alpha_i$ for $1\leq r \leq r-1$,
$f(\alpha_r)=\alpha_{r+1}$, $ f(\alpha_{r+1})=\alpha_r$: $D_{r+1}\rightarrow B_r$ }
\item{$A_{2r-1}$: $f(\alpha_i)=\alpha_{2r-i}$,
$f(\alpha_{2r-i})=\alpha_{i}$ ,
 for $1\leq i \leq r-1$,  and $f(\alpha_r)=\alpha_{r}$:  $A_{2r-1} \rightarrow C_r$;}
 \item{$E_6$:  $f(\alpha_1)=\alpha_{6}$, $f(\alpha_2)=\alpha_{5}$,
 $f(\alpha_3)=\alpha_{3}$, $f(\alpha_5)=\alpha_{2}$, $f(\alpha_6)=\alpha_{1}$,
 $f(\alpha4)=\alpha_{4}$:
 $E_6 \rightarrow F_4$;}
\item{$D_4$: $f(\alpha1)=\alpha_{3}$, $f(\alpha_2)=\alpha_{2}$,
$f(\alpha_3)=\alpha_{4}$,
 $f(\alpha4)=\alpha_{1}$: $D_4 \rightarrow G_2$. }
\end{enumerate}
Let us remark that the automorphisms in the first three cases are of
order 2, and the last case is of order 3.
We can use a similar algorithm to build a non--simply--laced Berger graph from
the simply--laced B(01111) graph (see Fig.~\ref{01111SL}). The corresponding 
non--affine Berger graph has a ${\mathbb Z}_3$ symmetry. Acting as in the Cartan 
case and using this symmetry we obtain from the simply--laced Berger graph of 
rank 12  a new non--simply and non--affine graph with rank 6.
On Fig.~\ref{01111SL} three consecutive steps to construct a new affine
non--simply--laced graph are shown: The first step, $ 1a \rightarrow 1b$,
corresponds to removing in the affine B(01111) graph  one zero node having
the Coxeter label equal to 1. In the second step, $1b \rightarrow 2b$, one
obtains the non--simply--laced graph. The third step $2b \rightarrow 2a$
corresponds to adding the zero node with the Coxeter label 1.
From the resulting new non--simply--laced graph one can construct
the corresponding Berger matrix with the determinant equal to zero:
{\small \begin{eqnarray}
 && \left ( \begin{array}{ccccccc}
2&-1&0&0&0&0&0\\
-1&2&-1&0&0&0&0\\
0&-1&2&-1&0&0&0\\
0&0&-1&3&-3&0&0\\
0&0&0&-1&2&-1&0\\
0&0&0&0&-1&2&-1\\
0&0&0&0&0&-1&2\\
\end{array} \right )
\nonumber
\end{eqnarray}}\\
with the set of eigenvalues $\{\, 0,\, 2,\, 3,\, 2 - \sqrt{2},\, 2 + \sqrt{2}, 3- \sqrt{3},\,  3 + \sqrt{3} \,\}$.
Two zero eigenvectors corresponding to the Berger matrix $B_{ij}$
and its transposed matrix $B_{ji}$ give us two sets of Coxeter labels:
\begin{eqnarray}
{\vec C}&=& \{\,1,\, 2,\, 3,\, 4,\, 3,\, 2,\, 1\,\}, \,\,\,
\tilde{\vec C} ~=~ \{\,1,\, 2,\, 3,\, 4,\, 9,\, 6,\, 3\,\}.
\end{eqnarray}
The corresponding non--affine Berger
matrix has the determinant equal to 1 similar to the $G_2$ graph in the
Cartan case ($D_4 \rightarrow G_2$).
This value for the determinant agrees with our previous arguments, when we
discussed the determinants for non--affine $B(01111)$ matrices and
also for the Cartan simply--laced cases.

In this way, using the symmetry of the
$B(0,...0,1)$, $B(0,..0,1,1)$, $B(0,..,0,1,1,1)$, $B(0,..,0,1,1,2)$, $B(0,..,0,1,2,3)$,
$B(0,1,1,2,2)$, $B(0,1,1,1,3)$, $B(0,1,2,2,5)$, $B(0,2,3,3,4)$ and
$B(0,1,3,3,4)$ Berger graphs
one can get new infinite series and some exceptional non--simply--laced
graphs. From this construction one can see that the generation of non--simply--laced Berger graphs takes place as in the $K3$ case. There we had just $A$--$D$--$E$ types of singularities but in $K3$ polyhedra one can also obtain 
non--simply--laced Dynkin graphs~\cite{CF,Bersh,AENV1}.  Moreover, in the 
arity--dimension approach in the $K3$ case we used only simply--laced
numbers $(1)[1]$, $(1,1)[2]$, $(1,1,1)[3]$, $(1,1,2)[4]$, $(1,2,3)[6]$
 but we also have non--simply--laced Dynkin diagrams. In this approach the 
non--simply--laced graphs appear as subgraphs of simply--laced ones.
To understand this more deeply it would be needed to use some additional 
dynamics~\cite{PS}. However in CY$_d$ cases with $d>2$ we have a new principle 
as we have briefly discussed here: We should take into consideration non only 
simply--laced numbers but also the set of non--simply--laced ones.